\documentclass[11pt]{article}

\usepackage[final]{acl}

\usepackage{times}
\usepackage{latexsym}

\usepackage[T1]{fontenc}
\usepackage[utf8]{inputenc}

\usepackage{microtype}

\usepackage{inconsolata}

\usepackage{graphicx}
\usepackage[table]{xcolor} 
\usepackage{amsthm}

\usepackage{graphicx}
\usepackage{xcolor}
\usepackage{mdframed}
\usepackage[most]{tcolorbox}
\usepackage{subcaption} % 用于插入子图
\usepackage{hyperref}
\usepackage{booktabs}
\usepackage{algorithm}
\usepackage[noend]{algorithmic}
\usepackage{amssymb}
\usepackage{CJKutf8}
\usepackage{colortbl}
\usepackage{multirow} 
\usepackage{amsmath}
\usepackage{makecell}
\usepackage{geometry}
\usepackage{forest} % <-- 树状图核心包
\usepackage{tikz}
\usepackage{longtable}
\usepackage{pifont} 
\usepackage{graphicx}
\usepackage{tabularx}
\definecolor{blueviolet}{RGB}{86, 49, 148}  % 自己可以微调
\definecolor{StrongOrange}{RGB}{230,120,0} % 可微调亮度/饱和度
\newcommand{\orangetext}[1]{\textcolor{StrongOrange}{#1}}
\definecolor{StrongBlue}{RGB}{0,92,184}
\newcommand{\bluetext}[1]{\textcolor{StrongBlue}{#1}}
\definecolor{StrongGray}{RGB}{130,130,130}  % 深灰稳定可读
\newcommand{\graytext}[1]{\textcolor{StrongGray}{#1}}
\definecolor{StrongGreen}{RGB}{0,150,0}
\definecolor{LightBlue}{RGB}{216,235,255}
\definecolor{LightOrange}{RGB}{255,230,204}
\definecolor{LightGray}{RGB}{240,240,240}
\definecolor{verylightyellow}{RGB}{255, 250, 205}

\newcommand{\cmark}{\ding{51}} % ✓
\newcommand{\xmark}{\ding{55}} % ✗
\newcommand{\tmark}{\ding{51}\rotatebox[origin=c]{0}{\kern-0.65em\ding{55}}} % half

\definecolor{verylightgreennew}{RGB}{220,255,220}
\definecolor{verylightrednew}{RGB}{255,230,230}
\definecolor{verylightreddarker}{HTML}{FFCBCB}
\definecolor{greencm}{RGB}{0,153,0} % 供 \cm 使用的绿色（你也可改为想要的绿）

\newcommand{\cm}{{\color{greencm}\normalsize\cmark}}                 % 绿色 ✓
\newcommand{\xm}{{\color{verylightreddarker}\normalsize\xmark}}      % 偏红 ✗
\providecommand{\cellyes}{%
  \cm
  \cellcolor{verylightgreennew}%
}
\providecommand{\cellno}{%
  \xm
  \cellcolor{verylightrednew}%
}

\definecolor{myred}{RGB}{160, 0, 0} 
\newif\ifshowred
\showredfalse   % 想关掉就改成 \showredfalse
\newcommand{\rev}[1]{\ifshowred\textcolor{myred}{#1}\else#1\fi}

\title{A Comprehensive Survey on Linguistic Steganography: \\ Methods, Countermeasures, Evaluation, and Challenges}

\author{
  Ruiyi Yan\textsuperscript{1} \quad
  Chenhui Chu\textsuperscript{1} \quad
  Zhongliang Yang\textsuperscript{2} \quad
  Yugo Murawaki\textsuperscript{1} \\[6pt]
  \textsuperscript{1}Graduate School of Informatics, Kyoto University \\
  \textsuperscript{2}School of Cyberspace Security, Beijing University of Posts and Telecommunications \\[4pt]
  \texttt{ruiyi@nlp.ist.i.kyoto-u.ac.jp}, \texttt{chu@i.kyoto-u.ac.jp} \\
  \texttt{yangzl@bupt.edu.cn}, \texttt{murawaki@i.kyoto-u.ac.jp}
}

\begin{document}
\maketitle
% \begin{abstract}
%     Linguistic steganography, which hides secret messages in natural language text, plays an important role in information protection. Historically, its development was constrained by limited embedding capacity and text quality, but large language models (LLMs) have brought dramatic advances.
%     Despite this progress, there is still a lack of a comprehensive survey that systematically covers linguistic steganography in the LLM era.
%     In this work, we present a comprehensive survey motivated by \textbf{LLM-era} developments, covering: (1) 148 modification-based and generative steganographic methods, (2) 60 countermeasures, i.e., linguistic steganalysis methods, (3) 23 evaluation metrics, and (4) 9 open challenges with future directions. For each of these four areas, we provide taxonomies, summaries, reviews, and adoption analyses.
%     This survey aims to provide a systematic understanding of linguistic steganography in the LLM era and to guide future research toward more practical system designs.
% \end{abstract}
\begin{abstract}
\rev{Linguistic steganography hides secret messages in natural language text. Large language models (LLMs) have reshaped the field, but a systematic account of how these scattered advances collectively reshape the field in this new era is still missing.
We provide one along four axes: 148~steganographic methods, 60~linguistic steganalysis countermeasures, 23~evaluation metrics, and 9~open challenges, each with taxonomies, reviews, and adoption analyses.
Cutting across these axes, we identify five specific \textbf{paradigm shifts in the LLM era}: (1) from covertext modification to prompt-only generation, (2) from heuristic to provable security, (3) from white-box symmetric LMs to black-box or asymmetric access, (4) from security-centric designs to joint optimization, and (5) from text-quality concerns to engineering issues. 
The survey aims to serve as both a reference and a roadmap for practical and responsible linguistic steganography in the LLM era.}
\end{abstract}

\section{Introduction}

% In recent decades, secure communication has been essential for protecting sensitive information and resisting surveillance~\cite{ sutton2002secure, 10.1145/2382196.2382212}. 
% Cryptography and encryption systems are traditionally and widely used to safeguard messages by transforming them into unreadable forms that can only be deciphered with the proper key~\cite{bhanot2015review, aumasson2024serious}.
% However, data transmitted by these tools can be easily identifiable as encrypted communication, which can be detected and likely blocked in repressive environments~\cite{ 10.1145/3460120.3484550}, because such communication can be deemed suspicious.

% To achieve both confidentiality and indistinguishability, researchers have explored alternative methods of covert communication that conceal not only the content of a message but also its existence. This pursuit has led to \textbf{steganography}, the practice of embedding hidden information within seemingly harmless carriers~\cite{7164679}.
\rev{While cryptography conceals the content of a message, encrypted traffic itself remains identifiable and can be blocked in repressive environments~\citep{10.1145/3460120.3484550}. Steganography instead conceals the \emph{existence} of the message by embedding it in innocuous carriers~\citep{7164679}.
% The roots of steganography can be traced back to antiquity, where civilizations such as the Greeks and Persians devised ingenious methods like carving messages on wooden tablets covered with wax or tattooing words on the scalp of messengers and waiting for the hair to regrow~\cite{10.5555/647594.728895}.
% During the medieval and Renaissance periods, written language became a common vehicle for hidden communication, with techniques ranging from invisible inks to acrostics~\cite{4655281}.
In the modern era, with the advancements of digital technology, steganography has been commonly based on digital carriers, including images~\cite{9335027}, audio~\cite{djebbar2012comparative}, video~\cite{LIU2019238}, and text~\cite{math9212829}. 
% Besides, steganographic techniques are also used in anonymous networks~\cite{269582, 601314}.
% , such as ScrambleSuit/obfs4~\cite{10.1145/2517840.2517856} and SkypeMorph~\cite{10.1145/2382196.2382210}.
% , StegoTorus~\cite{10.1145/2382196.2382211}, TapDance~\cite{266504, 184509}, and Format-Transforming Encryption~\cite{10.1145/2508859.2516657}.
}

\begin{figure}[!t]
    \centering
    % 第一个子图
    \begin{subfigure}[b]{0.48\textwidth}
        \centering
        \includegraphics[width=\textwidth]{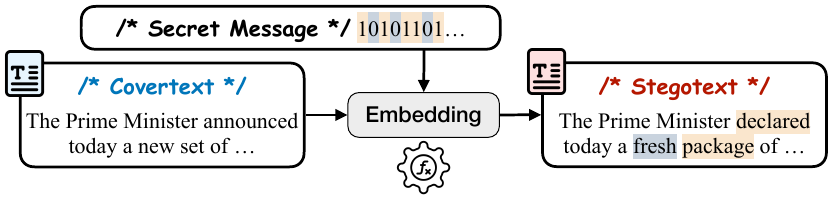}
        \caption{Modification-based steganography.}

        \label{fig:LS_modification}
    \end{subfigure}
    \begin{subfigure}[b]{0.48\textwidth}
        \centering
        \includegraphics[width=\textwidth]{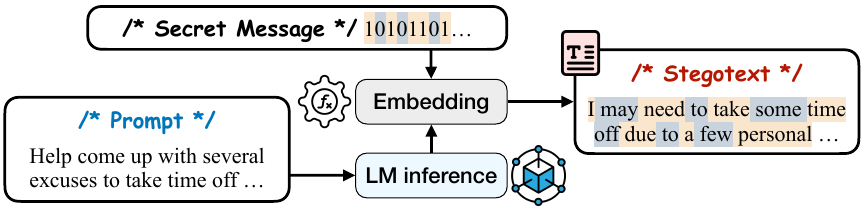}
        \caption{Generative steganography based on language model(s).}
        \label{fig:LS_generation}
    \end{subfigure}
    
    \caption{An overview of how to achieve linguistic steganography based on two main frameworks.}

    \label{fig: Linguistic_steganography}
\end{figure}

\begin{table*}[!t]
\renewcommand{\arraystretch}{1.0}
\centering
\scalebox{0.78}{
\begin{tabular}{m{3.3cm}|>{\centering\arraybackslash}m{1.2cm}|>{\centering\arraybackslash}m{1.2cm}|>{\centering\arraybackslash}m{3cm}|>{\centering\arraybackslash}m{1.2cm}|>{\centering\arraybackslash}m{2.7cm}|>{\centering\arraybackslash}m{1.2cm}|>{\centering\arraybackslash}m{2.7cm}}
\toprule[1.0pt]
\multirow{2}{*}{\textbf{Survey}} & \multirow{2}{*}{\textbf{Year}} & \multicolumn{2}{|c|}{\textbf{Steganographic methods}} &  \multicolumn{2}{|c|}{\textbf{Steganalysis methods}} & \multicolumn{2}{|c}{\textbf{Evaluation metrics}} \\ \cline{3-8}
 & & Number & Categories (L1/L2/L3) & Number  & Categories (L1/L2) & Number  & Categories (L1/L2)  \\
\midrule[1.0pt]
\citet{math9212829} & 2021 & 50 & 3 & -- & --  & --  & -- \\ \hline
\citet{xiang2022generative} & 2022 & 37  & 4/11/-- & --  & -- &  11 & 4/-- \\ \hline
\citet{10.1007/978-3-031-26876-2_61} & 2022 & 18  & --  & -- & -- & 12  & 3/--   \\ \hline
\citet{Wu2024} & 2023 & 28 & 2/4/8  & 29 & 2/7 & 8   & -- \\ \hline \rowcolor{blue!8} 
\textbf{Ours} & \textbf{2026} & \textbf{\rev{148}} & \textbf{2/6/17}  & \textbf{\rev{60}} & \textbf{2/8}  &  \textbf{23} & \textbf{3/6} \\ 
\bottomrule[1.0pt]
\end{tabular}}
\caption{Quantitative comparison of this survey with existing surveys, in which ``--'' indicates that relevant information on this dimension is not substantially or formally provided in the survey, and ``L1/L2/L3'' denotes the layer 1st/2nd/3rd for categorization.}
\label{table:Quantitative comparison of this survey with existing surveys}
\end{table*}

\begin{table*}[!t]
\renewcommand{\arraystretch}{1.0}
\centering

\scalebox{0.7}{
\begin{tabular}{m{3.3cm}|>{\centering\arraybackslash}m{2cm}|>{\centering\arraybackslash}m{1.8cm}|>{\centering\arraybackslash}m{2.3cm}|>{\centering\arraybackslash}m{2.7cm}|>{\centering\arraybackslash}m{2.7cm}|>{\centering\arraybackslash}m{2.2cm}|>{\centering\arraybackslash}m{2.2cm}}
\toprule[1.0pt]
 \textbf{Survey} & \textbf{Steganalysis taxonomy} &  \textbf{Evaluation taxonomy} & \textbf{Robustness consideration} & \textbf{Analyses of targeted metrics} & \textbf{Analyses of metric adoption} & \textbf{Analyses of steganalysis adoption}  &   \textbf{Features of the LLM era}    \\
\midrule[1.0pt]
\citet{math9212829} & \cellno & \cellno & \cellyes  & \cellyes & \cellno  & \cellno  & \cellno \\ \hline
\citet{xiang2022generative} & \cellno & \cellyes  & \cellno & \cellno & \cellno  & \cellno  & \cellno \\ \hline
\citet{10.1007/978-3-031-26876-2_61} & \cellno & \cellyes  & \cellno  & \cellno & \cellno  & \cellyes  & \cellno \\ \hline
\citet{Wu2024}& \cellyes  & \cellno & \cellno & \cellno & \cellno  & \cellno  & \cellno \\ \hline 
\textbf{Ours} & \cellyes  & \cellyes & \cellyes & \cellyes & \cellyes  & \cellyes  & \cellyes \\ 
\bottomrule[1.0pt]
\end{tabular}}
\caption{Qualitative comparison of this survey with existing surveys.}
\label{table:Qualitative comparison of this survey with existing surveys}
\end{table*}

\rev{
Among these steganographic techniques, \textbf{linguistic steganography} (currently the dominant category in text steganography) has attracted increasing research attention for several reasons:
(1) Textual content is ubiquitous on social networks~\cite{Aggarwal2011}.
(2) Recent advances in large language models (LLMs) have significantly improved the fluency, coherence, and naturalness of generated steganographic texts (\textit{stegotexts})~\cite{radford2019language, NEURIPS2020_1457c0d6, achiam2023gpt}.
(3) Compared to steganography based on images, audio, or video~\cite{650120}, linguistic steganography is often less affected by compression-related distortion during transmission~\cite{math9212829}.
Figure~\ref{fig: Linguistic_steganography} illustrates schematic diagrams of linguistic steganography, respectively, based on two primary frameworks: modification and generation. 
}

\rev{
In the era of large language models (LLMs), some influential open-weight models have been released~\cite{zhang2022optopenpretrainedtransformer, touvron2023llama}, which have demonstrated strong performance across a wide range of downstream tasks, thereby enabling new possibilities for linguistic steganography~\cite{10360487, 10.1007/978-981-97-9437-9_37}. 
However, a comprehensive survey of linguistic steganography in the LLM era is still missing. In particular, previous surveys~\cite{math9212829, xiang2022generative, 10.1007/978-3-031-26876-2_61, Wu2024} exhibit two main limitations: (1) the absence of LLM-era coverage and (2) coarse-grained, paper-narrow analyses.
}

\begin{figure*}[!t]
\centering
\begin{forest}
mytree
[ \textbf{A comprehensive survey on linguistic steganography}, rotate=90, child anchor=north, parent anchor=south, anchor=center
    [\textbf{Basics}: \S\ref{sec: Preliminaries and Background} Preliminaries \\ and background, text width=33mm
        [\S\ref{sec: Steganography System} Steganography system, text width=38mm
            [\S\ref{sec: Security of Steganography} Security of steganography, text width=40mm]
        ]
    ]
  [\textbf{Methods:} \S\ref{sec: Steganography Based on Modification} Steganography \\ based on modification, text width=38mm
    [Taxonomy (Figure~\ref{fig: The main content flow and categorization of steganographic methods based on modification}), text width=28mm
        [\S\ref{sec: Format-Based Methods}-\ref{sec: Rewriting-Based Methods} Detailed review, text width=34mm]
    ]
  ]
    [\textbf{Methods:} \S\ref{sec: Generative Linguistic Steganography} Generative \\ linguistic steganography, text width = 33mm
        [\S\ref{sec: Basics and notions of LM} Basics and \\ notions of LM, text width=21mm
            [\S\ref{sec: Paradigm of LM-based Steganography} Paradigm of LM-based \\ steganography (Figure~\ref{fig:GLS_framework}), text width=37mm]
        ]
        [\S\ref{sec: Taxonomy of Existing Methods} Taxonomy (Figure~\ref{fig: The main content flow and categorization of generative linguistic steganography}), text width=34mm
            [Analyses of methodology features (Table~\ref{table:GLS_taxonomy}), text width=57mm]
            [\S\ref{sec: Modern Generative Linguistic Steganographic Methods} Detailed review, text width=27mm]
        ]
    ]
  [\textbf{Countermeasures:} \\ \S\ref{sec: Linguistic Steganalysis}  Linguistic steganalysis, text width=34mm
    [
        \S\ref{sec: Basics and Notions of Steganalysis} Basics and Notions of Steganalysis, text width=53mm
    ]
    [\S\ref{sec: Taxonomy of steganalysis} Taxonomy (Figure~\ref{fig: The main content flow and categorization of linguistic steganalysis methods.}), text width=34mm
        [
        Analyses of methodology features (Table~\ref{table:Steganalysis_features}), text width=57mm
        ]
        [
        Analyses of steganalysis adoption (Table~\ref{table:GLS_evaluation_steganalysis}), text width=57mm
        ]
        [
        \S\ref{sec: Review of Linguistic Steganalysis Methods} Detailed review, text width=27mm
        ]
    ]
  ]
  [\textbf{Evaluation:} \S\ref{sec: Evaluation for Linguistic Steganography}  Evaluation \\ for linguistic steganography, text width=36mm 
    [Taxonomy (Figure~\ref{fig: The main content flow and categorization of evaluation metrics.}), text width=28mm 
        [Analyses of metric adoption (Figure~\ref{fig: Evaluation adoption}), text width=53mm]
    ]
  ]
  [\textbf{Challenges:} \S\ref{sec: Challenges and Future Directions} Challenges and Future Directions, text width=65mm]
]
\end{forest}
\caption{The main content flow and structure of this survey.}

\label{fig: The main content flow and structure of this survey}
\end{figure*}
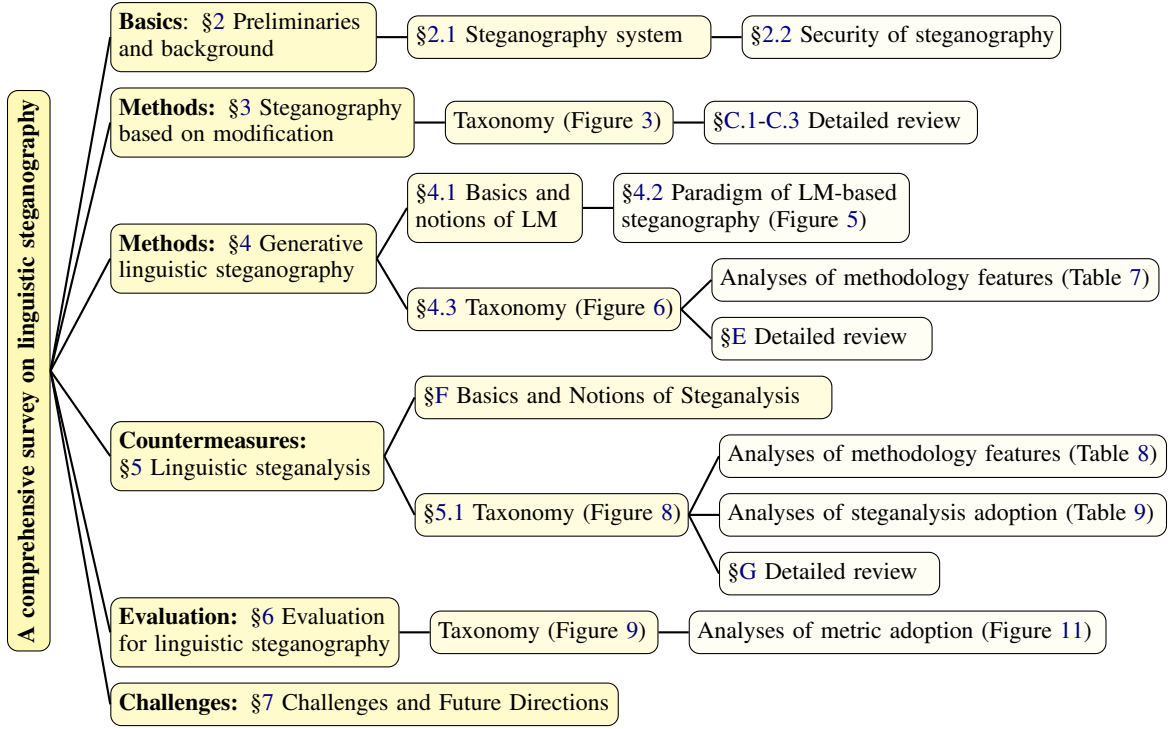

\rev{
Motivated by these limitations, we present \textbf{the first survey that systematically situates linguistic steganography in the LLM era}.
This survey provides a broad coverage and systematic analyses of current linguistic steganographic \textbf{methods} based on modification and generation, \textbf{countermeasures} (steganalysis methods), \textbf{evaluation} metrics, and \textbf{challenges}.
Table~\ref{table:Quantitative comparison of this survey with existing surveys} presents quantitative comparisons, highlighting our contribution on comprehensiveness.
% that covers about $3\times$ more steganographic methods, about $2\times$ more countermeasures (steganalysis), and about $2\times$ more evaluation metrics than prior work.
Table~\ref{table:Qualitative comparison of this survey with existing surveys} presents qualitative comparisons, revealing our in-depth and fine-grained features.
The main structure of this survey is shown in Figure~\ref{fig: The main content flow and structure of this survey}.
\textit{To ensure both timeliness and comprehensiveness, this survey is motivated by the LLM era rather than being confined exclusively to LLM-based methods.}
This broader scope is supported by an asymmetry discovered in this survey: Although most generative linguistic steganographic methods are adaptable to LLMs, most steganalysis methods have not yet.
Further, we summarize the paradigm shifts in the LLM era, which motivate the scope and structure of this survey, as follows:
}

\forestset{
  mytree/.style={
    for tree={
      draw,
      rounded corners,
      align=left,
      font=\small,
      edge={draw, line width=0.8pt},
      inner sep=3pt,
      grow'=0,
      parent anchor=east,
      child anchor=west,
      anchor=west,
      l sep=4mm,
      s sep=2mm
    },
    % 针对不同层级设置文本框宽度
    where level=0{minimum width=1mm, text width=72mm, fill=yellow!35}{},
    where level=1{minimum width=1mm, text width=32mm, fill=yellow!25}{},
    where level=2{minimum width=1mm, text width=20mm, fill=yellow!15}{},
    where level=3{minimum width=1mm, text width=75mm, fill=yellow!5}{}
  }
}

\begin{itemize}
    \item \rev{The shift from modifying a covertext to \textit{prompt-only generation}, removing the modification budget as a fundamental capacity bottleneck (especially since~\citet{8470163, ziegler-etal-2019-neural}, see Section~\ref{sec: Steganography Based on Modification} and~\ref{sec: Generative Linguistic Steganography}).}
    
    \item \rev{The shift from heuristic to \textit{provable security}, realized through information-theoretic mechanisms (since~\citet{zhang-etal-2021-provably}, see Section~\ref{sec: Taxonomy of Existing Methods} and Appendix~\ref{sec: Toward Higher Security}).}
    
    \item \rev{The shift from white-box, symmetric LMs to \textit{black-box} or \textit{asymmetric} access, including API-only sender scenarios (since~\citet{10.1145/3664647.3680562}, see Appendix~\ref{sec: Modern Generative Linguistic Steganographic Methods} and Table~\ref{table:GLS_taxonomy}).}
    
    \item \rev{The shift from security-centric designs to the \textit{joint optimization} of security, efficiency, and robustness (since~\citet{9714779, 10215094}, see Section~\ref{sec: Evaluation for Linguistic Steganography} and Appendix~\ref{sec: Review of Evaluation Metrics}).}
    
    \item \rev{The shift from concerns of pure text quality to LM \textit{engineering realities}, especially tokenization inconsistency and computational indeterminism (since ~\citet{nozaki-murawaki-2022-addressing}, see Section~\ref{sec: Challenges and Future Directions} and Appendix~\ref{sec: Introduction of Challenges and Future Directions}).}
\end{itemize}

\section{Preliminaries and Background}
\label{sec: Preliminaries and Background}
\subsection{Steganography System}
\label{sec: Steganography System}
A classic scenario used to illustrate steganography is Simmons' \textit{Prisoners' Problem}~\cite{Simmons1984}. In this setting, Alice and Bob (the steganographers) are confined in prison and attempt to hatch an escape plan. Their only way of communication is monitored by the warden Eve (the steganalyzer), who monitors all messages.
If Eve detects anything unusual or suspicious, she will block the exchange and impose punishment. Therefore, to evade detection, Alice and Bob must embed their secret messages into an \textit{innocent-looking} cover object to obtain a stego object.
Formally, a steganography system $\sum_{\mathcal{D}}$ with the channel distribution $\mathcal{D}$ is a triple of algorithms, $\sum_{\mathcal{D}} = (\text{KeyGen}_{\mathcal{D}}, \text{Embed}_{\mathcal{D}}, \text{Extract}_{\mathcal{D}})$, in which:
\begin{itemize}
    \item $\text{KeyGen}_{\mathcal{D}}(1^{\lambda})$ takes arbitrary input with length $\lambda$ and generates a shared key $K \in \{0,1\}^{k}$ with length $k$.
    \item $\text{Embed}_{\mathcal{D}}(K,\mathbf{m},\mathcal{H})$ takes a shared key $K$, a secret message $\mathbf{m} \in \{0,1\}^{*}$, and a channel history $\mathcal{H}$ as the input. It returns the steganographic content, $\mathbf{s}$.
    \item $\text{Extract}_{\mathcal{D}}(K,\mathbf{s},\mathcal{H})$ takes a shared key $K$, a steganographic content $\mathbf{s}$, and a channel history $\mathcal{H}$ as the input. It returns the message extracted from the steganographic content, $\mathbf{s}$.
\end{itemize}

To clarify the scope of steganography, we distinguish it from related concepts: cryptography and watermarking in Appendix~\ref{sec: Comparison with Related Concepts}.

\forestset{
  mytree/.style={
    for tree={
      draw,
      rounded corners,
      align=left,
      font=\small,
      edge={draw, line width=0.8pt},
      inner sep=3pt,
      grow'=0,
      parent anchor=east,
      child anchor=west,
      anchor=west,
      l sep=4mm,
      s sep=2mm
    },
    % 针对不同层级设置文本框宽度
    where level=0{minimum width=20mm, text width=56mm, fill=orange!35}{},
    where level=1{minimum width=20mm, text width=31mm, fill=orange!25}{},
    where level=2{minimum width=20mm, text width=25mm, fill=orange!15}{},
    where level=3{minimum width=20mm, text width=72mm, fill=orange!5, font=\scriptsize}{}
  }
}

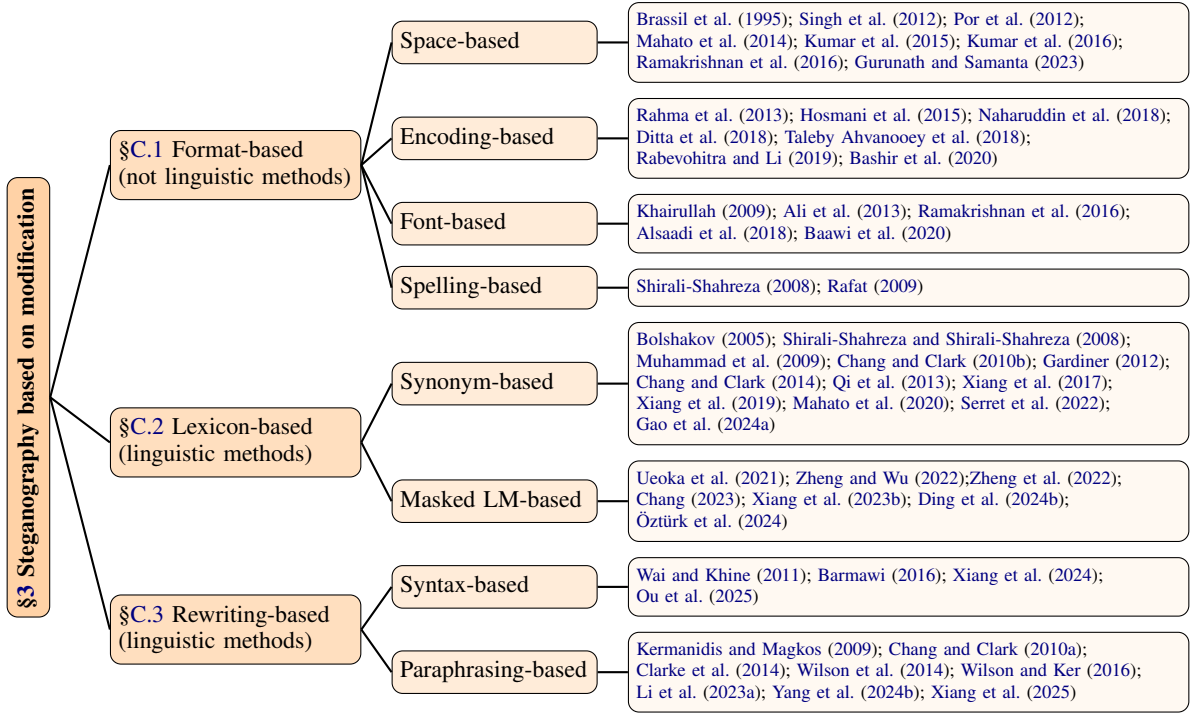
\begin{figure*}[!t]
\centering
\begin{forest}
mytree
[ \textbf{\S\ref{sec: Steganography Based on Modification} Steganography based on modification}, rotate=90, child anchor=north, parent anchor=south, anchor=center
  [\S\ref{sec: Format-Based Methods} Format-based \\ (not linguistic methods)
    [Space-based
        [\citet{464718}; \citet{singh2012novel}; \citet{POR20121075};\\
        \citet{10.1007/978-81-322-1665-0_107}; \citet{kumar2015efficient}; \citet{7813878};\\ \citet{ramakrishnan2016text}; \citet{gurunath2023new} ]
    ]
    [Encoding-based
        [\citet{rahma2013text}; \citet{10.1007/978-3-319-22915-7_26}; \citet{8711087};\\ \citet{ditta2018information}; \citet{8440030}; \\\citet{8947188};  \citet{9291576}]
    ]
    [Font-based
        [\citet{5380435}; \citet{ali2013new}; \citet{ramakrishnan2016text};\\ \citet{10.1145/3279996.3280006}; \citet{Baawi_2020}]
    ]
    [Spelling-based
        [\citet{4494159}; \citet{10.1145/1838002.1838082}]
    ]
  ]
  [\S\ref{sec: Lexicon-Based Methods} Lexicon-based \\ (linguistic methods)
    [Synonym-based
        [\citet{10.1007/978-3-540-30114-1_13}; \citet{4604331};\\ \citet{5224169}; \citet{chang-clark-2010-practical}; \citet{gardiner2012stegchat};\\ \citet{chang-clark-2014-practical}; \citet{6784896}; \citet{xiang2017novel};\\  \citet{xiang2019word}; \citet{MAHATO2020216}; \citet{serret2022linguistic}; \\ \citet{electronics13214216}]
    ]
    [Masked LM-based
        [\citet{ueoka-etal-2021-frustratingly}; \citet{zheng2022autoregressive};\citet{9948940};\\ \citet{9753668}; \citet{10315184}; \citet{10185651}; \\ \citet{10400450}]
    ]
  ]
  [\S\ref{sec: Rewriting-Based Methods} Rewriting-based \\ (linguistic methods)
    [Syntax-based
        [\citet{wai2011syntactic}; \citet{7740349}; \citet{10374202}; \\ \citet{Ou_Xiang_Liu_2025}]
    ]
    [Paraphrasing-based
        [\citet{10.1007/978-3-642-00382-0_43}; \citet{chang-clark-2010-linguistic};\\  \citet{7056859}; \citet{wilson2014linguistic}; \citet{wilson2016avoiding}; \\  \citet{10.1007/978-3-031-30637-2_41};  \citet{10049662}; \citet{XIANG20252375}]
    ]
  ]
]
\end{forest}
\caption{The main content flow and categorization of steganographic methods based on modification.}

\label{fig: The main content flow and categorization of steganographic methods based on modification}
\end{figure*}

\begin{figure*}[!t]
 \centering
 \includegraphics[width=\textwidth]{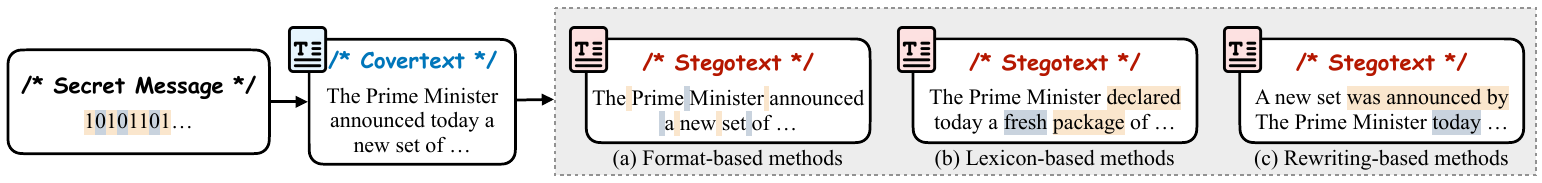} 
 \caption{Modification-based steganographic methods which are categorized into three types, where an example is shown in each type.}
 
 \label{fig: Modification_methods}
\end{figure*}

\subsection{Security of Steganography}
\label{sec: Security of Steganography}
To better understand linguistic steganography, it is important to first introduce a key property of steganographic systems: security. In the literature, two common definitions of steganographic security have been proposed. \citet{10.1007/3-540-49380-8_21} was the first to model steganographic security from an \textbf{information-theoretic} perspective. \citet{10.1007/3-540-45708-9_6} and~\citet{katzenbeisser2002defining} proposed the \textbf{complexity-theoretic} definition of steganographic security.
The detailed introduction of these two definitions is shown in Appendix~\ref{sec: Two Definitions of Steganographic Security}.

\section{Steganography Based on Modification}
\label{sec: Steganography Based on Modification}

Text steganographic methods based on modifying existing texts had dominated the field before the emergence of generative LMs.
These approaches embed secret messages through alterations to a covertext, while aiming to preserve its readability and naturalness.
We begin with \textbf{non-linguistic} format-based methods (e.g., manipulating spaces, punctuation, or text layout) and then move to \textbf{linguistic} methods, including lexicon-based (e.g., replacing words with synonyms or semantically similar alternatives) and rewriting-based methods (e.g., paraphrasing or restructuring sentences).
% In this section, we start by introducing traditional \textbf{non-linguistic} methods: format-based methods (e.g., manipulating spaces, punctuation, or text layout) to \textbf{linguistic} methods: lexicon-based methods (e.g., replacing words with synonyms or semantically similar alternatives) and rewriting-based methods (e.g., paraphrasing or restructuring sentences without changing their meaning).
% Figure~[] shows an overview of modification-based methods.
Figure~\ref{fig: The main content flow and categorization of steganographic methods based on modification} presents the content flow and taxonomy of modification-based methods in this section.
Figure~\ref{fig: Modification_methods} shows an example in each category of these modification-based steganographic methods.
\rev{The detailed introduction of methods based on modification is provided in Appendix~\ref{sec: Review of Modification-Based Methods}.
Although such methods are generally lightweight, they suffer from inherent limitations: the embedding capacity is low, and the modifications can be vulnerable to detection by statistical steganalysis.}

\begin{figure*}[!t]
    \centering
    % 第一个子图
    \begin{subfigure}[b]{0.9\textwidth}
        \centering
        \includegraphics[width=\textwidth]{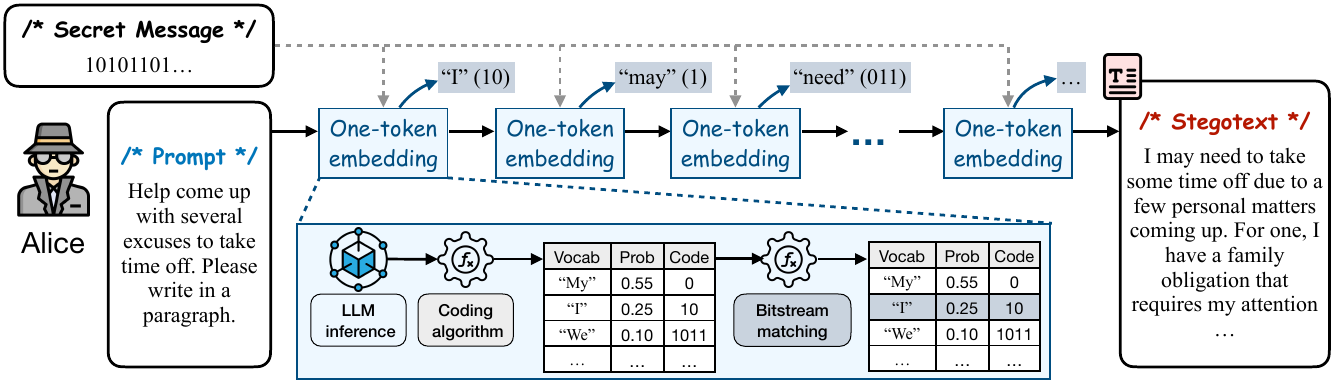}
        \caption{Embedding process of generative linguistic steganography.}
        
        \label{fig:GLS_embedding}
    \end{subfigure}
    
    \vspace{0.0cm} % 两个子图之间的竖直间距

    % 第二个子图
    \begin{subfigure}[b]{0.9\textwidth}
        \centering
        \includegraphics[width=\textwidth]{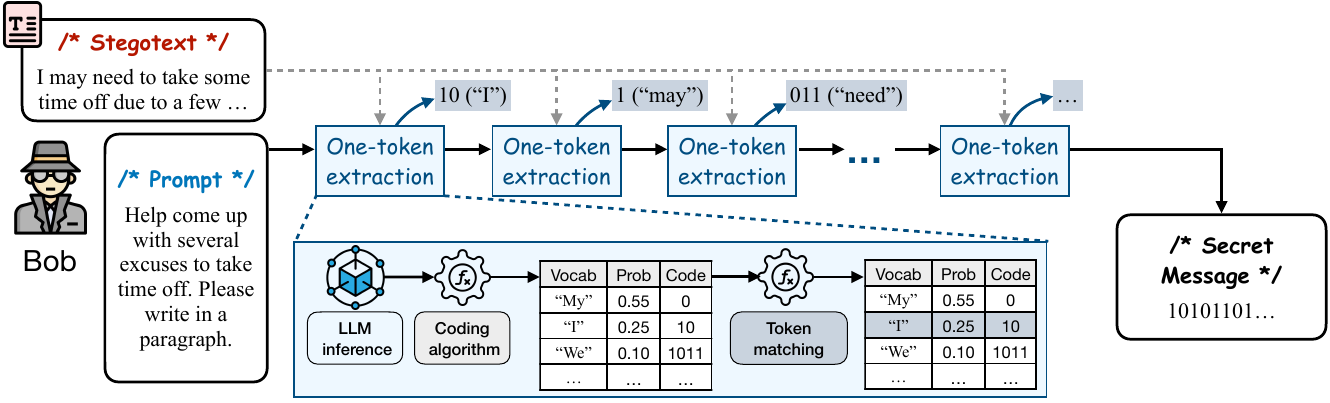}
        \caption{Extraction process of generative linguistic steganography.}
        
        \label{fig:GLS_extraction}
    \end{subfigure}
    
    \caption{Overall framework of generative linguistic steganography, in which the example is operated autoregressively based on the Huffman coding.}
    
    \label{fig:GLS_framework}
\end{figure*}

\section{Generative Linguistic Steganography}
\label{sec: Generative Linguistic Steganography}

As discussed in Section~\ref{sec: Steganography Based on Modification}, existing modification-based methods generally face the challenge of low embedding capacity, as the redundancy in the covertext is restricted.
% For example, in lexicon-based methods, the number of positions in the covertext that can be altered without significantly degrading text quality or changing semantics is limited~\cite{Wu2024}. 
Besides, in these methods, security is not proved or ensured according to the security definition discussed in Section~\ref{sec: Security of Steganography}.
\rev{These limitations have motivated \textbf{a shift toward generative approaches}, which offer greater flexibility and high imperceptibility.
% Some of them also provide information-theoretic security guarantees that focus mainly on achieving zero KL divergence caused by steganographic encoding in existing methods~\cite{zhang-etal-2021-provably, 10179287, WANG2025113101}.
Generative methods overcome or mitigate the capacity and security problems of modification-based ones. The detailed introduction of these reasons is provided in Appendix~\ref{sec: Reasons of Researching Generative Linguistic Steganography}.
% \footnote{Modification-based methods still laid the groundwork for later advances and remain a standard baseline for generative ones~\cite{ueoka-etal-2021-frustratingly}}
}

\subsection{Basics and Notions of LM}
\label{sec: Basics and notions of LM}
A language model (LM) has a vocabulary $\mathcal{V}$ consisting of $|\mathcal{V}|$ \textit{tokens}.
% The size of typical vocabularies ($|\mathcal{V}|$) is greater than $50{,}000$ tokens~\cite{radford2019language}.
Consider a sequence of LM-generated $T$ tokens $\{s^{(t)}\} \in \mathcal{V}^T$. 
Entries with negative indices, $[s^{(-N_p)},\dots,s^{(-1)}]$, represent a \textit{prompt} of length $N_p$ and $[s^{(0)},\dots,s^{(T-1)}]$ are tokens generated by an LM in response to the prompt.

Using an LM for the next-token prediction at position $t$, is a function defined as:
$f_{\mathrm{LM}}(\boldsymbol{s}^{(t-1)};\theta) = \boldsymbol{l}^{(t)}$,
in which $\boldsymbol{s}^{(t-1)}$ is the previous context $[s^{(-N_p)},\dots,s^{(t-1)}]$ at $t-1$, $\theta$ is the model parameters, and $\boldsymbol{l}^{(t)} = (l^{(t)}_1,\dots,l^{(t)}_{|\mathcal{V}|}) \in \mathbb{R}^{|\mathcal{V}|}$ is the logit vector, corresponding to each token 
% $t_i$ (where $i=1,\dots,|\mathcal{V}|$) 
in $\mathcal{V}$.
This discrete probability distribution can be calculated through the logit vector and the softmax function, i.e.,  $\boldsymbol{p}^{(t)} =(p^{(t)}_1,\dots,p^{(t)}_{|\mathcal{V}|}) = \text{softmax}(\boldsymbol{l}^{(t)}) = (\frac{\exp(l^{(t)}_1)}{\sum_{j=1}^{|\mathcal{V}|}\exp(l^{(t)}_j)},\dots,\frac{\exp(l^{(t)}_{|\mathcal{V}|})}{\sum_{j=1}^{|\mathcal{V}|}\exp(l^{(t)}_j)}) $ over the vocabulary.
The next token is then sampled from $\boldsymbol{p}^{(t)}$ (e.g., multinomial sampling). 

\rev{We provide only a limited overview of LMs, without detailing their internal structures, because most LM-based linguistic steganographic methods interact with the model solely through the post-softmax next-token distribution. This abstraction has become even more prominent in the LLM era, where access to model internals can be further restricted to API-only settings.}

\forestset{
  mytree/.style={
    for tree={
      draw,
      rounded corners,
      align=left,
      font=\small,
      edge={draw, line width=0.8pt},
      inner sep=3pt,
      grow'=0,
      parent anchor=east,
      child anchor=west,
      anchor=west,
      l sep=4mm,
      s sep=2mm
    },
    % 针对不同层级设置文本框宽度
    where level=0{minimum width=20mm, text width=54mm, fill=StrongBlue!35}{},
    where level=1{minimum width=18mm, text width=22mm, fill=StrongBlue!25}{},
    where level=2{minimum width=15mm, text width=28mm, fill=StrongBlue!15}{},
    where level=3{minimum width=20mm, text width=80mm, fill=StrongBlue!5, font=\scriptsize}{}
  }
}

\subsection{Paradigm of LM-based Steganography}
\label{sec: Paradigm of LM-based Steganography}
Alice (the sender) wants to communicate a secret message $\mathbf{m} \sim$ $U(\{0,1\}^l)$ with Bob (the receiver) by embedding it in a natural-language text $\mathbf{s}$ (a stegotext).
% The uniform distribution is chosen for $\mathbf{m}$ without loss of generality: if $\mathbf{m}$ has additional structure it can be further compressed to a uniformly distributed random variable~\cite{10.1109/TIT.2004.840860}. 
According to the notions and definitions of the steganographic system in Section~\ref{sec: Steganography System}, Alice and Bob have agreed on an embedding function $\text{Embed}_{\mathcal{D}}(K,\mathbf{m},\mathcal{H})$ and an extracting function $\text{Extract}_{\mathcal{D}}(K,\mathbf{s},\mathcal{H})$ that perform steganography.
Specifically, in the scenario of LM-based generative linguistic steganography, $\mathcal{D}$ is commonly defined as the same shared language model $\mathcal{M}^{o}$, $\mathcal{H}$ is the token sequence of the prompt $[s^{(-N_p)},\dots,s^{(-1)}]$, $\mathbf{s}$ is the generated stegotext.
These two functions are designed to be mutually consistent. In other words,
$\text{Embed}_{\mathcal{D}}(K,\mathbf{m},\mathcal{H}) = \mathbf{s}$ and
$\text{Extract}_{\mathcal{D}}(K,\mathbf{s},\mathcal{H}) = \mathbf{m}^{\prime}$,
% \footnote{In this work, we do not consider disambiguation methods~\cite{nozaki-murawaki-2022-addressing,yan2023A,qi2024provably} that focus on maintaining $m_s = m_s^{\prime}$, since this work is orthogonal to disambiguation.}
in which the ideal result is $\mathbf{m} = \mathbf{m}^{\prime}$.

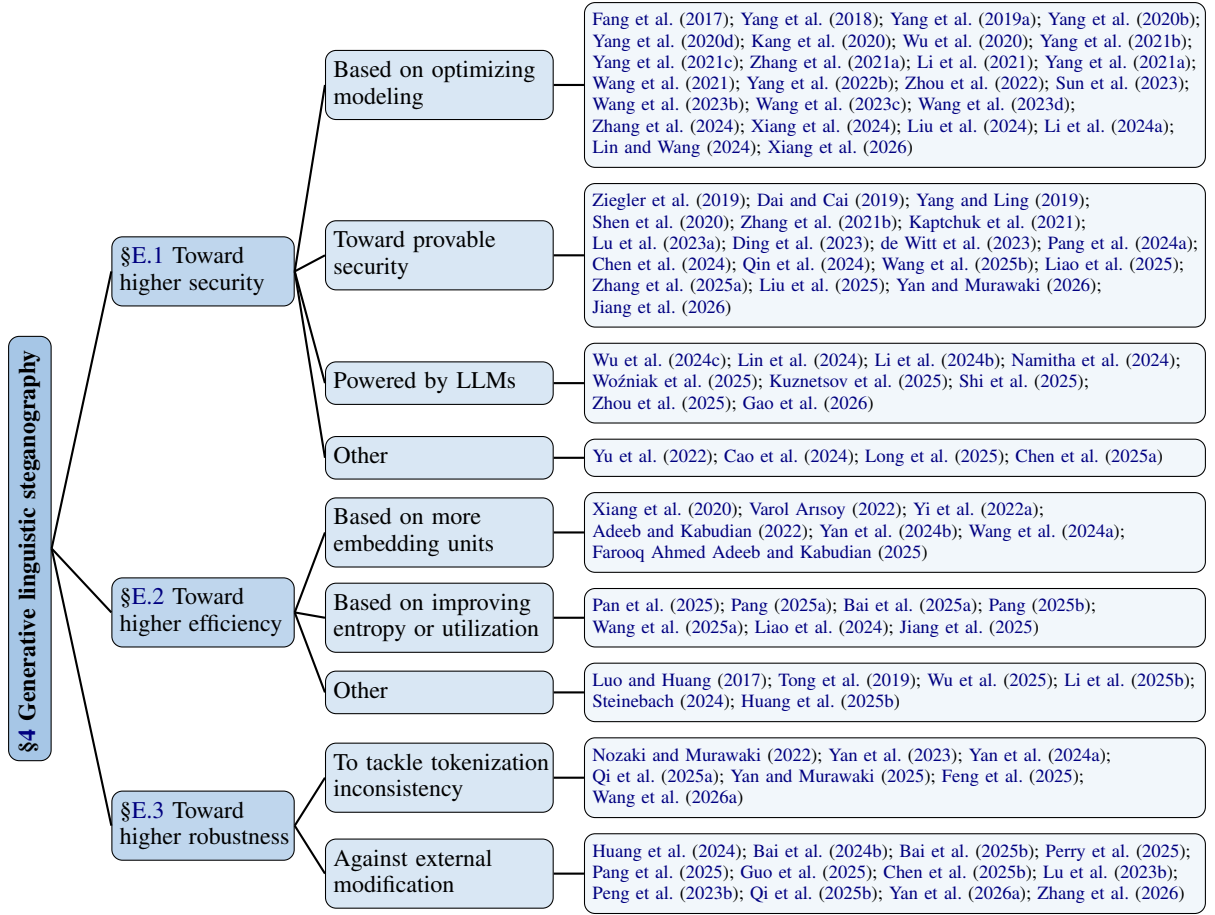
\begin{figure*}[!t]
\centering
\begin{forest}
mytree
[ \textbf{\S\ref{sec: Generative Linguistic Steganography} Generative linguistic steganography}, rotate=90, child anchor=north, parent anchor=south, anchor=center
  [\S\ref{sec: Toward Higher Security} Toward \\ higher security
    [Based on optimizing \\ modeling
        [\citet{fang-etal-2017-generating}; \citet{yang2018automaticallygeneratesteganographictext}; \citet{8470163}; \citet{yang2020graphstegasemanticcontrollablesteganographic}; \\ \citet{10.1007/978-3-030-43575-2_2}; \citet{doi:10.2352/ISSN.2470-1173.2020.4.MWSF-291}; \citet{doi:10.1177/1550147720914257}; \citet{9193914}; \\ \citet{9353234}; \citet{9280343}; \citet{10.1145/3418598}; \citet{yang2021generation}; \\ \citet{WANG20211375};  \citet{9975301}; \citet{9430708}; \citet{SUN2023157}; \\ \citet{10115433};   \citet{10.1007/978-3-031-44696-2_62};  \citet{10095722}; \\   \citet{10447545}; \citet{10374202};  \citet{10462497};   \citet{10687958}; \\ \citet{lin2024novel}; \citet{Xiang_Ou_He_Yang_Liu_2026} ]
    ]
    [Toward provable \\ security
        [\citet{ziegler-etal-2019-neural}; \citet{dai-cai-2019-towards}; \citet{9023313}; \\ \citet{shen-etal-2020-near};  \citet{zhang-etal-2021-provably}; \citet{10.1145/3460120.3484550};  \\  \citet{10191218};  \citet{10179287}; \citet{witt2023perfectly};   \citet{10446027}; \\  \citet{10.1007/978-981-99-8148-9_44};  \citet{qin2024adlmstegauniversal};  \citet{WANG2025113101}; \citet{liao2025framework}; \\  \citet{10919187};   \citet{10.1162/coli.a.22}; \citet{yan2026efficientprovablysecurelinguistic}; \\ \citet{11329500}]
    ]
    [Powered by LLMs
        [\citet{10.1145/3664647.3680562}; \citet{lin-etal-2024-zero}; \citet{10669737};  \citet{10551807}; \\ \citet{WOZNIAK2025113960}; \citet{computers14050165}; \citet{technologies13070264};  \\ \citet{zhou2025autostegaagentdrivenlifelongstrategy}; \citet{gao2026textsteganographydynamiccodebook}
        ]
    ]
    [Other
        [\citet{yu2022mts}; \citet{9778236}; \citet{10888762}; \citet{app15179663}
        ]
    ]
  ]
  [\S\ref{sec: Toward Higher Efficiency} Toward \\ higher efficiency
    [Based on more  \\  embedding units
        [\citet{math8091558}; \citet{varol2022lzw}; \citet{9714779}; \\ \citet{9864312}; \citet{10831652};  \citet{10.1007/978-981-99-8073-4_4};\\  \citet{11008605}]
    ]
    [Based on improving \\ entropy or utilization
        [\citet{10919130}; \citet{pang2025frestegaplugandplaymethodboosting}; \citet{bai2025shimmer}; \citet{pang2025provablesecuresteganographybased};\\  \citet{wang2025sparsampefficientprovablysecure}; \citet{10.1145/3658664.3659657}; \citet{NEURIPS2025_6c8e215a}]
    ]
    [Other
        [\citet{10.1145/3082031.3083240}; \citet{Text_steganography_on_RNN-Generated-lyrics}; \citet{10.1007/978-981-96-7005-5_12}; \citet{11017758}; \\ \citet{10.1145/3664476.3670930}; \citet{huang2025relativelysecurellmbasedsteganographyconstrained}]
    ]
  ]
  [\S\ref{sec: Toward Higher Robustness} Toward \\ higher robustness
    [To tackle tokenization \\ inconsistency
        [\citet{nozaki-murawaki-2022-addressing}; \citet{10215094}; \citet{10831370};\\ \citet{10804596}; \citet{yan-murawaki-2025-addressing}; \citet{feng2025highcapacitysecuredisambiguationalgorithm}; \\ \citet{wang2026retoksyncselfsynchronizingtokenizationdisambiguation}
        ]
    ]
    [Against external \\ modification [\citet{huang2024odstegallmbasednearimperceptiblesteganography};  \citet{bai2024semanticsteganographyframeworkrobust};  \citet{11023508}; \citet{perry2025robuststeganographylargelanguage}; \\\citet{10888944};  \citet{10888607}; \citet{sym17091416}; \citet{10191984}; \\ \citet{10075392};  \citet{qi2025stead}; \citet{yan2026anchoredslidingwindowrobust}; \citet{11363394}
    ]
    ]
  ]
]
\end{forest}
\caption{The main content flow and categorization of generative linguistic steganography.}

\label{fig: The main content flow and categorization of generative linguistic steganography}
\end{figure*}

Figures~\ref{fig:GLS_embedding} and~\ref{fig:GLS_extraction} respectively illustrate the embedding framework and the extraction framework of LM-based steganography, using the method based on the Huffman coding~\cite{8470163}.
From the figures, we can notice that the cover carrier, which is necessary in traditional steganography, is not required in generative linguistic steganography. Instead, the prompt is shared in the steganographic protocol.
Although generative linguistic steganographic methods generally follow the above frameworks, they can be classified according to the targeted performance metrics, methodology features, or the specific steganographic scenarios in which they are applied.

% \usepackage{multirow}

\subsection{Taxonomy of Existing Methods}
\label{sec: Taxonomy of Existing Methods}
In this section, we categorize existing generative linguistic steganographic methods into three types: (1) methods toward higher security, (2) methods toward higher efficiency, and (3) methods toward higher robustness. \textbf{Security} can be assessed using various text-quality metrics (such as perplexity~\cite{jelinek1977perplexity} and ROUGE~\cite{lin-2004-rouge}) or by measuring imperceptibility and indistinguishability that prevent humans or machine classifiers (steganalysis) from detecting stegotexts from normal texts. 
\textbf{Efficiency} refers to the embedding capacity of the stegotext as well as the embedding and extraction speeds.
\textbf{Robustness} evaluates whether, and to which extent, the extracted message $\mathbf{m}^{\prime}$ remains consistent with the embedded message $\mathbf{m}$, particularly under active attack scenarios, such as substitution, deletion, or insertion.

\begin{table*}[!t]
\centering
\resizebox{0.9\textwidth}{!}{%
\begin{tabular}{l|c|c|c}
\toprule
\textbf{Method} & \textbf{Dist. unchanged (KL = 0)?} & \textbf{Complexity (per token)} & \textbf{Mode} \\
\midrule
\textbf{ADG}~\cite{zhang-etal-2021-provably} & \cellno  & $O(N\log N)$  &  Stream \\\hline
\textbf{Meteor}~\cite{10.1145/3460120.3484550} & \cellno  & $O(N)$, $O(2^{\lceil H(P) \rceil}N)$ with reorder & Stream \\\hline
\textbf{iMEC}~\cite{witt2023perfectly} & \cellyes & $O(2^B N)$ & Block \\\hline
\textbf{Discop}~\cite{10179287} & \cellyes & $O(N)$, $O(N\log N)$ with reorder  & Stream \\\hline
\textbf{SparSamp}~\cite{wang2025sparsampefficientprovablysecure} & \cellyes & $O(2^B)$ & Block \\\hline
\textbf{RRC}~\cite{yan2026efficientprovablysecurelinguistic} &  \cellyes & $O(N)$, $O(N\log N)$ with reorder & Block \\
\bottomrule
\end{tabular}}
\caption{\rev{Methodology features of provably secure linguistic steganographic methods. The Dist. unchanged (KL = 0) column marks whether the method keeps the original distribution at each generation step. In the complexity column, $N$ denotes the number of candidate tokens considered at each step; $H(P)$ denotes the entropy of the next-token distribution $P$; $B$ denotes the block size used by block-mode methods.}}
\label{tab:methods toward provable security}
\end{table*}

\begin{table*}[!t]
\centering
\resizebox{\textwidth}{!}{%
\begin{tabular}{l|c|c|c}
\toprule
\textbf{Method} & \textbf{Strict 100\% correctness?} & \textbf{Dist. unchanged (KL = 0)?} & \textbf{Complexity (per token)} \\
\midrule
Basic~\cite{nozaki-murawaki-2022-addressing} & \cellyes  & \cellno &  $O(N^2)$\\\hline
MWIS~\cite{10215094} & \cellyes  & \cellno & $O(N^2)$ \\\hline
All-case verification~\cite{10831370} &  \cellno  & \cellyes & $O(N\times2^N)$\\\hline
SyncPool~\cite{10804596} & \cellyes  & \cellyes & $O(N\log N)$ \\\hline
Stepwise verification~\cite{yan-murawaki-2025-addressing} & \cellyes  & \cellno & $O(N)$ \\ \hline
ReTokSync~\cite{wang2026retoksyncselfsynchronizingtokenizationdisambiguation} & \cellno  & \cellyes & $O(L)$ \\
\bottomrule
\end{tabular}}
\caption{\rev{Methodology features of methods to tackle tokenization inconsistency. Strict 100\% correctness column marks whether the method can 100\% eliminate extraction errors caused by tokenization inconsistency. The Dist. unchanged (KL = 0) column marks whether the method keeps the original distribution at each generation step. In the complexity column, $N$ denotes the number of candidate tokens considered at each step; $L$ denotes the number of generated tokens in the real-time sequence.}}
\label{tab:methods to tackle tokenization inconsistency}
\end{table*}

Following~\citet{fang-etal-2017-generating}, which marked a milestone for RNN-based generative linguistic steganography, only modern methods (after~\citet{fang-etal-2017-generating}) are included for taxonomy.
Figure~\ref{fig: The main content flow and categorization of generative linguistic steganography} presents the main content flow and categorization of generative linguistic steganographic methods, organized according to the \textit{primary targeted metrics} and further methodology features.
% The review of early generative linguistic steganographic methods is provided in Appendix~\ref{sec: Early Attempts for Generative Linguistic Steganography}.
\rev{Among the three axes, security has undergone \textbf{a notable shift from heuristic to provable security}. Table~\ref{tab:methods toward provable security} compares the methodology features of provably secure generative methods.
We can observe a clear evolutionary trajectory from non-zero KL to KL = 0, and then toward lower complexity and higher efficiency.}
Table~\ref{tab:methods to tackle tokenization inconsistency} summarizes the features of these methods to tackle tokenization inconsistency, according to which several findings stand out:
\begin{itemize}
    \item \rev{Jointly achieving (a) distribution-preserving guarantees (KL = 0 or provable security), (b) strict 100\% extraction correctness, and (c) low complexity, remains an open challenge. SyncPool achieves this at the cost of a super-linear $O(N\log N)$  which is still heavier than the linear-time alternatives.}
    \item \rev{A clear trend from high to low computational complexity is emerging across successive works.}
\end{itemize}

The detailed review of modern generative linguistic steganographic methods is provided in Appendix~\ref{sec: Modern Generative Linguistic Steganographic Methods}.

\begin{figure*}[!t]
 \centering
 \includegraphics[width=0.8\textwidth]{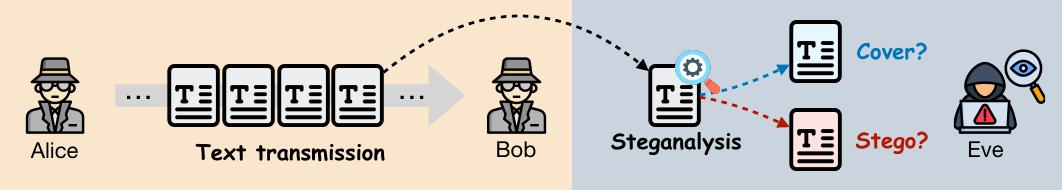} 
 \caption{A sketch of how Eve operates steganalysis.}
 
 \label{fig: Linguistic_steganalysis}
\end{figure*}

\forestset{
  mytree/.style={
    for tree={
      draw,
      rounded corners,
      align=left,
      font=\small,
      edge={draw, line width=0.8pt},
      inner sep=3pt,
      grow'=0,
      parent anchor=east,
      child anchor=west,
      anchor=west,
      l sep=4mm,
      s sep=2mm
    },
    % 针对不同层级设置文本框宽度
    where level=0{minimum width=20mm, text width=35mm, fill=blueviolet!20}{},
    where level=1{minimum width=18mm, text width=27mm, fill=blueviolet!15}{},
    where level=2{minimum width=15mm, text width=26mm, fill=blueviolet!10}{},
    where level=3{minimum width=20mm, text width=76mm, fill=blueviolet!5, font=\scriptsize}{}
  }
}

\begin{figure*}[!t]
\centering
\begin{forest}
mytree
[ \textbf{\S\ref{sec: Linguistic Steganalysis} Linguistic steganalysis}, rotate=90, child anchor=north, parent anchor=south, anchor=center
  [Based on self-trained \\ word representations
    [Based on RNNs
        [\citet{8727932}; \citet{10.1145/3369412.3395067}; \citet{9746219}]
    ]
    [Based on CNNs
        [\citet{10.1007/s11042-020-08716-w}; \citet{XUE2022140}; \citet{9732641};\\ \citet{xiang2023general}; \citet{10805412}]
    ]
    [Based on GNNs
        [\citet{9364681}; \citet{9844755}; \citet{9970400}; \\ \citet{10097161};  \citet{10.1007/978-981-99-8073-4_9}; \citet{huang2025gsdfusecapturingcognitiveinconsistencies}]
    ]
    [Hybrid
        [\citet{8903243}; \citet{9118410}; \citet{li2020two}; \\\citet{10038789}; \citet{10888159}; \citet{11460766}]
    ]
    [Other
        [\citet{8653856}; \citet{jiao2021neural}; \citet{10021965}]
    ]
  ]
  [Based on pre-trained \\ word representations
    [Based on \\ masked LM(s)
        [\citet{9484749}; \citet{10.1007/978-3-030-69449-4_7}; \citet{Xu_2021}; \citet{9591452}; \\ \citet{9913623}; \citet{9806187};   \citet{10.1145/3531536.3532963}; \\  \citet{9616404};  \citet{10.1145/3531536.3532949};  \citet{9956902}; \\  \citet{10268497}; \citet{WANG2023103512}; \citet{10.1145/3577163.3595111};  \\  \citet{10190738};    \citet{10299660}; \citet{YUAN2024123437}; \\  \citet{10582504};   \citet{YOU2024128260};  \citet{10916785}; \\  \citet{10849805}; \citet{10.1145/3746252.3760827};   \citet{10887993}; \\ \citet{10.1007/978-3-031-82907-9_10};  \citet{10.1145/3733102.3733114};  \citet{niu2025text}; \\ \citet{WANG2026129311}; \cite{11462507}]
    ]
    [Based on generative \\ LM(s) or LLM(s)
        [\citet{yang2024nextgenerationsteganalysisllmsunleash}; \citet{10.1145/3634814.3634837}; \citet{10752347}; \\ \citet{Wang_Zhou_Chen_Zhou_Yang_2025}; \citet{barlow2026large}]
    ]
    [Other
        [\citet{8625512}; \citet{10234544}; \citet{10596690}; \\ \citet{10870426};  \citet{10772620}]
    ]
  ]
]
\end{forest}
\caption{The main content flow and categorization of linguistic steganalysis methods.}

\label{fig: The main content flow and categorization of linguistic steganalysis methods.}
\end{figure*}
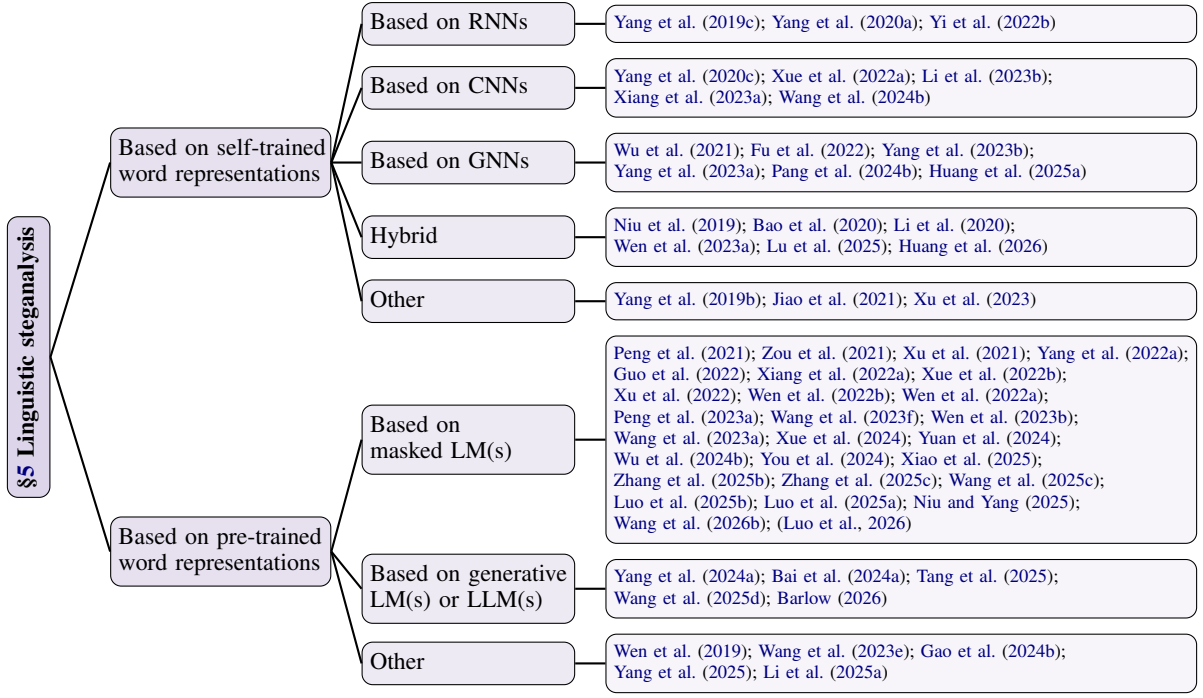

\begin{table*}[!t]
\centering
\resizebox{\textwidth}{!}{%
\begin{tabular}{l|c|c|c}
\toprule
\textbf{Method} & \textbf{Training-free?} & \textbf{LoRA-based fine tuning?} & \textbf{Technique keyword} \\
\midrule
\citet{yang2024nextgenerationsteganalysisllmsunleash} & \cellno  & \cellyes & Human-like perception \\\hline
\citet{10.1145/3634814.3634837} & \cellyes  & \cellno & Few shots; black box \\\hline
\citet{10752347} & \cellno  & \cellyes & Generation-classification hybrid \\\hline
\citet{Wang_Zhou_Chen_Zhou_Yang_2025} & \cellno  & \cellyes & Knowledge graphs \\ \hline
\citet{barlow2026large} & \cellno  & \cellyes & Multi-feature extraction \\
\bottomrule
\end{tabular}}
\caption{\rev{Methodology features of linguistic steganalysis methods which are adaptable to LLMs.}}
\label{tab:steganalysis based on LLMs}
\end{table*}

\section{Linguistic Steganalysis}
\label{sec: Linguistic Steganalysis}
Linguistic steganography is dual-use in nature, which can be used not only to protect personal privacy but also to enable malicious covert communication.
Its countermeasures, \textbf{linguistic steganalysis} methods, are essential for detecting the presence of steganography, particularly in security-critical contexts~\cite{10.1117/12.465263, NISSAR20101758}.
Figure~\ref{fig: Linguistic_steganalysis} shows a sketch of steganalysis. The  introduction of basics and notions of steganalysis is provided in Appendix~\ref{sec: Basics and Notions of Steganalysis}.

\forestset{
  mytree/.style={
    for tree={
      draw,
      rounded corners,
      align=left,
      font=\small,
      edge={draw, line width=0.8pt},
      inner sep=3pt,
      grow'=0,
      parent anchor=east,
      child anchor=west,
      anchor=west,
      l sep=4mm,
      s sep=2mm
    },
    % 针对不同层级设置文本框宽度
    where level=0{minimum width=20mm, text width=58mm, fill=red!35}{},
    where level=1{minimum width=18mm, text width=32mm, fill=red!25}{},
    where level=2{minimum width=15mm, text width=32mm, fill=red!15}{},
    where level=3{minimum width=20mm, text width=60mm, fill=red!5, font=\scriptsize}{}
  }
}

\begin{figure*}[!t]
\centering
\begin{forest}
mytree
[ \textbf{\S\ref{sec: Evaluation for Linguistic Steganography} Evaluation for linguistic steganography}, rotate=90, child anchor=north, parent anchor=south, anchor=center
  [\S\ref{sec: Security metrics} Security  metrics
    [Based on KL divergence
        [Text level KL divergence$\Downarrow$; Step-level KL divergence$\Downarrow$; \\ Variants of KL divergence$\Downarrow$]
    ]
    [Based on text quality
        [Perplexity$\Downarrow$; BLEU$\Uparrow$; ROUGE$\Uparrow$; METEOR$\Uparrow$; Diversity$\Uparrow$;\\ Human evaluation; LLM-as-a-judge]
    ]
    [Based on steganalysis
        [Accuracy$\Downarrow$; Precision$\Downarrow$; Recall$\Downarrow$; F1$\Downarrow$]
    ]
  ]
  [\S\ref{sec: Efficiency metrics} Efficiency  metrics
    [Embedding capacity
        [Bits per token$\Uparrow$; Entropy utilization$\Uparrow$; Embedding rate$\Uparrow$]
    ]
    [Operation speed
        [Average time to embed/extract one bit/an entire stegotext$\Downarrow$]
    ]
  ]
  [\S\ref{sec: Robustness metrics} Robustness metrics
    [Error rate
        [Message error rate$\Downarrow$; Bit error rate$\Downarrow$]
    ]
  ]
]
\end{forest}
\caption{An overview of evaluation metrics. $\Uparrow$/$\Downarrow$ indicate that higher/lower values correspond to better performance.}

\label{fig: The main content flow and categorization of evaluation metrics.}
\end{figure*}
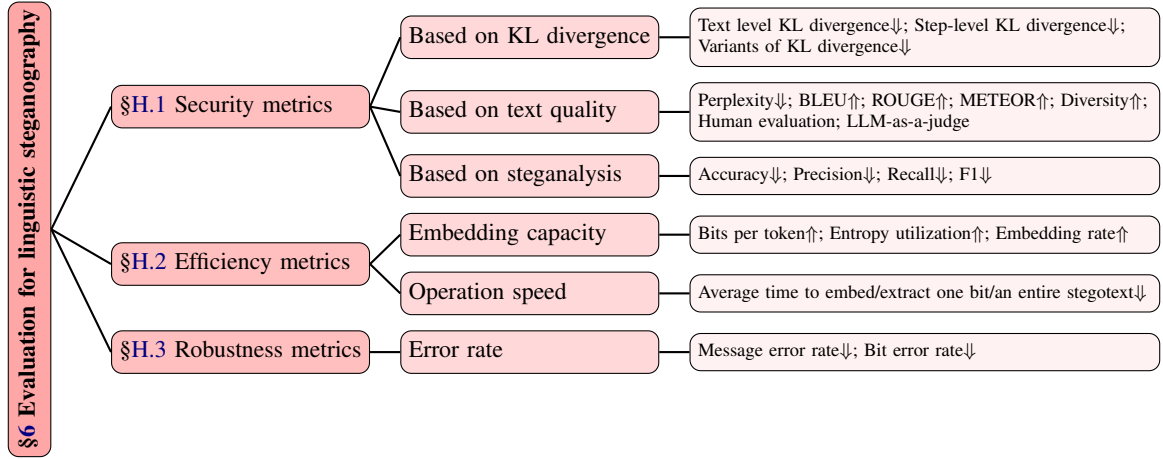

\subsection{Taxonomy of Existing Methods}
\label{sec: Taxonomy of steganalysis}
We categorize existing linguistic steganalysis methods along two axes.
First, we distinguish whether a method relies on pre-trained word representations or models, such as Word2Vec~\cite{mikolov2013efficientestimationwordrepresentations}, GloVe~\cite{pennington-etal-2014-glove}, or BERT~\cite{devlin-etal-2019-bert}.
Second, we further group methods by the type of backbone model they adopt.
Following these principles, Figure~\ref{fig: The main content flow and categorization of linguistic steganalysis methods.} presents the resulting taxonomy of linguistic steganalysis methods.
To keep the taxonomy focused on recent developments, our taxonomy and review only includes methods based on neural networks that are proposed after current representative methods, \textbf{FCN}~\cite{8653856} and \textbf{TS-RNN}~\cite{8727932}.
Table~\ref{table:Steganalysis_features} (in the Appendix) lists the features of these linguistic steganalysis methods, revealing the following trends:

(1) The \textbf{IMDB}~\cite{maas-etal-2011-learning}, \textbf{Twitter}~\cite{go2009twitter}, and \textbf{News}~\cite{kaggle_all_the_news} datasets have become mainstream for evaluation.

(2) Chronologically, methods based on \textbf{pre-trained} models (especially BERT~\cite{devlin-etal-2019-bert}) gradually \textbf{dominate} the field.

(3) \textbf{LLM-based} linguistic steganalysis methods have emerged only recently and remain relatively \textbf{underexplored}.

In addition, Table~\ref{table:GLS_evaluation_steganalysis} (in the Appendix) summarizes the steganalysis methods adopted to evaluate generative linguistic steganography. Among them, \textbf{FCN}~\cite{8653856}  \textbf{CNN}~\cite{8625512}, \textbf{R-BiLSTM-C}~\cite{8903243}, \textbf{TS-RNN}~\cite{8727932}, \textbf{TS-CSW}~\cite{10.1007/s11042-020-08716-w} are the five most commonly adopted steganalysis methods. 
% We also observe that most steganographic methods adopt \textbf{at least one} steganalysis model to assess imperceptibility.
The detailed review of linguistic steganalysis methods is provided in Section~\ref{sec: Review of Linguistic Steganalysis Methods}.
\rev{In particular, Table~\ref{tab:steganalysis based on LLMs} summarizes the features of LLM-adaptable linguistic steganalysis methods, as they constitute a promising yet still underexplored direction in linguistic steganalysis.}

\section{Evaluation for Linguistic Steganography}
\label{sec: Evaluation for Linguistic Steganography}

Following the taxonomy of generative linguistic steganography in Section~\ref{sec: Taxonomy of Existing Methods}, this section details the main metrics for evaluating linguistic steganography, which are categorized into security metrics, efficiency metrics, and robustness metrics.

Figure~\ref{fig: The main content flow and categorization of evaluation metrics.} presents an overview and their taxonomy of these evaluation metrics.
Figure~\ref{fig: Evaluation adoption} (in the Appendix) summarizes the chronological trend on adoption rates of the major evaluation approaches in \rev{95} papers of modern generative linguistic steganographic methods.
Tables~\ref{table:GLS_evaluation_security} and~\ref{table:GLS_evaluation_others} (in the Appendix) present metrics chosen to evaluate these generative methods.
From these tables, we draw the following key observations:

% (1) PPL, metrics based on KL divergence, and BPT are used \textbf{most frequently}.

(1) Entropy utilization is gaining emphasis because model-agnostic methods compatible with diverse LLMs require \textbf{fair cross-model assessments} of embedding capacity.

(2) Metrics based on error rates are used increasingly with the growing focus on \textbf{robustness}.

\rev{Chronologically, the increasing focus on evaluating efficiency and robustness reflects \textbf{the shift from security-centric to joint optimization}.}
More details about evaluation are shown in Appendix~\ref{sec: Review of Evaluation Metrics}.

\section{Challenges and Future Directions}
\label{sec: Challenges and Future Directions}
Despite the rapid progress surveyed above, the field still lacks ready-to-use software and engineering-aware designs. We identify the open challenges and group them into three themes (details shown in Appendix~\ref{sec: Introduction of Challenges and Future Directions}):
\begin{itemize}
    \item \rev{\textit{Engineering realities and issues}: (1)~computational indeterminism, (2)~tokenization inconsistency, (3)~abrupt termination, and (4)~challenges of synchronizing prompts.}
    \item \rev{\textit{Impractical assumptions and evaluations}: (5)~limited embedding capacity, (6)~high dependence on white-box LLMs, and (7)~unrealistic security evaluations.}
    \item \rev{\textit{Community and ethical concerns}: (8)~lack of unity in benchmarks and (9)~ethics and use restrictions of LLM licenses.}
\end{itemize}

\section{Conclusion}
\rev{In this survey, we provide a systematic overview of linguistic steganography in the LLM era along three aspects: (1)~modification-based and generative steganographic methods, (2)~linguistic steganalysis methods, and (3)~evaluation metrics. Across these axes we identify five paradigm shifts that mark the LLM era, most notably the rise of \textbf{provably secure generative} methods, \textbf{robustness-oriented} designs, and emerging \textbf{black-box} and \textbf{asymmetric} schemes. Besides, we identify and group the remaining open challenges into three themes spanning (1) engineering realities, (2) impractical assumptions and evaluations, and (3) community and ethical concerns.
This survey aims to catalyze further work on linguistic steganography that is practical to deploy and responsible to use.}

\section*{Limitations}
% This work is intended to be a comprehensive survey motivated by the LLM era, rather than a survey restricted exclusively to the LLM era. We recognize that this broader scope may, from the perspective of some readers, reduce the perceived timeliness or readability of the paper. However, we chose this scope for the following reasons:

% (1) Existing pre-LLM surveys do not provide a truly comprehensive review of linguistic steganographic methods, as shown in Tables~\ref{table:Quantitative comparison of this survey with existing surveys} and~\ref{table:Qualitative comparison of this survey with existing surveys}. This gap motivates us to systematically survey and review the full body of work, including both pre-LLM and LLM-era methods.

% (2) Linguistic steganalysis and evaluation methods have not yet been broadly reshaped by LLMs, as analyzed in Figures~\ref{fig: The main content flow and categorization of linguistic steganalysis methods.} and~\ref{fig: Evaluation adoption}. Restricting the survey to LLM-era methods exclusively would therefore result in an overly narrow scope.

% (3) A substantial portion of linguistic steganography methods are \textit{model-agnostic} and can be applied to both small and large language models. As a result, distinguishing methods solely by whether they are ``LLM-era'' is often of limited significance. Instead, we use the  ``LLM-adaptable'' feature to describe these steganographic and steganalysis methods in Tables~\ref{table:GLS_taxonomy} and~\ref{table:Steganalysis_features}.
\rev{We acknowledge the following limitations of our survey.
(1)~We focus exclusively on linguistic (text) steganography and do not cover image/audio/video or cross-modal schemes.
(2)~We synthesize results as reported by the original authors and do not re-implement or re-benchmark methods under a unified protocol; building such a benchmark is itself one of the open challenges identified in Section~\ref{sec: Challenges and Future Directions}.}
% (3)~We treat text watermarking as related but distinct work and do not aim to comprehensively cover it, despite its overlapping technical machinery with generative steganography.

\section*{Ethical Considerations}
Although linguistic steganography can serve legitimate purposes, such as embedding and tracing copyright information, it can also be misused to evade legal censorship. This dual-use nature underscores the urgent need for effective regulation.
In this survey, we highlight the ethics and use restrictions (Section~\ref{sec: Challenges and Future Directions} and Appendix~\ref{sec: Introduction of Challenges and Future Directions}) and advocate for research on both responsible defensive capabilities and legitimate privacy-preserving uses, thereby promoting responsible use of steganography.

\section*{Acknowledgments}

We express our gratitude to the anonymous reviewers for their valuable and insightful comments. This work was supported by JSPS KAKENHI Grant Number JP26KJ1382.

\bibliography{custom}

\appendix

\begin{table*}[!t]
\renewcommand{\arraystretch}{1.0}
\centering

\scalebox{0.8}{
\begin{tabular}{m{2cm}|m{6cm}|m{4cm}|m{6cm}}
\toprule[1.0pt]
\textbf{Feature} & \textbf{Steganography} & \textbf{Cryptography} & \textbf{Watermarking} \\
\midrule[1.0pt]
\textbf{Goal} & Hide the existence of communication & Protect the content of communication & Assert ownership or authenticity \\
\hline
\textbf{Carrier} & Multimedia & Plaintext $\rightarrow$ ciphertext & Multimedia \\
\hline
\textbf{Visibility} & Definitely not  (ideally undetectable) & Obvious  (ciphertext is visible) & Optional  (depending on the watermarking type)  \\
\hline
\textbf{Adversary} & Detect whether a given object contains hidden information & Recover the plaintext or secret key from the ciphertext & Destroy, alter, or remove the embedded watermark without significantly degrading the carrier quality \\
\hline
\textbf{Robustness} & Important (but optional in some work) & Strong against brute force  & Designed to withstand distortion \\
\hline
\textbf{Uses} & Covert messaging, censorship evasion & Secure communication, authentication & Copyright protection, forensic tracking,  digital rights management (DRM) \\
\bottomrule[1.0pt]
\end{tabular}}
\caption{Comparison of steganography, cryptography, and watermarking.}
\label{table:steg-crypto-watermark}
\end{table*}

\begin{figure*}[!t]
    \centering
    % 第一个子图
    \begin{subfigure}[b]{0.33\textwidth}
        \centering
        \includegraphics[width=\textwidth]{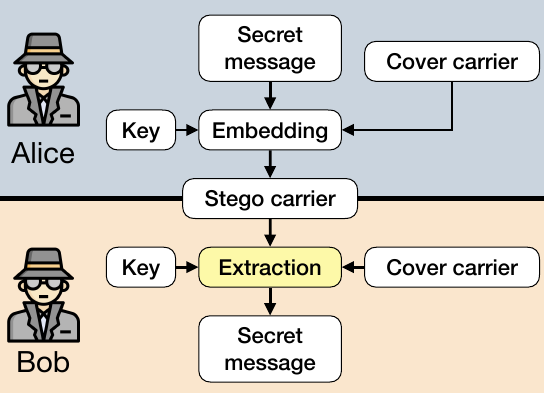}
        \caption{Steganography.}
        
        \label{fig: sketch_steganography}
    \end{subfigure}
    \begin{subfigure}[b]{0.2115\textwidth}
        \centering
        \includegraphics[width=\textwidth]{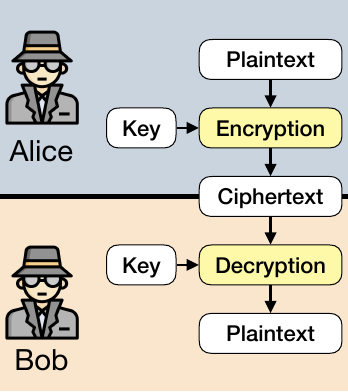}
        \caption{Cryptography.}
        
        \label{fig:sketch_cryptography}
    \end{subfigure}
    \begin{subfigure}[b]{0.33\textwidth}
        \centering
        \includegraphics[width=\textwidth]{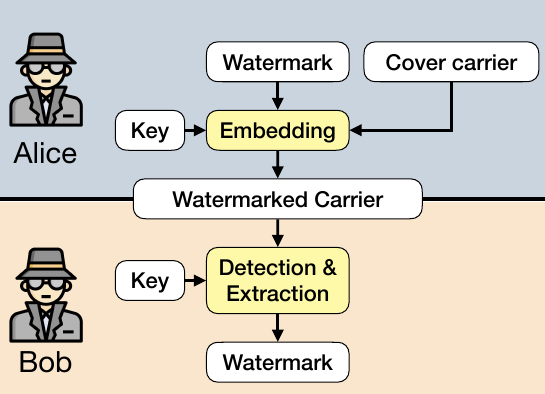}
        \caption{Watermarking.}
        
        \label{fig:sketch_wryptography}
    \end{subfigure}
    \caption{Common procedures of both sides (Alice and Bob) in steganography, cryptography, and watermarking.}
    
    \label{fig: Steganography_Cryptography_Watermarking}
\end{figure*}

% \begin{figure*}[!t]
%  \centering
%  \includegraphics[width=\textwidth]{Steganography_Cryptography_Watermarking.pdf} 
%  \caption{Common procedures of both sides (Alice and Bob) in steganography, cryptography, and watermarking.}
 
%  \label{fig: Steganography_Cryptography_Watermarking}
% \end{figure*}

\section{Comparison with Related Concepts}
\label{sec: Comparison with Related Concepts}

(1) \textbf{Cryptography} protects the content of a message, ensuring confidentiality, integrity, and authenticity, by transforming plaintext into ciphertext using mathematical algorithms and keys~\cite{RIVEST1990717}.

(2) \textbf{Watermarking} embeds information into a carrier to prove ownership, authenticity, or traceability, by inserting robust marks or patterns that survive common transformations (e.g., compression, distortion)~\cite{10.1117/1.1494075}.

In summary, Table~\ref{table:steg-crypto-watermark} compares the features of steganography, cryptography, and watermarking. 
In addition, Figure~\ref{fig: Steganography_Cryptography_Watermarking} illustrates their procedures, respectively.

\section{Two Definitions of Steganographic Security}
\label{sec: Two Definitions of Steganographic Security}

(1)~\citet{10.1007/3-540-49380-8_21} was the first to model steganographic security from an \textbf{information-theoretic} perspective. Given an observation $\mathbf{x}$, the security of a steganographic system can be quantified by the Kullback–Leibler (KL) divergence between the cover distribution $P_c$ and the stego distribution $P_s$:
\begin{equation}
    D_{\text{KL}}(P_c || P_s) = \sum_{\textbf{x} \in \mathcal{D}} P_c(\mathbf{x})\log \frac{P_c(\mathbf{x})}{P_s(\mathbf{x})},
\end{equation}
which measures the statistical difference between the two distributions. When $D_{\text{KL}}(P_c || P_s) = 0$, the steganographic system is considered to be \textit{perfectly secure}, as $P_c = P_s$ and thus the steganalyzer cannot distinguish the stego object from the cover object better than by random guessing.

(2)~\citet{10.1007/3-540-45708-9_6} and~\citet{katzenbeisser2002defining} proposed the \textbf{complexity-theoretic} definition of steganographic security: A steganographic system is computationally secure if, for all probabilistic polynomial-time (PPT) adversaries $\mathcal{A}$ in distinguishing between the output of steganography sampler $\mathcal{O}_s$ and the output of random sampler $\mathcal{O}_r$ is negligible.
\begin{equation}
    \left|\text{Pr}[\mathcal{A}^{\mathcal{O}_s}(1^{\lambda}) = 1] -  \text{Pr}[\mathcal{A}^{\mathcal{O}_r}(1^{\lambda}) = 1]\right| < \text{negl}(\lambda),
\end{equation}
% where $\text{negl}(\lambda)$ is the negligible function that correlates the length of the input of $\text{KeyGen}_{\mathcal{D}}(1^{\lambda})$.
where $\lambda$ is the security parameter, $1^{\lambda}$ denotes its unary encoding, and $\text{KeyGen}_{\mathcal{D}}(1^{\lambda})$ is the key-generation algorithm run on this parameter. The function $\text{negl}(\lambda)$ is negligible in $\lambda$, i.e., for every polynomial $p(\cdot)$ there exists an $N \in \mathbb{N}$ such that for all $\lambda > N$ it holds that $\text{negl}(\lambda) < \frac{1}{p(\lambda)}$. Intuitively, this means that any PPT adversary can only distinguish the two oracles with at most negligible advantage as a function of the security parameter.

\section{Review of Modification-Based Methods}
\label{sec: Review of Modification-Based Methods}

\subsection{Format-Based Methods}
\label{sec: Format-Based Methods}
\rev{Format-based methods exploit the physical features of text symbols (e.g., spaces, encoding, fonts, spelling) without altering the linguistic content; they are therefore visually unobtrusive but, strictly speaking, lie outside the scope of linguistic steganography. We summarize them here for completeness.}

\rev{\textbf{Space-based} methods hide bits in line/character/word/inter-paragraph spacing or invisible whitespace, by either inserting null spaces, shifting text vertically/horizontally, or jointly mapping multiple spacing operations to symbol blocks~\citep{464718, singh2012novel, POR20121075, 10.1007/978-81-322-1665-0_107, kumar2015efficient, 7813878, ramakrishnan2016text, gurunath2023new}.}

\rev{\textbf{Encoding-based} methods exploit the fact that visually identical glyphs may correspond to different code points, e.g., Unicode homoglyphs of English/Arabic letters, ASCII variants, and zero-width characters or joiners~\citep{rahma2013text, 10.1007/978-3-319-22915-7_26, 8711087, ditta2018information, 8440030, 8947188, 9291576}.}

\rev{\textbf{Font-based} methods carry bits via font color, mixed-case typesetting, character spacing, font-color formatting in spreadsheets, or specially designed fonts with hiding-friendly properties~\citep{5380435, ali2013new, ramakrishnan2016text, 10.1145/3279996.3280006, Baawi_2020}.}

\rev{\textbf{Spelling-based} methods substitute spellings (e.g., US/UK variants) to embed bits~\citep{4494159, 10.1145/1838002.1838082}.}

\rev{Despite their simplicity, format-based methods are (i)~fragile under text normalization, document conversion, or copy--paste; (ii)~detectable by statistical steganalysis on spacing, punctuation, or font usage~\citep{4299822}; and (iii)~highly platform- and language-dependent (e.g., spelling-based methods only apply to selected languages).}

\subsection{Lexicon-Based Methods}
\label{sec: Lexicon-Based Methods}
\rev{Lexicon-based methods modify individual words or word forms while preserving semantics and fluency. They divide into synonym-based and masked-LM-based methods.}

\rev{\textbf{Synonym-based} methods exploit lexical redundancy: bits are embedded by selecting among synonyms or regional variants subject to contextual, frequency, or grammatical constraints. Representative work uses collocation-verified synonyms~\citep{10.1007/978-3-540-30114-1_13}, US/UK variants~\citep{4604331}, language-specific grammar~\citep{5224169}, $n$-gram-checked contextual substitution~\citep{chang-clark-2010-practical, 6784896, chang-clark-2014-practical}, WordNet-based substitution~\citep{gardiner2012stegchat}, frequency-preserving encodings to defeat statistical steganalysis~\citep{xiang2017novel, xiang2019word, MAHATO2020216}, messaging-oriented covertext extension~\citep{serret2022linguistic}, and IoT-oriented ambiguous-token selection~\citep{electronics13214216}.}

\rev{\textbf{Masked-LM-based} methods mask tokens in the covertext and use a masked language model (typically BERT~\citep{devlin-etal-2019-bert}) to predict context-appropriate replacements that encode bits, going well beyond classical synonym substitution~\citep{ueoka-etal-2021-frustratingly, zheng2022autoregressive, 9948940, 9753668, 10315184, 10185651, 10400450}.}

\rev{Lexicon-based methods preserve the surface semantics of the covertext, but suffer from (i)~limited capacity, since only words with valid lexical variants can be substituted under context constraints; and (ii)~poor language generalizability, as many methods rely on language-specific resources (thesauri, WordNet) unavailable for low-resource languages.}

\subsection{Rewriting-Based Methods}
\label{sec: Rewriting-Based Methods}
\rev{Rewriting-based methods alter the syntactic structure of sentences or paraphrase them while preserving meaning, e.g., reordering constituents, switching voice, or inserting subordinate clauses, thus operating beyond the lexical level. They divide into syntax-based and paraphrasing-based methods.}

\rev{\textbf{Syntax-based} methods choose among alternative grammatical structures of the same sentence, e.g., via syntax banks combined with Shannon--Fano coding or signatures~\citep{wai2011syntactic}, joint lexical-syntactic transformations to lift capacity~\citep{7740349}, or multi-granularity hiding across syntactic and symbolic spaces~\citep{Ou_Xiang_Liu_2025}.}

\rev{\textbf{Paraphrasing-based} methods select among alternative phrasings of the same meaning. Representative ideas include rule-based paraphrasing followed by supervised filtering~\citep{10.1007/978-3-642-00382-0_43}, paraphrase dictionaries with $n$-gram and CCG-parser checks~\citep{chang-clark-2010-linguistic}, hash-chain-driven interactive rewriting~\citep{7056859}, hierarchical-LM unilingual translation for Twitter~\citep{wilson2014linguistic, wilson2016avoiding}, BART-based token-level rewriting~\citep{10.1007/978-3-031-30637-2_41}, Seq2Seq2Seq pivot-translation rewriting~\citep{10049662}, and attribute-controlled paraphrase generation~\citep{XIANG20252375}.}

\rev{Rewriting-based methods are particularly strong at preserving semantics and readability, often yielding the best imperceptibility among modification-based approaches. Their main drawback is \textbf{low embedding capacity}, as redundancy must come from entire sentences or syntactic structures rather than localized lexical slots.}

\section{Reasons of Researching Generative Linguistic Steganography}
\label{sec: Reasons of Researching Generative Linguistic Steganography}

(1) \textbf{High capacity.} Generative language models can utilize the redundancy of the whole vocabulary and flexibly choose a token conditioned on the targeted secret message at each inference position, enabling a much higher embedding rate than modification-based approaches that rely only on limited redundant positions in existing text.

(2) \textbf{Advances of language models.} 
% Early generative steganography methods relied on traditional statistical language models such as n-gram algorithms~\cite{5741643} and hidden Markov models~\cite{moraldo2014approachtextsteganographybased}, which were severely limited in terms of grammatical correctness, fluency and semantic richness.
% The introduction of neural language models, particularly recurrent neural networks (RNNs)~\cite{8470163}, such as long short-term memory (LSTM)~\cite{fang-etal-2017-generating} and gated recurrent unit (GRU) architectures~\cite{9975301}, enhanced the coherence and contextual consistency of stegotexts. 
% With the advent of pre-trained generative models built on the transformer architecture~\cite{NIPS2017_3f5ee243}, such as GPT-2~\cite{radford2019language}, a decisive breakthrough was achieved over earlier RNN-based models. Unlike sequential architectures that struggled with long-range dependencies and often produced repetitive or incoherent text, transformers leverage self-attention mechanisms to capture global contextual relationships efficiently. This advancement enabled researchers to handle flexible and diverse prompts to generate more natural and human-like stegotexts.
% The progress to large language models (LLMs), including the OPT~\cite{zhang2022optopenpretrainedtransformer}, LLaMA~\cite{touvron2023llama}, and Qwen~\cite{bai2023qwentechnicalreport} families, has further transformed the landscape.
A milestone marking the entry into the era of LLMs was the release of OpenAI’s GPT-3~\cite{NEURIPS2020_1457c0d6}, the first widely recognized model to demonstrate that scaling transformer architectures to hundreds of billions of parameters and training them on massive corpora could achieve unprecedented performance across various tasks.
Building on it, subsequent open-weight LLMs are used to generate stegotexts that are both indistinguishable from human writing and adaptable across domains and languages.
% As a result, generative linguistic steganography has entered a new era in which the capabilities of LLMs directly support higher imperceptibility, increased capacity, and broader applicability than was possible with earlier generations of models.

(3) \textbf{Imperceptibility and security.}
The capabilities of LLMs directly support higher imperceptibility. 
Moreover, recent advances have shown that provably secure steganographic methods (e.g.,~\citet{zhang-etal-2021-provably} and~\citet{10179287}) can be realized within generative frameworks. These approaches equip generative steganography with strict security guarantees, thereby advancing it beyond empirical validation toward theoretical resilience against steganalysis.

\section{Review of Modern Generative Linguistic Steganographic Methods}
\label{sec: Modern Generative Linguistic Steganographic Methods}
Table~\ref{table:GLS_taxonomy} presents a detailed taxonomy of modern generative linguistic steganographic methods, which supplements the rough taxonomy shown in Figure~\ref{fig: The main content flow and categorization of generative linguistic steganography}. It is based on the following seven dimensions:
\begin{itemize}
    \item \rev{\textbf{Targeted metric(s).} This dimension refers to what kind of improvements are aimed at by the proposed method. Explanations of these three types (security, efficiency, and robustness) are introduced in Section~\ref{sec: Taxonomy of Existing Methods}.}
    \item \rev{\textbf{Scenario(s).} This dimension refers to the designated scenarios in which the steganographic method is proposed. The methods that can be applied to general generation scenarios are labeled \textit{General}.}
    \item \rev{\textbf{Backbone model(s).} This dimension refers to each method's language-model requirements or assumptions. Methods that apply across generative autoregressive LMs (for example, those optimizing embedding algorithm and extraction algorithm without relying on a particular architecture) are labeled \textit{Model-agnostic}.}
    \item \rev{\textbf{LLM-adaptable.} This dimension indicates whether the method proposed is implemented based on LLM(s) or it has the model-agnostic feature can be transferred to LLMs. \textit{Notably, methods which are LLM-adaptable but not model-agnostic can be considered as LLM-specific methods.}}
    \item \rev{\textbf{Training-free.} This dimension indicates whether the method operates without additional training or fine-tuning of the language model (i.e., using an off-the-shelf LM).}
    \item \rev{\textbf{Asymmetric.} This dimension indicates whether embedding and extraction are not mirror operations, i.e., the extraction procedure fundamentally differs from embedding. For example, embedding requires LM inference, while extraction does not.}
    \item \rev{\textbf{Black-box.} This dimension indicates whether the method operates without access to an LM’s internal states (e.g., next-token probabilities) and can be implemented through an API-only interaction.}
\end{itemize}
    
\rev{Organized chronologically, Table~\ref{table:GLS_taxonomy} reveals the following trends in this field:}

\rev{(1) \textbf{Efficiency}, \textbf{robustness}, or multi-metric optimization has received growing emphasis. Earlier studies mainly focused on improving security, while more recent methods increasingly consider practical performance under multiple objectives.}

\rev{(2) \textbf{Model-agnostic}, \textbf{LLM-adaptable} and \textbf{training-free} methods started dominating since 2024. This trend is closely related to the rapid adoption of large language models, where methods are expected to work across different backbones without costly retraining or architecture-specific modification.}

\rev{(3)  \textbf{Asymmetric} and \textbf{black-box} methods enable lightweight implementations, but almost remain underexplored. These settings better reflect practical interactions with commercial LLM APIs, where users often only have access to generated text rather than to the model internals.}

\rev{(4) Most methods fall under the \textbf{general scenarios}, which means that they are not tied to a single language and can support multiple languages based on the underlying language model. Therefore, beyond a few language-specific studies on Arabic, Malay, and Chinese poetry, the majority of recent methods can be regarded as multilingual in scope. This suggests that multilingual applicability has become an implicit feature of modern generative linguistic steganography.}

In this section, we review several representative methods for each target metric.

\subsection{Toward Higher Security}
\label{sec: Toward Higher Security}

(1) \textbf{Methods based on optimizing modeling.}
These methods mark the first wave of neural linguistic steganography emerging alongside neural language models.
\citet{fang-etal-2017-generating} are one of the earliest attempts to directly use LSTMs for language modeling to generate stegotexts, and its embedding and extraction are based on block coding.
Building on this line, \citet{8470163} formalize the RNN-based embedding paradigm and improve the coding strategy: It ranks next-token probabilities to form a candidate pool, but embeds bits using fixed-length or variable-length Huffman coding. 
Further, \citet{9193914} discover the \textit{Psic effect}, i.e., perceptual-imperceptibility and statistical-imperceptibility conflict effect in linguistic steganography, and propose a new linguistic steganography based on Variational Auto-Encoder to guarantee both the perceptual-imperceptibility and statistical-imperceptibility.

(2) \textbf{Methods toward provable security.} 
These methods aiming at provable security largely pursue distribution-preserving at each generative step after steganographic encoding. Without considering language modeling, these methods are generally \textbf{training-free}.
The early baseline is arithmetic coding (AC)~\cite{ziegler-etal-2019-neural}, which maps a bitstream to successive token intervals so that sampling closely follows the LM's next-token distribution, yielding low per-step KL distortion and thus strong statistical imperceptibility. 
Building on this, self-adjusting arithmetic coding~\cite{shen-etal-2020-near} dynamically trims/reshapes the candidate pool to meet a user-specified imperceptibility (KL divergence) threshold, improving the security–capacity trade-off over vanilla AC.
ADG (Adaptive Dynamic Grouping)~\cite{zhang-etal-2021-provably} aims at provably perfect matching by recursively grouping tokens into equal-probability bins and selecting within bins.
For broader cryptographic guarantees under realistic settings, Meteor~\cite{10.1145/3460120.3484550} provides a symmetric-key scheme that achieves computational security while sampling from generative approximations of natural language.
iMEC~\cite{witt2023perfectly} shows that a steganography procedure is perfectly secure under an information-theoretic model of steganography if and only if it is induced by a coupling.
To make perfect security practical, Discop~\cite{10179287} constructs multiple ``distribution copies'' and samples from the chosen copy indexed by message bits, preserving the original distribution (\textbf{zero KL divergence}) while improving efficiency.
Recently, SparSamp~\cite{wang2025sparsampefficientprovablysecure} further reduces overhead via sparse, message-derived sampling intervals, maintaining provable security but significantly boosting efficiency of embedding and extraction.
\rev{Recently, \citet{yan2026efficientprovablysecurelinguistic} proposes an efficient provably secure method based on rotation range coding (RRC) of which approximate 100\% entropy utilization for embedding capacity is empirically validated.}
% Table~\ref{tab:methods toward provable security} lists the features of existing provably secure methods. We can observe a clear evolutionary trajectory from non-zero KL to KL = 0, and then toward lower complexity.

(3) \textbf{Methods powered by LLMs.}
These methods are marked by black-box access, prompt-driven control, or higher-level strategy design beyond the basic token-by-token generation.
\citet{10.1145/3664647.3680562} propose LLM-Stega, the first method tailored to \textbf{API-only} LLMs: instead of relying on internal logits, it builds keyword sets and an encrypted mapping at the interface level, showing that practical steganography is feasible even without probability access (\textbf{black-box}).
\citet{lin-etal-2024-zero} propose a \textbf{prompt-driven} linguistic steganography with an \textbf{in-context learning} framework, explicitly targeting both perceptual and statistical imperceptibility.
Most recently, \citet{zhou2025autostegaagentdrivenlifelongstrategy} frame LLM steganography as a \textbf{self-evolving agent system}: an agent loop automatically discovers, composes, and updates a library of embedding strategies at inference time, allowing adaptation to domains and constraints.

\subsection{Toward Higher Efficiency}
\label{sec: Toward Higher Efficiency}

(1) \textbf{Methods based on more embedding units.}
These methods mainly push linguistic steganography below the word or token level, exploiting characters and script-specific symbols as basic embedding units to boost capacity.
\citet{math8091558}  use an LSTM-based character-level language model to embed at least one bit per character, achieving faster generation and significantly higher embedding rates than earlier word-level neural methods.
\citet{9864312}  hide multiple bits per Arabic letter by modulating fine-grained script features such as elongation, pseudo-spaces, and dotted/diacritic patterns, realizing high-capacity, letter-based embedding tailored to Arabic and related scripts.
\citet{10831652} apply a tokenization-free character-level scheme with entropy-based candidate construction and lexical emphasis, which substantially reduces out-of-vocabulary cases and improves imperceptibility.  
\citet{11008605} combine character-level embedding with foundation LLMs, designing a stegosystem that operates at the character granularity and reports large gains in bits-per-word and bits-per-character over prior neural methods while maintaining comparable security and imperceptibility.
\citet{yan2026hitmshighthroughputmultistreamlinguistic} propose the High-Throughput Multi-Stream (HiTMS) framework, which distributes a secret across multiple responses, thereby amortizing the cost of model invocation and substantially improving throughput.

(2) \textbf{Methods based on improving entropy or utilization.} These methods mainly aim to make full use of the entropy provided by generative models rather than creating more embedded positions.
\citet{10919130} revisit prefix-based schemes (e.g., AC/Meteor) and propose two enhancements:  security-enhanced and efficiency-enhanced variants. The first optimizes quantization distortion to minimize KL divergence. The second redesigns the sampling mechanism via distribution coupling to improve the steganographic capacity. 
\citet{bai2025shimmer} introduce a mechanism of entropy collection that gathers residual entropy left unused in each embedding step and recycles it in subsequent steps, resulting in higher capacity than prior distribution-preserving methods.
\citet{pang2025provablesecuresteganographybased} improves entropy utilization in \textbf{black-box} settings: instead of requiring access to model's logits, it  embeds messages across multiple candidate samples while keeping the generator's output behavior statistically unchanged, achieving provable security and competitive capacity with only API-level access.
\citet{wang2025sparsampefficientprovablysecure} propose sparse, message-driven sampling, which combines the message with pseudo-random numbers to form sampling indices and increasing sampling intervals. It preserves the original probability distribution and introduces minor additional complexity per sampling step and nearly full entropy utilization, enabling the fastest embedding speed without compromising generation speed. 
% \citet{10.1145/3658664.3659657}; 
% \citet{10.1145/3082031.3083240}

% \citet{Text_steganography_on_RNN-Generated-lyrics}

% \citet{math8091558} propose CLLS, a character-level LSTM steganography scheme that embeds bits at every time by operating on next-character probabilities rather than words, so that this method can achieve much higher embedding capacity than traditional word-level or token-level methods.

% \citet{9864312} propose a way to hide secret data inside newly formulated Arabic poetry based on previous Arabic poetic texts. It embeds particularly at the level of letters, so this work is targeted at higher embedding capacity.

% \citet{9714779}

% \citet{varol2022lzw}

% \citet{10831652}

% \citet{10.1145/3664476.3670930}

% \citet{10.1145/3658664.3659657}

% \citet{10.1007/978-981-99-8073-4_4}

% \citet{bai2024semanticsteganographyframeworkrobust}

% \citet{10.1007/978-981-96-7005-5_12}

% \citet{10919130} revisit provably secure \textit{prefix} schemes (which embed bits by mapping them to shared probability intervals during LM sampling) and contributes two upgrades: a security-enhanced quantization that optimizes the partitioning distortion to minimize KL divergence and an efficiency-enhanced sampling redesign via distribution coupling that raises payload without extra computational overhead.

% \citet{pang2025frestegaplugandplaymethodboosting}

% \citet{huang2025relativelysecurellmbasedsteganographyconstrained}

% \citet{11008605}

% \citet{pang2025provablesecuresteganographybased}

% \citet{bai2025shimmer}

% \citet{11017758}

% \citet{wang2025sparsampefficientprovablysecure}

\subsection{Toward Higher Robustness}
\label{sec: Toward Higher Robustness}

(1) \textbf{Methods to tackle tokenization inconsistency.} These methods address incorrect results or failures extraction when the sent tokens are different from received tokens after the \textbf{detokenization-retokenization} pipeline between Alice and Bob.
\citet{nozaki-murawaki-2022-addressing} first show that subword-based generators can produce multiple valid segments for the same surface text, causing occasional extraction errors (called \textbf{segmentation ambiguity}). They propose simple, language-agnostic disambiguation tricks that prune prefix-confusable tokens so that detokenization is unique without heavily modifying existing systems.
\citet{10215094} design a distribution-aware pruning strategy that resolves segmentation ambiguity while preserving the probability mass of remaining candidates, improving both security and imperceptibility. The method is realized by finding out a maximum weight independent set (MWIS) in each candidate graph.
\citet{10831370} further reduce distortion by adding a verification stage: instead of actively modifying candidate pools, an all-case extraction method is proposed to obtain possible true extracted results. Further, verification is adopted to filter wrong extracted results caused by segmentation ambiguity.
\citet{10804596} (SyncPool) then show how to eliminate segmentation ambiguity without sacrificing provable security by grouping all tokens with prefix relationships into ambiguity pools and using a shared CSPRNG to synchronously sample from these pools, keeping both candidate set size and token distribution exactly unchanged.
\citet{yan-murawaki-2025-addressing}  generalize from segmentation ambiguity to full tokenization inconsistency between Alice and Bob's models.
Recently, \citet{wang2026retoksyncselfsynchronizingtokenizationdisambiguation} propose ReTokSync, a self-synchronizing disambiguation framework, which triggers a corrective reset only when ambiguity actually occurs to mitigate the performance reduction caused by disambiguation.

(2) \textbf{Methods against external modification.}
% \citet{10191984};
% \citet{10075392};
These methods target robustness when a stegotext is perturbed by active tampering.
\citet{10888944} propose an adaptive robust enhancement framework using a sliding window approach, that can be applied on top of existing generative linguistic steganography schemes. 
\citet{11023508} present an \textbf{asymmetric} framework in which only the encoder has access to the model. Specifically, the encoder uses various permutations of distribution to hide secret bits, while the decoder relies on a sampling function to extract the hidden bits by guessing the permutation used.
% \citet{sym17091416};
\citet{10888607} integrate classical error-correcting codes into a generative linguistic stegosystem. 
% \citet{perry2025robuststeganographylargelanguage};
\citet{qi2025stead} leverage \textbf{diffusion language models} to achieve robust and provable security. Their method identifies robust positions in the partially parallel denoising trajectory. Furthermore, they introduce an error correction strategies, including pseudo-random error correction and neighborhood search correction, during steganographic extraction.

\section{Basics and Notions of Steganalysis}
\label{sec: Basics and Notions of Steganalysis}
According to Simmons' \textit{Prisoners' Problem}~\cite{Simmons1984} discussed in Section~\ref{sec: Steganography System}, in the steganographic scenarios, Eve (the steganalyzer) uses steganalysis methods to distinguish stegotexts from covertexts to determine whether or not she needs to block Alice and Bob's communication.
Linguistic steganalysis is a text binary classifier that maps an observed text to a decision on whether it is steganographic or not.

Formally, let $\mathcal{V}$ denote the vocabulary and $\mathcal{X} = \mathcal{V}^{*}$ the set of all finite-length texts, i.e., $x_i \in \mathcal{X}$.
Let $\mathcal{Y} = \{0,1\}$ be the label set, i.e., $y_i \in \mathcal{Y}$ where $y_i = 0$ indicates that $x_i$ is a covertext and $y_i = 1$ indicates that $x_i$ is a stegotext.
The linguistic steganalysis is then modeled as a binary classifier:
\begin{equation}
    g(\theta): \mathcal{X} \rightarrow \mathcal{Y},\ g(x;\theta) = \mathbf{1}[f(x;\theta) \geq \tau],
\end{equation}
in which $f(x;\theta)$ is a scoring function, $\tau$ is a decision threshold, and $\theta$ is the learnable parameters.
Figure~\ref{fig: Linguistic_steganalysis} shows a sketch of steganalysis. Note that Eve does not need to extract the embedded message and is unable to edit the sent text.
Among the classical typology of steganalysis attacks (e.g., stego-only, known-cover, known-stego, and chosen-stego attacks), the online decision problem faced by Eve in our setting is stego-only: at test time she only observes a single text $x \in \mathcal{X}$ and must decide whether $y=0$ or $y=1$ without access to the corresponding cover or the embedded message.

\section{Review of Linguistic Steganalysis Methods}
\label{sec: Review of Linguistic Steganalysis Methods}
In this section, we review representative linguistic steganalysis methods and introduce the features of various categories. 

\subsection{Methods Based on Self-Trained Word Representations}
This category mainly includes early methods based on neural modeling methods to extract and analyze text features, consisting of recurrent neural networks (RNNs), convolutional neural networks (CNNs), graph neural networks (GNNs), and hybrid modeling methods.
RNN-based detectors typically model the conditional distribution of the next token given previous context, and then use likelihood-based or perplexity-based criteria to identify abnormal sequences introduced by embedding operations. For example, LSTM~\cite{8903243, 8727932, 10.1145/3369412.3395067} or GRU~\cite{10021965} architectures can be trained on large collections of clean carrier texts to learn typical syntactic and semantic patterns.
% Early works in this category typically initialize word embeddings randomly and learn them jointly with the classifier. RNN-based models, such as LSTMs and GRUs, encode sequences by iteratively updating hidden states, enabling capture of temporal dependencies and contextual information throughout the text.
CNN-based models~\cite{10.1007/s11042-020-08716-w, XUE2022140, 10021965}, on the other hand, focus more on local text fragments and are effective at capturing stego-induced changes in short n-grams, such as unusual phrase constructions or statistically rare word combinations.
With the introduction of GNNs into linguistic steganalysis, researchers began to represent texts as graphs, where nodes may correspond to words, sentences, or documents, and edges encode semantic similarity, syntactic dependencies, or user–post relationships in social media platforms~\cite{9364681, 10097161, 10.1007/978-981-99-8073-4_9}. GNN-based steganalysis then performs message passing over these structures to aggregate multi-hop contextual information, enabling the detection of more global, structure-level anomalies that are difficult to capture using purely sequential models.
% Message-passing mechanisms are then used to propagate information along edges, allowing the model to aggregate signals from relevant neighbors and capture higher-order dependency structures that are difficult for purely sequential or convolutional models. 

\subsection{Methods Based on Pre-Trained Word Representations}
This category mainly includes modern methods based on pre-trained language models, consisting of masked LMs and generative LMs (or LLMs).
Method based on masked language modeling~\cite{9484749, 10.1007/978-3-030-69449-4_7, 9591452, 9806187, 10190738}, such as those built on BERT-like architectures, typically exploit the contextualized token representations and token-level prediction probabilities learned from large-scale unlabeled corpora. In a common pipeline, the pre-trained model is either frozen as a feature extractor or lightly fine-tuned on clean carrier texts, and then a simple classifier (e.g., MLP or shallow CNN) is trained on top of its hidden states to discriminate between stegotexts and covertexts.
Generative LMs (or LLMs) extend this idea by modeling the full left-to-right generation process. Instead of relying on hand-crafted or shallow neural features, \citet{yang2024nextgenerationsteganalysisllmsunleash} treat linguistic steganalysis as an LLM-based generative task. They fine-tune open LLMs with LoRA~\cite{hu2022lora} for stego vs. cover detection and analyze text distributions via length, perplexity and log-probability.
\citet{10.1145/3634814.3634837} treat ChatGPT as a black-box few-shot classifier: supply a prompt with up to 32 labeled examples and ask it whether a new sentence is steganographic or not.
\citet{10752347} target strongly concealed stegotexts, especially those generated by LLM-based steganography, and proposes an LLM-based steganalysis framework with two operation modes: generation and classification.
\citet{Wang_Zhou_Chen_Zhou_Yang_2025}  move from isolated sentences to social media settings, and propose STLC-KG, which fuses an LLM with a commonsense knowledge graph to catch logical inconsistencies introduced by steganography.

\subsection{Dataset Resources}
\rev{In this section, we introduce the three datasets which are mainstream for evaluating linguistic steganalysis: \textbf{IMDB}~\cite{maas-etal-2011-learning}, \textbf{Twitter}~\cite{go2009twitter}, and \textbf{News}~\cite{kaggle_all_the_news}.}

\rev{\textbf{IMDB} is applied by 43 out of 60 steganalysis methods (shown in Table~\ref{table:Steganalysis_features}). It contains up to 50,000 movie reviews. This dataset is initially designed for binary sentiment classification.}

\rev{\textbf{Twitter} is applied by 46 out of 60 steganalysis methods (shown in Table~\ref{table:Steganalysis_features}). It is a short-text dataset collected from the Twitter platform, and contains 1.6 million training tweets, including 800,000 positive and 800,000 negative samples. 
Twitter is widely used to evaluate the robustness of text analysis models in noisy short-text scenarios.}

\rev{\textbf{News} is applied by 31 out of 60 steganalysis methods (shown in Table~\ref{table:Steganalysis_features}). It consists of up to 2.7 million news articles and essays from 27 American publications.}

\rev{Generally, the evaluation protocol uses sentences from the dataset as covertexts and uses prefixes of these covertexts to prompt steganographic models to generate stegotexts. Although these three datasets are not modern, they remain the only widely adopted benchmarks. We suggest a limitation that \textit{although a degree of benchmark consistency is shown, the evolution of contemporary text domains and generation scenarios is lacked in steganalysis}.}

\section{Review of Evaluation Metrics}
\label{sec: Review of Evaluation Metrics}

Following the terminology in Section~\ref{sec: Basics and notions of LM}, we denote the generated stegotext by the token sequence $[s^{(0)},\dots,s^{(T-1)}]$.

\subsection{Security}
\label{sec: Security metrics}
\subsubsection{Metrics Based on KL Divergence}
According to the information-theoretic perspective of steganography security in Section~\ref{sec: Security of Steganography}, the KL divergence between covertexts and stegotexts matters~\cite{9193914, zhang-etal-2021-provably}. One typical way to calculate this \textbf{text-level} KL divergence is based on a third-party sentence vectorization tool~\cite{pmlr-v32-le14} to map all covertexts and stegotexts to fixed length dense vectors $\boldsymbol{v}_c$ and $\boldsymbol{v}_s$. Assuming that these vectors obey isotropic Gaussian distribution, the divergence is denoted as:
\begin{align}
& D_{\text{KL-text}}  =  D_{\text{KL}}(p(\boldsymbol{v}_c) \,\|\, p(\boldsymbol{v}_s)) \notag \\
 & \approx 
\sum (
    \log\frac{\boldsymbol\sigma_s}{\boldsymbol\sigma_c}
    + \frac{\boldsymbol\sigma_c^2 + (\boldsymbol\mu_c - \boldsymbol\mu_s)^2}{2\boldsymbol\sigma_s^2}
    - \frac{1}{2}
),
\label{eq:kl_text_level}
\end{align}
in which $\boldsymbol{\mu}$ and $\boldsymbol{\sigma}$ are the mean and standard deviation of sentence vectors. 
The lower text-level KL divergence indicates greater security.

In addition to the above text-level KL divergence, a formal statistical notion of near-imperceptibility is proposed based on \textbf{step-level} KL divergence~\cite{dai-cai-2019-towards}. Specifically, the average step-level KL divergence is denoted as:
\begin{equation}
    D_{\text{KL-step}} = \frac{1}{T}\sum_{t = 0}^{T-1}D_{\text{KL}}(\boldsymbol{p}^{(t)}||\boldsymbol{\hat{p}}^{(t)}),
\end{equation}
in which $\boldsymbol{p}^{(t)}$ is the original probability distribution at $t^{th}$ step, and $\boldsymbol{\hat{p}}^{(t)}$ is transformed from $\boldsymbol{p}^{(t)}$ via steganographic sampling and encoding.
The lower step-level KL divergence indicates a higher statistical imperceptibility. Its calculation has two variants: (1) inference based on random-sampling sequences (e.g., theoretical analyses in~\citet{dai-cai-2019-towards}), and (2) inference based on steganographic sequences (that including its reverse versions are commonly applied)~\cite{ziegler-etal-2019-neural}.

Besides, several studies also use KL-derived divergences, most notably the Jensen–Shannon divergence ($\text{JSD}$), to quantify the gap between stegotext and covertext~\cite{9193914, lin-etal-2024-zero}. 

\subsubsection{Metrics Based on Text Quality}
First, one of the most common metrics to evaluate text quality and fluency is perplexity ($\text{PPL}$)~\cite{jelinek1977perplexity}. The calculation of $\text{PPL}$ is:
\begin{align}
    & \text{PPL}([s^{(0)},\dots,s^{(T-1)}]) \notag \\ &= \exp\left\{-\frac{1}{T-1} \sum_{t=1}^{T-1}\log p^{(t)}_{s^{(t)}} \right\},
\end{align}
in which $\log p^{(t)}_{s^{(t)}}$ is the log-likelihood of the $t$-th token conditioned on the preceding tokens. The calculated result depends on the adopted evaluation model, and a lower score is better.

Besides, text quality can be measured via similarity between the stegotext (candidate) and its covertext (reference) using BLEU~\cite{Papineni02bleu:a, lin-och-2004-orange}, ROUGE~\cite{lin-2004-rouge}, and METEOR~\cite{banarjee2005}. 
BLEU computes the geometric mean of modified n-gram precisions to estimate how similar a candidate text is to one or more reference texts at the lexical n-gram level.
ROUGE measures content overlap by counting matching n-grams, skip-bigrams, or longest common subsequences between a candidate and reference to assess how much of the reference’s content the candidate preserves.
METEOR aligns candidate and reference at the unigram level, then computes a weighted mean of precision and recall with a fragmentation penalty, aiming to better capture semantic adequacy and correlate with human judgments than n-gram precision alone.
For these metrics, higher scores indicate greater similarity.

In addition, to evaluate the diversity of stegotexts, the metric called \textit{distinct-}$n$ computes $\frac{|\text{unique } n\text{-grams}|}{\text{|all } n\text{-grams}|}$ over the generated text to quantify lexical diversity~\cite{li-etal-2016-diversity}. Higher values indicate less repetition and more varied output.

Beyond traditional automatic metrics, human evaluation and LLM-as-a-judge~\cite{gu2025surveyllmasajudge} are used to rate various dimensions, such as text fluency, naturalness, and logic, which provide human or near-human perceptual assessments.
% \begin{itemize}
%     \item PPL
%     \item BLEU
%     \item BERTScore
%     \item Rouge
%     \item ...
% \end{itemize}

Note that several studies not only use absolute values to denote text quality, but also use the metric differences between stegotexts and covertexts to denote imperceptibility. For example, the perplexity difference $\Delta \text{PPL}$ is calculated when the perceptual-imperceptibility and statistical-imperceptibility conflict effect (\textit{Psic Effect}) is taken into account~\cite{9193914}.

\subsubsection{Metrics Based on Steganalysis}
Following the concepts and introduction in Section~\ref{sec: Linguistic Steganalysis}, steganalysis equips Eve with a binary classifier to distinguish stegotext from covertext. The Lower classifier $\text{Accuracy}$, $\text{Precision}$, $\text{Recall}$, or $\text{F1}$ implies a higher imperceptibility of the stegosystem. Let $\text{TP}$, $\text{TN}$, $\text{FP}$, $\text{FN}$ be true/false positives/negatives. $\text{Accuracy} = \frac{\text{TP} + \text{TN}}{\text{TP} + \text{TN} + \text{FP} + \text{FN}}$, $\text{Precision} = \frac{\text{TP}}{\text{TP} + \text{FP}}$, $\text{Recall} = \frac{\text{TP}}{\text{TP} + \text{FN}}$, and $\text{F1} = \frac{2\text{TP}}{2\text{TP} + \text{FP} + \text{FN}}$. These four indicators are suitable for scenarios in which the positive and negative samples are (approximately) balanced.

\subsection{Efficiency}
\label{sec: Efficiency metrics}
\subsubsection{Embedding Capacity}
Bits per token ($\text{BPT}$) is the standard measure of embedding capacity (\textbf{bit-rate efficiency}), defined as the number of embedded bits divided by the number of tokens in the stegotext: $\text{BPT} = \frac{|\mathbf{m}|}{T}$, where $|\mathbf{m}|$ is the bit number of the embedded message $\mathbf{m}$ and $T$ is the token number of the generated stegotext.
Some studies report variants, e.g., bits per word~\cite{fang-etal-2017-generating, 8470163}, to express capacity at a different granularity.

Besides, entropy utilization is assessed, which normalizes the payload by the generator’s information budget and thus enables fair cross-model comparisons. It is defined as
\begin{align}
   & \text{Entropy utilization}  = \frac{|\mathbf{m}|}{\sum_{t = 0}^{T-1}\text{Entropy}(\boldsymbol{p}^{(t)})} \notag \\ & = \frac{|\mathbf{m}|}{\sum_{t = 0}^{T-1}\sum_{i = 1}^{|\mathcal{V}|}- p^{(t)}_i\log_2 p^{(t)}_i},
\end{align}
in which $\boldsymbol{p}^{(t)}$ is the (possibly truncated) distribution used to sample the $t$-th token, and $|\mathcal{V}|$ is the vocabulary size.

Additionally, the embedding rate $\text{ER}$ is used to indicate the percentage of payload bits relative to the total number of bits in the generated stegotext (e.g., under UTF-8 encoding):
\begin{equation}
    \text{ER} = \frac{|\mathbf{m}|}{\text{bits}([s^{(0)},\dots,s^{(T-1)}])} \times 100\%.
\end{equation}

\subsubsection{Operation Speed}
To assess computational complexity and \textbf{runtime efficiency}, some studies report operation-speed metrics, particularly (i) the average time to embed/extract one bit (e.g., ms/bit) and (ii) the average time to embed/extract an entire stegotext (e.g., ms/message).

\subsection{Robustness}
\label{sec: Robustness metrics}
As there are possible tokenization inconsistencies between Alice and Bob, active attacks, and even computational indeterminism (discussed in Section~\ref{sec: Challenges and Future Directions}), it is significant to assess whether the extracted message $\mathbf{m}^{\prime}$ is equal to the embedded message $\mathbf{m}$.
The standard metric is the message error rate ($\text{MER}$), which is the ratio of $\mathbf{m}^{\prime} \neq \mathbf{m}$ cases to all cases: $\text{MER} = \frac{1}{N}\sum_{i = 1}^{N}\mathbf{1}\{\mathbf{m}^{\prime} \neq \mathbf{m}\}$, in which $N$ is the number of stegotext samples.
Several variants, such as the bit error rate ($\text{BER}$), are also reported, but length mismatches between $\mathbf{m}^{\prime}$ and $\mathbf{m}$ can arise from the chosen encoding scheme or from various types of active attack, and length mismatches should be handled explicitly.

\begin{figure*}[!t]
 \centering
 \includegraphics[width=1.0\textwidth]{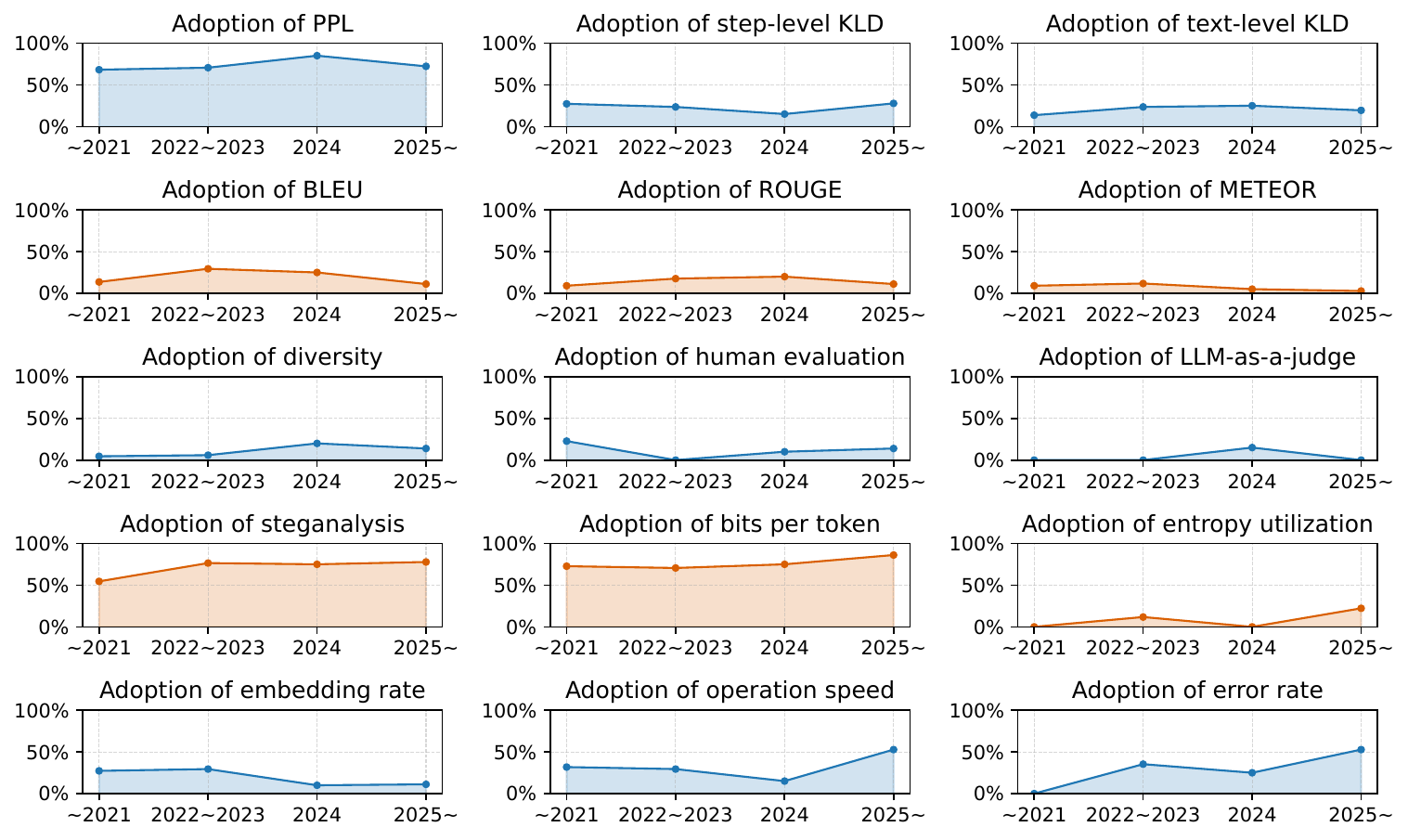} 
 \caption{Adoption rates of the 15 major evaluation methods in \rev{95} papers on modern generative linguistic steganography, presented in chronological order. Among these \rev{95} papers, 22 were published before 2021, 17 between 2022 and 2023, 20 in 2024, and 36 after 2025.}
 
 \label{fig: Evaluation adoption}
\end{figure*}

\section{Introduction of Challenges and Future Directions}
\label{sec: Introduction of Challenges and Future Directions}
(1) \textbf{Computational indeterminism.} \citet{atil2025nondeterminismdeterministicllmsettings} report that LLMs in hosted environments exhibit computational indeterminism, i.e., the probability distribution over next-token predictions is not fixed even for the same prompt and the same model. When applied to linguistic steganography, this phenomenon substantially undermines the robustness (i.e., extraction success rate) of API-only or black-box steganographic methods.

(2) \textbf{Tokenization inconsistency.} 
Even though tokenization inconsistency between Alice and Bob can be prevented from affecting extraction (see Section~\ref{sec: Toward Higher Robustness}), existing countermeasures cannot simultaneously achieve provable security and high capacity. In particular, for the provably secure disambiguation method SyncPool~\cite{10804596}, the achievable capacity drops to zero when it is required to operate over the entire vocabulary.

(3) \textbf{Abrupt termination.}
In current experiments, generation usually stops abruptly once the hidden message has been fully embedded (e.g., the samples reported in~\citet{ziegler-etal-2019-neural}), producing obviously unnatural stegotext, and the decoded messages often include extra bits. 
% A length-prefixed framing scheme would address this.

(4) \textbf{Challenges of synchronizing prompts.}
Fixing a single prompt as a hyperparameter makes steganography inflexible and poorly adapted to different topics, styles, or contexts. On the other hand, synchronizing prompts dynamically (e.g., by sharing or updating them each time) implicitly opens an additional side channel outside the stegotext itself, and this auxiliary channel must also be secured and authenticated. Designing schemes that allow prompt adaptation while avoiding new vulnerabilities therefore remains a non-trivial challenge.

(5) \textbf{Limited embedding capacity.} Compared to steganography based on images, audio, or video, linguistic steganography suffers from an inherently lower embedding capacity, since the smallest manipulable unit is typically a token, word, or character, carrying much more semantic load than a single pixel. Beyond improving entropy utilization, several directions could offer breakthroughs in capacity: (a) integrating steganography with \textbf{information retrieval}~\cite{10.1007/978-981-99-8073-4_4}, (b) enhancing \textbf{diversity}~\cite{10.1145/3658664.3659657}, and (c) using \textbf{diffusion language models}~\cite{10.1007/978-981-96-7005-5_12, qi2025stead}.

(6) \textbf{High dependence on white-box LLMs.} According to Table~\ref{table:GLS_taxonomy}, most generative steganographic methods are symmetric, requiring both Alice and Bob to access the same white-box language model for embedding and extraction. As LLMs continue to scale up, their computational and memory demands also grow, pushing such symmetric white-box assumptions further away from realistic deployment scenarios. This suggests that \textbf{black-box} and \textbf{asymmetric} designs (where only the sender has access to a powerful model, or where interaction is limited to API calls) are particularly promising directions for improving computational efficiency and practicality.

(7) \textbf{Unrealistic security evaluations.}
Current provably secure methods~\cite{10179287, wang2025sparsampefficientprovablysecure} focus on minimizing the divergence between stegotexts and LM-generated covertexts, but such distributional guarantees alone do not ensure imperceptibility in realistic, human-facing scenarios. 
LM-based evaluations (e.g., KL divergence and perplexity) should be regarded only as approximations, which is only justified by the lack of access to the ground-truth distribution.
Moreover, as summarized in Figure~\ref{fig: Evaluation adoption}, \textbf{human evaluation} and \textbf{LLM-as-a-judge} metrics are increasingly crucial in the LLM era, yet they are still missing from the majority of existing works.

(8) \textbf{Lack of unity in benchmarks.}
For current works of linguistic steganography, there are remarkable inconsistencies in selecting (a) datasets or prompts, (b) language models, (c) steganalysis methods, (d) other metrics.

(9) \textbf{Ethics and use restrictions of LLM licenses.}
Linguistic steganography faces an unlikely existential threat: LLM licenses. A family of use-restricted licenses, including OpenRAIL and Meta's Llama licenses, explicitly prohibit certain forms of deception, such as representing model use or model outputs as human-generated, which lies at the core of linguistic steganography. The critical perspectives on these licenses come largely from outside the steganography literature. For example, \citet{CuiAraujo2024UseRestricted} argue that use-restricted licenses sit uneasily with established open-weight principles and are unlikely to provide effective governance on their own. In parallel, work on model openness and on an explicit Open Source AI Definition treats Llama-style and OpenRAIL-style licenses as source-available or open-weight rather than genuinely open source, precisely because they impose behavioral or field-of-use restrictions~\cite{WhiteEtAl2025ModelOpenness,OSAID1.0}. We believe that the steganography research community also needs to add its voice to this debate.

\rowcolors{2}{gray!10}{white}
\begin{table*}[!t]
\renewcommand{\arraystretch}{0.6}
\centering

\scalebox{0.615}{
\begin{tabular}{m{1cm}|m{4.5cm}|>{\centering\arraybackslash}m{3.2cm}|>{\centering\arraybackslash}m{2.2cm}|>{\centering\arraybackslash}m{5cm}|>{\centering\arraybackslash}m{1.8cm}|>{\centering\arraybackslash}m{1.5cm}|>{\centering\arraybackslash}m{2cm}|>{\centering\arraybackslash}m{0.75cm}}
\toprule[1.0pt]
\textbf{Year} & \textbf{Reference} &  \textbf{Targeted metric(s)} & \textbf{Scenario(s)} &  \textbf{Backbone model(s)} &\textbf{\rev{LLM-adaptable?}} &\textbf{Training-free?} & \textbf{Asymmetric?} & \textbf{Black-box?} \\
\hline
2017  & \citet{fang-etal-2017-generating} &  Security  & General & \graytext{LSTM} & \cellno & \cellno & \cellyes & \cellno \\
2017 & \citet{10.1145/3082031.3083240}  & \bluetext{Efficiency}  & \graytext{Lyric} & \graytext{LSTM} & \cellno & \cellno & \cellno & \cellno \\

\hline
% \citet{8440030} & Security  & \cellcolor{gray!50} Social media & \cellno & \cellno & \cellno \\
2018 & \citet{yang2018automaticallygeneratesteganographictext} &  Security  & General & \graytext{Markov} & \cellno & \cellno & \cellno & \cellno \\

\hline
2019 & \citet{8470163} &  Security  & General & \graytext{LSTM} & \cellno & \cellno & \cellno & \cellno \\
2019 & \citet{ziegler-etal-2019-neural} & Security  & General & Model-agnostic & \cellyes &  \cellyes & \cellno & \cellno \\
2019 & \citet{dai-cai-2019-towards} & Security  & General & Model-agnostic & \cellyes & \cellyes & \cellno & \cellno \\
2019 & \citet{9023313} & Security  & General & Model-agnostic & \cellyes & \cellyes & \cellno & \cellno \\
2019 & \citet{Text_steganography_on_RNN-Generated-lyrics} & \bluetext{Efficiency}  & \graytext{Lyric} & \graytext{LSTM} & \cellno & \cellno & \cellno & \cellno \\

\hline
2020 & \citet{yang2020graphstegasemanticcontrollablesteganographic} & Security  & General & \graytext{LSTM} & \cellno & \cellno & \cellno & \cellno \\
2020 & \citet{10.1007/978-3-030-43575-2_2} & Security  & General & \graytext{GRU} & \cellno & \cellno & \cellno & \cellno \\
2020 & \citet{shen-etal-2020-near} & Security  & General & Model-agnostic & \cellyes & \cellyes & \cellno & \cellno \\
2020 & \citet{doi:10.2352/ISSN.2470-1173.2020.4.MWSF-291} & Security  & General & \graytext{LSTM} & \cellno & \cellno & \cellno & \cellno \\
2020 & \citet{doi:10.1177/1550147720914257} & Security  & General & \graytext{Markov} & \cellno & \cellno & \cellno & \cellno \\
2020 & \citet{math8091558} & \bluetext{Efficiency}  & General & \graytext{LSTM} & \cellno & \cellno & \cellno & \cellno \\

\hline
2021 & \citet{9193914} &   Security  & General & \graytext{LSTM, BERT} & \cellno & \cellno & \cellno & \cellno \\
2021 & \citet{zhang-etal-2021-provably} &  Security  & General & Model-agnostic & \cellyes  &  \cellyes & \cellno & \cellno \\
2021 & \citet{9353234} &  Security  & General & \graytext{LSTM, GRU, Transformer} & \cellno & \cellno & \cellno & \cellno \\
2021 & \citet{10.1145/3460120.3484550} & Security  & General & Model-agnostic & \cellyes & \cellyes & \cellno & \cellno \\
2021 & \citet{9280343} & Security  & General & \graytext{Transformer} & \cellno & \cellno & \cellno & \cellno \\
2021 & \citet{10.1145/3418598} & Security  & General & \graytext{LSTM, Transformer} & \cellno & \cellno & \cellno & \cellno \\
2021 & \citet{yang2021generation} & Security  & General & \graytext{LSTM} & \cellno & \cellno & \cellno & \cellno \\
2021 & \citet{WANG20211375} & Security  & \graytext{Lyric} & \graytext{LSTM} & \cellno & \cellno & \cellno & \cellno \\

\hline
2022  & \citet{9975301} &  Security  & \graytext{Lyric} & \graytext{GRU} & \cellno & \cellno & \cellno & \cellno \\
2022 & \citet{yu2022mts} & Security  & General & Model-agnostic & \cellyes & \cellyes & \cellno & \cellno \\
2022 & \citet{9430708} & Security  & General & \graytext{LSTM, CNN} & \cellno & \cellno & \cellno & \cellno \\
2022 & \citet{9714779} & Security/\bluetext{Efficiency}  & General & \graytext{BERT} & \cellno & \cellyes & \cellno & \cellno \\
2022 & \citet{varol2022lzw} & \bluetext{Efficiency}  & General & \graytext{LSTM} & \cellno & \cellno & \cellno & \cellno \\
2022 & \citet{9864312} & \bluetext{Efficiency}  & \graytext{Arabic text} & \graytext{LSTM} & \cellno & \cellno & \cellno & \cellno \\
2022 & \citet{nozaki-murawaki-2022-addressing} & \orangetext{Robustness}  & General & Model-agnostic & \cellyes & \cellyes & \cellno & \cellno \\

\hline
2023 & \citet{SUN2023157} &  Security  & General & \graytext{T5} & \cellno & \cellno & \cellno & \cellno \\
2023 & \citet{10191218} & Security  & General & Model-agnostic & \cellyes & \cellyes & \cellno & \cellno \\
2023 & \citet{10115433} & Security  & General & \graytext{Transformer} & \cellno & \cellno & \cellno & \cellno \\
2023 & \citet{10179287} & Security  & General & Model-agnostic & \cellyes & \cellyes & \cellno & \cellno \\
2023 & \citet{10.1007/978-3-031-44696-2_62} & Security  & General & \graytext{BART, GPT-2} & \cellno & \cellyes & \cellno & \cellno \\
2023 & \citet{10095722} & Security  & \graytext{Image-to-text} & \graytext{ClipCap} & \cellno & \cellno & \cellno & \cellno \\
2023 & \citet{witt2023perfectly} & Security  & General & Model-agnostic & \cellyes & \cellyes & \cellno & \cellno \\
2023 & \citet{10215094} & Security/\orangetext{Robustness}  & General & Model-agnostic & \cellyes & \cellyes & \cellno & \cellno \\
2023 & \citet{10191984} & Security/\orangetext{Robustness}  & \graytext{Translation} & \graytext{Transformer} & \cellno & \cellno & \cellno & \cellno \\
2023 & \citet{10075392} & \orangetext{Robustness}  & \graytext{Image-to-text} & \graytext{ResNet-101, LSTM} & \cellno & \cellno & \cellno & \cellno \\

\hline
2024 & \citet{10447545} &  Security  & \graytext{Summarization}  & \graytext{BERT, Conditional Random Field (CRF)} & \cellno & \cellyes & \cellno & \cellno \\
2024 & \citet{10.1145/3664647.3680562} & Security  & General & Model-agnostic & \cellyes & \cellyes & \cellyes & \cellyes \\
2024 & \citet{10374202}  & Security  & \graytext{Translation} & \graytext{BERT, LSTM} & \cellno & \cellno & \cellno & \cellno \\
2024 & \citet{lin-etal-2024-zero} & Security  & General & Model-agnostic & \cellyes & \cellyes & \cellno & \cellno \\
2024 & \citet{10669737} & Security  & General & Model-agnostic & \cellyes & \cellyes & \cellno & \cellno \\
2024 & \citet{10462497} & Security  & \graytext{Live comments} & \graytext{ResNet, LSTM} & \cellno & \cellno & \cellno & \cellno \\
2024 & \citet{10551807} & Security  & General & Model-agnostic & \cellyes & \cellyes & \cellno & \cellno \\
2024 & \citet{10446027} & Security  & General & Model-agnostic & \cellyes & \cellyes & \cellno & \cellno \\
2024 & \citet{10687958} & Security  & \graytext{Group chat} & \graytext{BERT, Transformer} & \cellno & \cellyes & \cellno & \cellno \\
2024 & \citet{9778236} & Security  & General & Model-agnostic & \cellyes & \cellyes & \cellno & \cellno \\
2024 & \citet{10.1007/978-981-99-8148-9_44} & Security  & General & Model-agnostic & \cellyes  & \cellno & \cellno & \cellno \\
2024 & \citet{qin2024adlmstegauniversal} & Security  & General & Model-agnostic & \cellyes & \cellyes & \cellno & \cellno \\
2024 & \citet{lin2024novel} & Security  & \graytext{Translation} & \graytext{LSTM} & \cellno & \cellno & \cellno & \cellno \\

2024 & \citet{10831652} & Security/\bluetext{Efficiency}  & General & \graytext{ByGPT5} & \cellno & \cellyes & \cellno & \cellno \\
2024 & \citet{10.1145/3664476.3670930}   & \bluetext{Efficiency}  & General & Model-agnostic & \cellyes & \cellyes & \cellyes & \cellyes \\
2024 & \citet{10.1007/978-981-99-8073-4_4} &  \bluetext{Efficiency}  & \graytext{Social media} & Model-agnostic & \cellyes & \cellyes & \cellno & \cellno \\
2024 & \citet{10.1145/3658664.3659657} & \bluetext{Efficiency}  & \graytext{Social media} & Model-agnostic & \cellyes & \cellyes & \cellno & \cellno \\
2024 & \citet{huang2024odstegallmbasednearimperceptiblesteganography} & Security/\orangetext{Robustness}  & General & Model-agnostic & \cellyes & \cellyes & \cellno & \cellno \\
2024 & \citet{10831370} & Security/\orangetext{Robustness}  & General & Model-agnostic & \cellyes & \cellyes & \cellno & \cellno \\
2024 & \citet{bai2024semanticsteganographyframeworkrobust} & \bluetext{Efficiency}/\orangetext{Robustness}  & General & Model-agnostic & \cellyes & \cellyes & \cellno & \cellyes \\

\hline
2025  & \citet{WANG2025113101} &  Security  & General & Model-agnostic & \cellyes & \cellyes & \cellno & \cellno \\
2025 & \citet{liao2025framework} & Security  & General & Model-agnostic & \cellyes & \cellyes & \cellno & \cellno \\
2025 & \citet{WOZNIAK2025113960} & Security  & General & Model-agnostic & \cellyes & \cellyes & \cellno & \cellno \\
2025 & \citet{10888762} & Security  & General &  Model-agnostic & \cellyes & \cellyes & \cellno & \cellno \\
2025 & \citet{10919187} & Security  & General & Model-agnostic & \cellyes & \cellyes & \cellno & \cellno \\
2025 & \citet{app15179663} & Security  & General & Model-agnostic & \cellyes & \cellyes & \cellno & \cellno \\
2025 & \citet{technologies13070264} & Security  & General & \graytext{XLM-RoBERTa, Transformer} & \cellno & \cellyes & \cellno & \cellno \\
2025 & \citet{10.1162/coli.a.22} & Security  & General & Model-agnostic & \cellyes & \cellyes & \cellno & \cellno \\
2025 & \citet{computers14050165} & Security  & General & Model-agnostic & \cellyes & \cellyes & \cellno & \cellno \\
2025 & \citet{zhou2025autostegaagentdrivenlifelongstrategy} & Security  & General & \graytext{Multi-agent} & \cellyes & \cellyes & \cellyes & \cellno \\
2025 & \citet{10919130}  &  Security/\bluetext{Efficiency}  & General & Model-agnostic & \cellyes & \cellyes & \cellno & \cellno \\
2025 & \citet{pang2025frestegaplugandplaymethodboosting} &  Security/\bluetext{Efficiency}  & General & Model-agnostic & \cellyes & \cellyes & \cellno & \cellno \\
2025 & \citet{wang2025sparsampefficientprovablysecure} &  Security/\bluetext{Efficiency}  & General & Model-agnostic & \cellyes & \cellyes & \cellno & \cellno \\
2025 & \citet{bai2025shimmer} &  Security/\bluetext{Efficiency}  & General & Model-agnostic & \cellyes & \cellyes & \cellno & \cellno \\
2025 & \citet{NEURIPS2025_6c8e215a} &  Security/\bluetext{Efficiency}  & General & \graytext{Multi-agent} & \cellyes & \cellno & \cellyes & \cellno \\
2025 & \citet{pang2025provablesecuresteganographybased} & \bluetext{Efficiency}  & General & Model-agnostic & \cellyes & \cellyes & \cellno & \cellyes \\
2025 & \citet{huang2025relativelysecurellmbasedsteganographyconstrained} & \bluetext{Efficiency}  & General & Model-agnostic & \cellyes & \cellyes & \cellno & \cellno \\
2025 & \citet{10.1007/978-981-96-7005-5_12} & \bluetext{Efficiency}  & General & \graytext{Diffusion language model} & \cellyes & \cellno & \cellno & \cellno \\
2025 & \citet{11017758} & \bluetext{Efficiency}  & General & \graytext{Multi-agent} & \cellyes & \cellno & \cellno & \cellyes \\
2025 & \citet{11008605} & \bluetext{Efficiency}  & General & Model-agnostic & \cellyes & \cellyes & \cellno & \cellno \\
2025 & \citet{yan-murawaki-2025-addressing} & Security/\orangetext{Robustness}  & General & Model-agnostic & \cellyes & \cellyes & \cellno & \cellno \\
2025 & \citet{11023508} &  Security/\orangetext{Robustness}  & General & Model-agnostic & \cellyes & \cellyes & \cellyes & \cellno \\
2025 & \citet{10804596} & Security/\orangetext{Robustness}  & General & Model-agnostic & \cellyes & \cellyes & \cellno & \cellno \\
2025 & \citet{qi2025stead} & Security/\orangetext{Robustness}  & General & \graytext{Diffusion language model} & \cellyes & \cellyes & \cellno & \cellno \\
2025 & \citet{feng2025highcapacitysecuredisambiguationalgorithm} & \bluetext{Efficiency}/\orangetext{Robustness}  & General & Model-agnostic & \cellyes & \cellyes & \cellno & \cellno \\
2025 & \citet{perry2025robuststeganographylargelanguage} & \orangetext{Robustness}  & General & Model-agnostic & \cellyes & \cellyes & \cellno & \cellno \\
2025 & \citet{10888944} & \orangetext{Robustness}  & General & Model-agnostic & \cellyes  & \cellyes & \cellno & \cellno \\

2025 & \citet{10888607} & \orangetext{Robustness}  & General & Model-agnostic & \cellyes & \cellyes & \cellno & \cellno \\

2025 & \citet{sym17091416} & \orangetext{Robustness}  & General & Model-agnostic & \cellyes & \cellyes & \cellno & \cellno \\
\hline
2026 & \citet{Xiang_Ou_He_Yang_Liu_2026} & Security  & General & \graytext{BERT} & \cellno  & \cellno  & \cellyes & \cellno \\
2026 & \citet{gao2026textsteganographydynamiccodebook} & Security & General & \graytext{Multimodal LLM} & \cellyes & \cellyes  & \cellyes & \cellyes \\
2026 & \citet{yan2026efficientprovablysecurelinguistic} & Security/\bluetext{Efficiency}  & General & Model-agnostic & \cellyes & \cellyes  & \cellno & \cellno \\
2026 & \citet{11329500} & Security/\bluetext{Efficiency}  & General & Model-agnostic & \cellyes & \cellyes  & \cellyes & \cellno \\
2026 & \citet{11363394} & Security/\orangetext{Robustness}  & General & \graytext{BERT, GPT-2} & \cellno & \cellno  & \cellno & \cellno \\
2026 & \citet{wang2026retoksyncselfsynchronizingtokenizationdisambiguation} & 
\bluetext{Efficiency}/\orangetext{Robustness}  & General & Model-agnostic & \cellyes & \cellyes  & \cellyes & \cellno \\
2026 & \citet{yan2026anchoredslidingwindowrobust} & \orangetext{Robustness}  & General & Model-agnostic & \cellyes & \cellno  & \cellno & \cellno \\
\bottomrule[1.0pt]
\end{tabular}}
\caption{\rev{Taxonomy for generative linguistic steganographic methods based on methodology features.}}
\label{table:GLS_taxonomy}
\end{table*}
\rowcolors{0}{}{}

\rowcolors{2}{gray!10}{white}
\begin{table*}[!t]
\renewcommand{\arraystretch}{0.6}
\centering

\scalebox{0.6}{
\begin{tabular}{m{0.7cm}|m{3.5cm}|>{\centering\arraybackslash}m{3cm}|>
{\centering\arraybackslash}m{2cm}|>{\centering\arraybackslash}m{2cm}|>{\centering\arraybackslash}m{3cm}|>{\centering\arraybackslash}m{3cm}|>{\centering\arraybackslash}m{3cm}|>{\centering\arraybackslash}m{2.3cm}}
\toprule[1.0pt]
\textbf{Year} & \textbf{Reference} &  \textbf{Backbone model} & \textbf{Based on pre-trained model(s)?} & \textbf{\rev{LLM-adaptable?}} &\textbf{Experiments on IMDB dataset~\cite{maas-etal-2011-learning}?} & \textbf{Experiments on Twitter dataset~\cite{go2009twitter}?} &\textbf{Experiments on News dataset~\cite{kaggle_all_the_news}?} & \textbf{Experiments on other dataset(s)?}  \\
\midrule[1.0pt]
2019 & \citet{8653856}  & Feed-forward neural networks & \cellno & \cellno & \cellno & \cellno & \cellno & \cellyes  \\
2019 & \citet{8903243}  & Bi-LSTMs; CNNs & \cellno & \cellno & \cellyes & \cellyes & \cellno & \cellno  \\
2019 & \citet{8727932}  & Bi-RNNs; LSTMs & \cellno & \cellno & \cellyes & \cellyes & \cellno & \cellno  \\
2019 & \citet{8625512}  & Word2vec; CNNs & \cellyes & \cellno & \cellno & \cellno & \cellno & \cellyes  \\
\hline

2020 & \citet{10.1007/s11042-020-08716-w}  & CNNs & \cellno & \cellno & \cellno & \cellno & \cellno & \cellyes  \\
2020 & \citet{10.1145/3369412.3395067}  & Bi-LSTMs & \cellno & \cellno & \cellyes & \cellyes & \cellyes & \cellno  \\
2020 & \citet{9118410}  & Bi-LSTMs; CNNs & \cellno & \cellno & \cellyes & \cellyes & \cellyes & \cellno  \\
2020 & \citet{li2020two}  & Bi-LSTMs; GNNs & \cellno & \cellno & \cellno & \cellno & \cellno & \cellyes  \\
\hline

2021 & \citet{9364681}  & GNNs & \cellno & \cellno & \cellyes & \cellyes & \cellno & \cellno  \\
2021 & \citet{jiao2021neural}  & Attention & \cellno & \cellno & \cellyes & \cellyes & \cellyes & \cellno  \\
2021 & \citet{9484749}  & BERT; Bi-LSTMs; CNNs & \cellyes & \cellno & \cellyes & \cellyes & \cellno & \cellno  \\
2021 & \citet{10.1007/978-3-030-69449-4_7}  & BERT; Bi-LSTMs & \cellyes & \cellno & \cellno & \cellyes & \cellyes & \cellno  \\
2021 & \citet{Xu_2021}  & BERT; Bi-LSTMs & \cellyes & \cellno & \cellyes & \cellyes & \cellyes & \cellno  \\
\hline
 
2022 & \citet{9844755}  & GNNs & \cellno & \cellno & \cellyes & \cellyes & \cellno & \cellno  \\
2022 & \citet{9746219}  & LSTMs & \cellno & \cellno & \cellyes & \cellyes & \cellno & \cellno  \\
2022 & \citet{XUE2022140}  & CNNs & \cellno & \cellno & \cellyes & \cellyes & \cellyes & \cellno  \\
2022 & \citet{9591452}  & BERT; GNNs & \cellyes & \cellno & \cellyes & \cellyes & \cellyes & \cellno  \\
2022 & \citet{9913623}  & BERT & \cellyes & \cellno & \cellyes & \cellyes & \cellyes & \cellno  \\
2022 & \citet{9806187}  & BERT; GNNs & \cellyes & \cellno & \cellyes & \cellyes & \cellno & \cellno  \\
2022 & \citet{10.1145/3531536.3532963}  & BERT; CNNs & \cellyes & \cellno & \cellyes & \cellyes & \cellyes & \cellno  \\
2022 & \citet{9616404}  & BERT; Bi-GRUs; CNNs & \cellyes & \cellno & \cellyes & \cellyes & \cellyes & \cellno  \\
2022 & \citet{10.1145/3531536.3532949}  & BERT; Bi-LSTMs & \cellyes & \cellno & \cellyes & \cellyes & \cellyes & \cellno  \\
2022 & \citet{9956902}  & BERT & \cellyes & \cellno & \cellyes & \cellyes & \cellyes & \cellno  \\
\hline

2023 & \citet{9970400}  & GNNs & \cellno & \cellno & \cellno & \cellno & \cellno & \cellyes  \\
2023 & \citet{10021965}  & CNNs; Bi-GRU & \cellno & \cellno & \cellno & \cellno & \cellno & \cellyes  \\
2023 & \citet{10097161}  & GNNs & \cellno & \cellno & \cellyes & \cellyes & \cellno & \cellno  \\
2023 & \citet{9732641}  & Attention; CNNs & \cellno & \cellno & \cellyes & \cellyes & \cellno & \cellno  \\
2023 & \citet{10038789}  & Bi-LSTMs; CNNs & \cellno & \cellno & \cellyes & \cellyes & \cellyes & \cellyes  \\
2023 & \citet{xiang2023general}  & CNNs & \cellno & \cellno & \cellno & \cellno & \cellno & \cellyes  \\
2023 & \citet{10268497}  & BERT & \cellyes & \cellno & \cellyes & \cellyes & \cellno & \cellno  \\
2023 & \citet{10234544}  & GloVe; LSTMs & \cellyes & \cellno & \cellno & \cellno & \cellno & \cellyes  \\
2023 & \citet{WANG2023103512}  & BERT & \cellyes & \cellno & \cellyes & \cellyes & \cellno & \cellno  \\
2023 & \citet{10.1145/3577163.3595111}  & BERT & \cellyes & \cellno & \cellyes & \cellyes & \cellyes & \cellno  \\
2023 & \citet{10190738}  & BERT & \cellyes & \cellno & \cellyes & \cellyes & \cellyes & \cellno  \\
\hline

2024 & \citet{10.1007/978-981-99-8073-4_9}  & GNNs & \cellno & \cellno & \cellno & \cellyes & \cellno & \cellyes \\
2024 & \citet{10805412}  & CNNs; Attention & \cellno & \cellno & \cellyes & \cellyes & \cellyes & \cellno \\
2024 & \citet{yang2024nextgenerationsteganalysisllmsunleash}  & LLM (model-agnostic); LoRA & \cellyes & \cellyes & \cellyes & \cellyes & \cellyes & \cellno  \\
2024 & \citet{10.1145/3634814.3634837}  & LLM (model-agnostic) & \cellyes & \cellyes & \cellyes & \cellyes & \cellno & \cellno  \\
2024 & \citet{10299660}  & BERT; CNNs & \cellyes & \cellno & \cellyes & \cellyes & \cellyes & \cellno  \\
2024 & \citet{YUAN2024123437}  & BERT; Bi-LSTMs & \cellyes & \cellno & \cellyes & \cellyes & \cellyes & \cellno \\
2024 & \citet{10582504}  & BERT & \cellyes & \cellno & \cellyes & \cellyes & \cellyes & \cellno \\
2024 & \citet{10596690}  & ELECTRA~\cite{pmlr-v9-gutmann10a} & \cellyes & \cellno & \cellyes & \cellyes & \cellyes & \cellno \\
2024 & \citet{YOU2024128260}  & BERT & \cellyes & \cellno & \cellyes & \cellyes & \cellno & \cellno \\
\hline

2025 & \citet{huang2025gsdfusecapturingcognitiveinconsistencies}  & GNNs & \cellno & \cellno & \cellno & \cellyes & \cellno & \cellyes \\
2025 & \citet{10888159}  & CNNs; GNNs; Attention & \cellno & \cellno & \cellno & \cellno & \cellno & \cellyes \\
2025 & \citet{10752347}  & LLM (model-agnostic); LoRA & \cellyes & \cellyes & \cellyes & \cellyes & \cellyes & \cellno  \\
2025 & \citet{Wang_Zhou_Chen_Zhou_Yang_2025}  & LLM (model-agnostic); GNNs & \cellyes & \cellyes & \cellyes & \cellyes & \cellyes & \cellno  \\
2025 & \citet{10916785}  & RoBERTa; CNNs & \cellyes & \cellno & \cellyes & \cellyes & \cellyes & \cellno  \\
2025 & \citet{10849805}  & BERT; CNNs & \cellyes & \cellno & \cellyes & \cellyes & \cellyes & \cellno  \\
2025 & \citet{10870426}  & ERNIE~\cite{sun2019ernieenhancedrepresentationknowledge}; Attention & \cellyes & \cellno & \cellyes & \cellyes & \cellyes & \cellno  \\
2025 & \citet{10772620}  & Word2vec; LSTMs; CNNs & \cellyes & \cellno & \cellyes & \cellyes & \cellyes & \cellno  \\
2025 & \citet{10.1145/3746252.3760827}  & BERT & \cellyes & \cellno & \cellyes & \cellyes & \cellyes & \cellno  \\
2025 & \citet{10887993}  & BERT; GNNs & \cellyes & \cellno & \cellno & \cellno & \cellno & \cellyes  \\
2025 & \citet{10.1007/978-3-031-82907-9_10}  & BERT; Bi-LSTMs & \cellyes & \cellno & \cellyes & \cellyes & \cellyes & \cellno  \\
2025 & \citet{10.1145/3733102.3733114}  & BERT; Bi-LSTMs & \cellyes & \cellno & \cellyes & \cellyes & \cellyes & \cellno  \\
2025 & \citet{niu2025text}  & RoBERTa; Bi-RNNs & \cellyes & \cellno & \cellno & \cellno & \cellno & \cellyes  \\
\hline

2026 & \citet{WANG2026129311}  & BERT & \cellyes & \cellno & \cellno & \cellno & \cellno & \cellyes  \\
2026 & \citet{11462507} & BERT  & \cellyes & \cellno & \cellyes & \cellyes & \cellyes &  \cellno  \\
2026 & \citet{barlow2026large}  & LLM (model-agnostic) & \cellyes & \cellyes & \cellno & \cellno & \cellno & \cellyes  \\
2026 & \citet{11460766}  & GNNs & \cellno & \cellno & \cellno & \cellno & \cellno & \cellyes  \\

\bottomrule[1.0pt]
\end{tabular}}
\caption{\rev{Methodology features and experiment datasets of linguistic steganalysis methods.}}
\label{table:Steganalysis_features}
\end{table*}
\rowcolors{0}{}{}

\rowcolors{2}{gray!10}{white}
\begin{table*}[!t]
\renewcommand{\arraystretch}{0.6}
\centering

\scalebox{0.63}{
\begin{tabular}{m{1cm}|m{4.5cm}|>{\centering\arraybackslash}m{3cm}|>{\centering\arraybackslash}m{2cm}|>{\centering\arraybackslash}m{3cm}|>{\centering\arraybackslash}m{3cm}|>{\centering\arraybackslash}m{3cm}|>{\centering\arraybackslash}m{1cm}}
\toprule[1.0pt]
\textbf{Year} & \textbf{Reference} &   \textbf{FCN~\cite{8653856}}& \textbf{CNN~\cite{8625512}} & \textbf{R-BiLSTM-C~\cite{8903243}} & \textbf{TS-RNN~\cite{8727932}} & \textbf{TS-CSW~\cite{10.1007/s11042-020-08716-w}} &  \textbf{Other} \\
\midrule[1.0pt]
2017 & \citet{fang-etal-2017-generating}  & \cellno & \cellno & \cellno & \cellno & \cellno & \cellno  \\
2017 & \citet{10.1145/3082031.3083240} & \cellno & \cellno & \cellno & \cellno & \cellno & \cellno  \\

\hline
2018 & \citet{yang2018automaticallygeneratesteganographictext}   & \cellno & \cellno & \cellno & \cellno & \cellno & \cellyes  \\

\hline
2019 & \citet{8470163}   & \cellno & \cellno & \cellno & \cellno & \cellno & \cellyes  \\
2019 & \citet{ziegler-etal-2019-neural} & \cellno & \cellno & \cellno & \cellno & \cellno & \cellno  \\
2019 & \citet{dai-cai-2019-towards} & \cellno & \cellno & \cellno & \cellno & \cellno & \cellno  \\
2019 & \citet{9023313} & \cellno & \cellno & \cellno & \cellno & \cellno & \cellyes  \\
2019 & \citet{Text_steganography_on_RNN-Generated-lyrics} & \cellno & \cellno & \cellno & \cellno & \cellno & \cellno  \\

\hline
2020 & \citet{yang2020graphstegasemanticcontrollablesteganographic} & \cellyes & \cellno & \cellno & \cellno & \cellno & \cellno  \\
2020 & \citet{10.1007/978-3-030-43575-2_2} & \cellno & \cellno & \cellno & \cellno & \cellno & \cellno  \\
2020 & \citet{shen-etal-2020-near} & \cellno & \cellno & \cellno & \cellno & \cellno & \cellno  \\
2020 & \citet{doi:10.2352/ISSN.2470-1173.2020.4.MWSF-291} & \cellno & \cellno & \cellno & \cellno & \cellno & \cellyes  \\
2020 & \citet{doi:10.1177/1550147720914257} & \cellno & \cellno & \cellno & \cellno & \cellno & \cellyes  \\
2020 & \citet{math8091558} & \cellno & \cellno & \cellno & \cellno & \cellno & \cellno  \\

\hline
2021 & \citet{9193914} &  \cellyes & \cellyes & \cellno & \cellno & \cellno & \cellyes  \\
2021 & \citet{zhang-etal-2021-provably} & \cellyes & \cellno & \cellno & \cellno & \cellyes & \cellno  \\
2021 & \citet{9353234} & \cellyes & \cellno & \cellno & \cellyes & \cellyes & \cellno  \\
2021 & \citet{10.1145/3460120.3484550} & \cellno & \cellno & \cellno & \cellno & \cellno & \cellno  \\
2021 & \citet{9280343} & \cellyes & \cellno & \cellno & \cellno & \cellyes & \cellno  \\
2021 & \citet{10.1145/3418598} & \cellyes & \cellno & \cellno & \cellno & \cellno & \cellno  \\
2021 & \citet{yang2021generation} & \cellno & \cellyes & \cellno & \cellyes & \cellno & \cellno  \\
2021 & \citet{WANG20211375} & \cellno & \cellno & \cellno & \cellno & \cellno & \cellno  \\

\hline
2022 & \citet{9975301} &  \cellno & \cellno & \cellno & \cellno & \cellno & \cellno  \\
2022 & \citet{yu2022mts} & \cellno & \cellno & \cellno & \cellno & \cellno & \cellno  \\
2022 & \citet{9430708} & \cellno & \cellyes & \cellyes & \cellyes & \cellno & \cellno  \\
2022 & \citet{9714779} & \cellno & \cellno & \cellno & \cellyes & \cellno & \cellyes  \\
2022 & \citet{varol2022lzw} & \cellyes & \cellyes & \cellno & \cellno & \cellno & \cellyes  \\
2022 & \citet{9864312} & \cellno & \cellno & \cellno & \cellno & \cellno & \cellno  \\
2022 & \citet{nozaki-murawaki-2022-addressing} & \cellno & \cellno & \cellno & \cellno & \cellno & \cellyes  \\

\hline
2023 & \citet{SUN2023157} & \cellyes & \cellyes & \cellyes & \cellno & \cellno & \cellno  \\
2023 & \citet{10191218} & \cellno & \cellno & \cellyes & \cellyes & \cellyes & \cellno  \\
2023 & \citet{10115433} & \cellno & \cellno & \cellno & \cellno & \cellno & \cellyes  \\
2023 & \citet{10179287} & \cellyes & \cellno & \cellyes & \cellno & \cellno & \cellyes  \\
2023 & \citet{10.1007/978-3-031-44696-2_62} & \cellno & \cellno & \cellyes & \cellno & \cellno & \cellyes  \\
2023 & \citet{10095722} & \cellyes & \cellno & \cellno & \cellno & \cellno & \cellno  \\
2023 & \citet{witt2023perfectly} & \cellno & \cellno & \cellno & \cellno & \cellno & \cellno  \\
2023 & \citet{10215094} & \cellno & \cellno & \cellno & \cellno & \cellno & \cellyes  \\
2023 & \citet{10191984} & \cellno & \cellno & \cellyes & \cellyes & \cellyes & \cellno  \\
2023 & \citet{10075392} & \cellno & \cellyes & \cellyes & \cellno & \cellno & \cellyes  \\

\hline
2024 & \citet{10447545} & \cellno & \cellno & \cellno & \cellno & \cellyes & \cellyes  \\
2024 & \citet{10.1145/3664647.3680562} & \cellno & \cellyes & \cellno & \cellno & \cellno & \cellyes  \\
2024 & \citet{10374202} & \cellno & \cellyes & \cellyes & \cellyes & \cellno & \cellno  \\
2024 & \citet{lin-etal-2024-zero} & \cellno & \cellno & \cellyes & \cellno & \cellno & \cellyes  \\
2024 & \citet{10669737} & \cellno & \cellno & \cellyes & \cellno & \cellyes & \cellno  \\
2024 & \citet{10462497} & \cellno & \cellno & \cellno & \cellno & \cellno & \cellno  \\
2024 & \citet{10551807} & \cellno & \cellno & \cellno & \cellno & \cellno & \cellyes  \\
2024 & \citet{10446027} & \cellno & \cellno & \cellno & \cellno & \cellyes & \cellno  \\
2024 & \citet{10687958} & \cellno & \cellno & \cellno & \cellno & \cellyes & \cellno  \\
2024 & \citet{9778236} & \cellno & \cellno & \cellno & \cellyes & \cellno & \cellno  \\
2024 & \citet{10.1007/978-981-99-8148-9_44} & \cellno & \cellno & \cellno & \cellno & \cellyes & \cellno  \\
2024 & \citet{qin2024adlmstegauniversal} & \cellno & \cellno & \cellno & \cellno & \cellno & \cellyes  \\
2024 & \citet{lin2024novel} & \cellno & \cellyes & \cellno & \cellno & \cellno & \cellyes  \\

2024 & \citet{10831652} & \cellno & \cellno & \cellno & \cellno & \cellno & \cellyes  \\
2024 & \citet{10.1145/3664476.3670930} & \cellno & \cellno & \cellno & \cellno & \cellno & \cellno  \\
2024 & \citet{10.1007/978-981-99-8073-4_4} & \cellno & \cellno & \cellno & \cellno & \cellno & \cellyes  \\
2024 & \citet{10.1145/3658664.3659657} & \cellno & \cellno & \cellno & \cellno & \cellno & \cellno  \\
2024 & \citet{huang2024odstegallmbasednearimperceptiblesteganography} & \cellno & \cellno & \cellno & \cellno & \cellno & \cellno  \\
2024 & \citet{10831370} & \cellno & \cellno & \cellno & \cellno & \cellno & \cellyes  \\
2024 & \citet{bai2024semanticsteganographyframeworkrobust} & \cellno & \cellno & \cellno & \cellno & \cellno & \cellno  \\

\hline
2025 & \citet{WANG2025113101} & \cellno & \cellyes & \cellno & \cellno & \cellyes & \cellyes  \\
2025 & \citet{liao2025framework} & \cellno & \cellno & \cellno & \cellno & \cellyes & \cellno  \\
2025 & \citet{WOZNIAK2025113960} & \cellno & \cellno & \cellno & \cellno & \cellno & \cellno  \\
2025 & \citet{10888762} & \cellyes & \cellyes & \cellyes & \cellno & \cellno & \cellno  \\
2025 & \citet{10919187} & \cellyes & \cellno & \cellyes & \cellno & \cellno & \cellyes  \\
2025 & \citet{app15179663} & \cellyes & \cellyes & \cellyes & \cellyes & \cellno & \cellyes  \\
2025 & \citet{technologies13070264} & \cellno & \cellyes & \cellno & \cellyes & \cellno & \cellyes  \\
2025 & \citet{10.1162/coli.a.22} & \cellno & \cellno & \cellno & \cellno & \cellno & \cellyes  \\
2025 & \citet{computers14050165} & \cellno & \cellno & \cellno & \cellyes & \cellno & \cellyes  \\
2025 & \citet{zhou2025autostegaagentdrivenlifelongstrategy} & \cellno & \cellyes & \cellno & \cellno & \cellno & \cellyes  \\
2025 & \citet{10919130} & \cellyes & \cellno & \cellyes & \cellno & \cellno & \cellyes  \\
2025 & \citet{pang2025frestegaplugandplaymethodboosting} & \cellno & \cellno & \cellyes & \cellyes & \cellyes & \cellyes  \\
2025 & \citet{wang2025sparsampefficientprovablysecure} & \cellyes & \cellno & \cellyes & \cellno & \cellno & \cellyes  \\
2025 & \citet{bai2025shimmer} & \cellno & \cellno & \cellno & \cellno & \cellno & \cellno  \\
2025 & \citet{NEURIPS2025_6c8e215a} & \cellyes & \cellno & \cellno & \cellno & \cellno & \cellyes  \\
2025 & \citet{pang2025provablesecuresteganographybased} & \cellno & \cellno & \cellno & \cellno & \cellyes & \cellno  \\
2025 & \citet{huang2025relativelysecurellmbasedsteganographyconstrained} & \cellno & \cellno & \cellno & \cellno & \cellno & \cellno  \\
2025 & \citet{10.1007/978-981-96-7005-5_12} & \cellno & \cellyes & \cellno & \cellyes & \cellno & \cellyes  \\
2025 & \citet{11017758} & \cellyes & \cellno & \cellno & \cellno & \cellno & \cellyes  \\
2025 & \citet{11008605} & \cellno & \cellno & \cellno & \cellno & \cellno & \cellyes  \\
2025 & \citet{yan-murawaki-2025-addressing} & \cellno & \cellno & \cellno & \cellno & \cellno & \cellyes  \\
2025 & \citet{11023508} & \cellno & \cellno & \cellno & \cellno & \cellno & \cellno  \\
2025 & \citet{10804596} & \cellno & \cellno & \cellno & \cellno & \cellno & \cellno  \\
2025 & \citet{qi2025stead} & \cellyes & \cellno & \cellyes & \cellno & \cellno & \cellyes  \\
2025 & \citet{feng2025highcapacitysecuredisambiguationalgorithm} & \cellno & \cellno & \cellno & \cellno & \cellno & \cellno  \\
2025 & \citet{perry2025robuststeganographylargelanguage} & \cellno & \cellno & \cellno & \cellno & \cellno & \cellno  \\
2025 & \citet{10888944} & \cellno & \cellno & \cellno & \cellno & \cellyes & \cellno  \\
2025 & \citet{10888607} & \cellno & \cellno & \cellno & \cellno & \cellno & \cellno  \\
2025 & \citet{sym17091416} & \cellyes & \cellno & \cellno & \cellyes & \cellno & \cellyes  \\
\hline
2026 & \citet{Xiang_Ou_He_Yang_Liu_2026} & \cellno & \cellno & \cellno & \cellno & \cellno & \cellyes  \\
2026 & \citet{gao2026textsteganographydynamiccodebook} & \cellno & \cellno & \cellno & \cellno & \cellno & \cellyes  \\
2026 & \citet{yan2026efficientprovablysecurelinguistic} & \cellno & \cellno & \cellno & \cellno & \cellno & \cellyes  \\
2026 & \citet{11329500} & \cellyes & \cellno & \cellyes & \cellno & \cellno & \cellyes  \\
2026 & \citet{11363394} & \cellyes & \cellno & \cellno & \cellno & \cellno & \cellyes  \\
2026 & \citet{wang2026retoksyncselfsynchronizingtokenizationdisambiguation} & \cellyes & \cellno & \cellno & \cellno & \cellno & \cellyes  \\
2026 & \citet{yan2026anchoredslidingwindowrobust} & \cellno & \cellno & \cellno & \cellno & \cellno & \cellyes  \\
\bottomrule[1.0pt]
\end{tabular}}
\caption{\rev{Adopted steganalysis methods to evaluate imperceptibility of the generative linguistic steganographic methods.}}
\label{table:GLS_evaluation_steganalysis}
\end{table*}
\rowcolors{0}{}{}

\rowcolors{2}{gray!10}{white}
\begin{table*}[!t]
\renewcommand{\arraystretch}{0.6}
\centering

\scalebox{0.63}{
\begin{tabular}{m{1cm}|m{5.1cm}|>{\centering\arraybackslash}m{0.6cm}|>{\centering\arraybackslash}m{1.8cm}|>{\centering\arraybackslash}m{1.8cm}|>{\centering\arraybackslash}m{0.9cm}|>{\centering\arraybackslash}m{1.4cm}|>{\centering\arraybackslash}m{1.6cm}|>{\centering\arraybackslash}m{1.4cm}|>{\centering\arraybackslash}m{1.75cm}|>{\centering\arraybackslash}m{1.45cm}|>{\centering\arraybackslash}m{0.75cm}}
\toprule[1.0pt]
\textbf{Year} & \textbf{Reference} &  \textbf{PPL} & \textbf{KLD (step-level)} & \textbf{KLD (text-level)} &\textbf{BLEU} & \textbf{ROUGE}  & \textbf{METEOR} & \textbf{Diversity}  & \textbf{Human evaluation} & \textbf{LLM-as-a-judge} & \textbf{Other} \\
\midrule[1.0pt]
2017 & \citet{fang-etal-2017-generating}  & \cellyes & \cellno & \cellno & \cellno & \cellno & \cellno & \cellno & \cellno & \cellno & \cellno \\
2017 & \citet{10.1145/3082031.3083240} & \cellno & \cellno & \cellno & \cellno & \cellno & \cellno & \cellno & \cellyes & \cellno & \cellno \\

\hline
2018 & \citet{yang2018automaticallygeneratesteganographictext}  & \cellyes & \cellno & \cellno & \cellno & \cellno & \cellno & \cellno & \cellno & \cellno & \cellno \\

\hline
2019 & \citet{8470163} & \cellyes & \cellno & \cellno & \cellno & \cellno & \cellno & \cellno & \cellno & \cellno & \cellyes \\
2019 & \citet{ziegler-etal-2019-neural} & \cellno & \cellyes & \cellno & \cellno & \cellno & \cellno & \cellno & \cellyes & \cellno & \cellno \\
2019 & \citet{dai-cai-2019-towards} & \cellno & \cellyes & \cellno & \cellno & \cellno & \cellno & \cellno & \cellno & \cellno & \cellno \\
2019 & \citet{9023313} & \cellyes & \cellyes & \cellno & \cellno & \cellno & \cellno & \cellno & \cellyes & \cellno & \cellno \\
2019 & \citet{Text_steganography_on_RNN-Generated-lyrics} & \cellno & \cellno & \cellno & \cellno & \cellno & \cellno & \cellno & \cellno & \cellno & \cellno \\

\hline
2020 & \citet{yang2020graphstegasemanticcontrollablesteganographic}  & \cellyes & \cellno & \cellno & \cellyes & \cellyes & \cellno & \cellno & \cellno & \cellno & \cellyes \\
2020 & \citet{10.1007/978-3-030-43575-2_2} & \cellyes & \cellno & \cellno & \cellno & \cellno & \cellno & \cellno & \cellno & \cellno & \cellno \\
2020 & \citet{shen-etal-2020-near} & \cellno & \cellyes & \cellno & \cellno & \cellno & \cellno & \cellno & \cellyes & \cellno & \cellno \\
2020 & \citet{doi:10.2352/ISSN.2470-1173.2020.4.MWSF-291} & \cellyes & \cellno & \cellno & \cellno & \cellno & \cellno & \cellno & \cellno & \cellno & \cellyes \\
2020 & \citet{doi:10.1177/1550147720914257} & \cellyes & \cellno & \cellno & \cellno & \cellno & \cellno & \cellno & \cellno & \cellno & \cellno \\
2020 & \citet{math8091558} & \cellyes & \cellno & \cellno & \cellno & \cellno & \cellno & \cellno & \cellno & \cellno & \cellno \\

\hline
2021 & \citet{9193914}  & \cellyes & \cellno & \cellyes & \cellno & \cellno & \cellno & \cellno & \cellyes & \cellno & \cellno \\
2021 & \citet{zhang-etal-2021-provably} & \cellno & \cellyes & \cellyes & \cellno & \cellno & \cellno & \cellno & \cellno & \cellno & \cellno \\
2021 & \citet{9353234} & \cellyes & \cellno & \cellno & \cellno & \cellyes & \cellyes & \cellno & \cellno & \cellno & \cellno \\
2021 & \citet{10.1145/3460120.3484550} & \cellyes & \cellyes & \cellno & \cellno & \cellno & \cellno & \cellno & \cellno & \cellno & \cellno \\
2021 & \citet{9280343} & \cellyes & \cellno & \cellno & \cellno & \cellno & \cellno & \cellno & \cellno & \cellno & \cellno \\
2021 & \citet{10.1145/3418598} & \cellyes & \cellno & \cellno & \cellyes & \cellno &  \cellyes & \cellno & \cellno & \cellno & \cellno \\
2021 & \citet{yang2021generation} & \cellyes & \cellno & \cellyes & \cellyes & \cellno & \cellno & \cellyes & \cellno & \cellno & \cellno \\
2021 & \citet{WANG20211375} & \cellno & \cellno & \cellno & \cellno & \cellno & \cellno & \cellno & \cellno & \cellno & \cellno \\

\hline
2022 & \citet{9975301} & \cellyes & \cellno & \cellno & \cellno & \cellno & \cellno & \cellno & \cellno &\cellno & \cellyes \\
2022 & \citet{yu2022mts} & \cellyes & \cellno & \cellno & \cellno & \cellno & \cellno & \cellno & \cellno &\cellno & \cellno \\
2022 & \citet{9430708} & \cellyes & \cellno & \cellno & \cellno & \cellno & \cellno & \cellno & \cellno &\cellno & \cellyes \\
2022 & \citet{9714779} & \cellyes & \cellno & \cellno & \cellno & \cellno & \cellno & \cellno & \cellno &\cellno & \cellno \\
2022 & \citet{varol2022lzw} & \cellyes & \cellno & \cellyes & \cellno & \cellno & \cellno & \cellno &\cellno & \cellno & \cellno \\
2022 & \citet{9864312} & \cellno & \cellno & \cellno & \cellno & \cellno & \cellno & \cellno & \cellno &\cellno & \cellno \\
2022 & \citet{nozaki-murawaki-2022-addressing} & \cellno & \cellno & \cellno & \cellno & \cellno & \cellno & \cellno & \cellno & \cellno & \cellno \\

\hline
2023 & \citet{SUN2023157} & \cellyes & \cellno & \cellno & \cellyes & \cellyes & \cellno & \cellno & \cellno & \cellno & \cellyes \\
2023 & \citet{10191218} & \cellyes & \cellyes & \cellno & \cellno & \cellno & \cellno & \cellno & \cellno & \cellno & \cellno \\
2023 & \citet{10115433} & \cellyes & \cellno & \cellno & \cellno & \cellno & \cellno & \cellno & \cellno & \cellno & \cellyes \\
2023 & \citet{10179287} & \cellyes & \cellyes & \cellno & \cellno & \cellno & \cellno & \cellno & \cellno & \cellno & \cellno \\
2023 & \citet{10.1007/978-3-031-44696-2_62} & \cellyes & \cellno & \cellyes & \cellyes & \cellyes & \cellno & \cellno & \cellno & \cellno & \cellyes \\
2023 & \citet{10095722} & \cellyes & \cellno & \cellno & \cellyes & \cellno & \cellyes & \cellyes & \cellno & \cellno & \cellno \\
2023 & \citet{witt2023perfectly} & \cellno & \cellyes & \cellno & \cellno & \cellno & \cellno & \cellno & \cellno & \cellno & \cellno \\
2023 & \citet{10215094} & \cellyes & \cellyes & \cellyes & \cellno & \cellno & \cellno & \cellno & \cellno & \cellno & \cellno \\
2023 & \citet{10191984} & \cellno & \cellno & \cellno & \cellyes & \cellno & \cellno & \cellno & \cellno & \cellno & \cellno \\
2023 & \citet{10075392} & \cellno & \cellno & \cellyes & \cellyes & \cellyes & \cellyes & \cellno & \cellno & \cellno & \cellno \\

\hline
2024 & \citet{10447545}  & \cellno & \cellno & \cellno &  \cellno  & \cellyes & \cellyes & \cellno & \cellno & \cellno & \cellyes \\
2024 & \citet{10.1145/3664647.3680562} & \cellyes & \cellno & \cellyes & \cellno & \cellno & \cellno & \cellno & \cellyes & \cellno & \cellyes \\
2024 & \citet{10374202} & \cellyes & \cellyes & \cellno & \cellyes & \cellno & \cellno & \cellno & \cellno & \cellno & \cellyes \\
2024 & \citet{lin-etal-2024-zero} & \cellyes & \cellno & \cellyes & \cellno & \cellno & \cellno & \cellno & \cellno & \cellyes & \cellno \\
2024 & \citet{10669737} & \cellyes & \cellno & \cellno & \cellno & \cellyes & \cellno & \cellno & \cellno & \cellno & \cellyes \\
2024 & \citet{10462497} & \cellyes & \cellno & \cellno & \cellyes & \cellyes & \cellno & \cellno & \cellno & \cellno & \cellno \\
2024 & \citet{10551807} & \cellyes & \cellno & \cellno & \cellno & \cellno & \cellno & \cellno & \cellyes & \cellno & \cellno \\
2024 & \citet{10446027} & \cellyes & \cellno & \cellno & \cellno & \cellno & \cellno & \cellyes & \cellno & \cellno & \cellno \\
2024 & \citet{10687958} & \cellyes & \cellno & \cellno & \cellyes & \cellyes & \cellno & \cellno & \cellno & \cellno & \cellno \\
2024 & \citet{9778236} & \cellyes & \cellno & \cellyes & \cellyes & \cellno & \cellno & \cellno & \cellno & \cellno & \cellno \\
2024 & \citet{10.1007/978-981-99-8148-9_44} & \cellyes & \cellno & \cellyes & \cellno & \cellno & \cellno & \cellno & \cellno & \cellno & \cellno \\
2024 & \citet{qin2024adlmstegauniversal} & \cellyes & \cellno & \cellno & \cellno & \cellno & \cellno & \cellyes & \cellno & \cellno  & \cellno \\
2024 & \citet{lin2024novel} & \cellyes & \cellno & \cellno & \cellyes & \cellno & \cellno & \cellno & \cellno & \cellno & \cellno \\

2024 & \citet{10831652} & \cellyes & \cellno & \cellyes & \cellno & \cellno & \cellno & \cellno & \cellno & \cellno & \cellyes \\
2024 & \citet{10.1145/3664476.3670930} & \cellno & \cellno & \cellno & \cellno & \cellno & \cellno & \cellno & \cellno & \cellno & \cellno \\
2024 & \citet{10.1007/978-981-99-8073-4_4} & \cellyes & \cellno & \cellno & \cellno & \cellno & \cellno & \cellyes & \cellno & \cellno & \cellyes \\
2024 & \citet{10.1145/3658664.3659657} & \cellyes & \cellno & \cellno & \cellno & \cellno & \cellno & \cellno & \cellno & \cellno & \cellyes \\
2024 & \citet{huang2024odstegallmbasednearimperceptiblesteganography} & \cellno & \cellyes & \cellno & \cellno & \cellno & \cellno & \cellno & \cellno & \cellyes & \cellno \\
2024 & \citet{10831370} & \cellyes & \cellyes & \cellno & \cellno & \cellno & \cellno & \cellno & \cellno & \cellno & \cellno \\
2024 & \citet{bai2024semanticsteganographyframeworkrobust} & \cellyes & \cellno & \cellno & \cellno & \cellno & \cellno & \cellyes & \cellno & \cellyes & \cellno \\

\hline
2025 & \citet{WANG2025113101}  & \cellyes & \cellno & \cellyes & \cellyes & \cellno & \cellno & \cellno & \cellno & \cellno & \cellyes \\
2025 & \citet{liao2025framework} & \cellno & \cellyes & \cellno & \cellno & \cellno & \cellno & \cellno & \cellno & \cellno & \cellno \\
2025 & \citet{WOZNIAK2025113960} & \cellyes & \cellno & \cellno & \cellno & \cellno & \cellno & \cellno & \cellno & \cellno & \cellyes \\
2025 & \citet{10888762} & \cellyes & \cellno & \cellno & \cellyes & \cellyes & \cellyes & \cellno & \cellno & \cellno & \cellno \\
2025 & \citet{10919187} & \cellno & \cellno & \cellno & \cellno & \cellno & \cellno & \cellno & \cellno & \cellno & \cellno \\
2025 & \citet{app15179663} & \cellyes & \cellno & \cellyes & \cellno & \cellno & \cellno & \cellyes & \cellno & \cellno & \cellyes \\
2025 & \citet{technologies13070264} & \cellyes & \cellyes & \cellno & \cellyes & \cellno & \cellno & \cellno & \cellno & \cellno & \cellno \\
2025 & \citet{10.1162/coli.a.22} & \cellyes & \cellno & \cellyes & \cellno & \cellno & \cellno & \cellno & \cellyes & \cellno & \cellno \\
2025 & \citet{computers14050165} & \cellyes & \cellno & \cellno & \cellno & \cellno & \cellno & \cellno & \cellno & \cellno & \cellyes \\
2025 & \citet{zhou2025autostegaagentdrivenlifelongstrategy} & \cellyes & \cellno & \cellyes & \cellno & \cellno & \cellno & \cellno & \cellyes & \cellno & \cellyes \\
2025 & \citet{10919130} & \cellno & \cellyes & \cellno & \cellno & \cellno & \cellno & \cellno & \cellno & \cellno & \cellno \\
2025 & \citet{pang2025frestegaplugandplaymethodboosting} & \cellyes & \cellno & \cellno & \cellno & \cellno & \cellno & \cellyes & \cellno & \cellno & \cellyes \\
2025 & \citet{wang2025sparsampefficientprovablysecure} & \cellno & \cellyes & \cellno & \cellno & \cellno & \cellno & \cellno & \cellno & \cellno & \cellno \\
2025 & \citet{bai2025shimmer} & \cellyes & \cellno & \cellno & \cellno & \cellno & \cellno & \cellno & \cellno & \cellno & \cellno \\
2025 & \citet{NEURIPS2025_6c8e215a} & \cellno & \cellno & \cellno & \cellno & \cellyes & \cellno & \cellno & \cellno & \cellno & \cellyes \\
2025 & \citet{pang2025provablesecuresteganographybased} & \cellyes & \cellno & \cellno & \cellno & \cellno & \cellno & \cellyes & \cellno & \cellno & \cellno \\
2025 & \citet{huang2025relativelysecurellmbasedsteganographyconstrained} & \cellno & \cellno & \cellno & \cellno & \cellno & \cellno & \cellno & \cellno & \cellno & \cellno \\
2025 & \citet{10.1007/978-981-96-7005-5_12} & \cellno & \cellno & \cellyes & \cellno & \cellno & \cellno & \cellno & \cellno & \cellno & \cellno \\
2025 & \citet{11017758} & \cellno & \cellno & \cellno & \cellno & \cellno & \cellno & \cellno & \cellno & \cellno & \cellno \\
2025 & \citet{11008605} & \cellyes & \cellno & \cellno & \cellno & \cellno & \cellno & \cellno & \cellno & \cellno & \cellno \\
2025 & \citet{yan-murawaki-2025-addressing} & \cellyes & \cellyes & \cellno & \cellno & \cellno & \cellno & \cellno & \cellno & \cellno & \cellno \\
2025 & \citet{11023508} & \cellyes & \cellno & \cellno & \cellno & \cellno & \cellno & \cellno & \cellno & \cellno & \cellno \\
2025 & \citet{10804596} & \cellyes & \cellyes & \cellno & \cellno & \cellno & \cellno & \cellno & \cellno & \cellno & \cellno \\
2025 & \citet{qi2025stead} & \cellyes & \cellno & \cellno & \cellno & \cellno & \cellno & \cellno & \cellno & \cellno & \cellno \\
2025 & \citet{feng2025highcapacitysecuredisambiguationalgorithm} & \cellyes & \cellyes & \cellno & \cellno & \cellno & \cellno & \cellno & \cellno & \cellno & \cellno \\
2025 & \citet{perry2025robuststeganographylargelanguage} & \cellno & \cellno & \cellno & \cellno & \cellno & \cellno & \cellno & \cellno & \cellno & \cellyes \\
2025 & \citet{10888944} & \cellyes & \cellno & \cellno & \cellno & \cellyes & \cellno & \cellyes & \cellno & \cellno & \cellno \\

2025 & \citet{10888607} & \cellno & \cellno & \cellno & \cellno & \cellno & \cellno & \cellno & \cellno & \cellno & \cellno \\

2025 & \citet{sym17091416} & \cellyes & \cellno & \cellyes & \cellno & \cellno & \cellno & \cellyes & \cellno & \cellno & \cellyes \\
\hline

2026 & \citet{Xiang_Ou_He_Yang_Liu_2026} & \cellyes & \cellno & \cellno & \cellno & \cellno & \cellno & \cellno & \cellno & \cellno & \cellno \\
2026 & \citet{gao2026textsteganographydynamiccodebook} & \cellyes & \cellno & \cellyes & \cellno & \cellno & \cellno & \cellno & \cellyes & \cellno & \cellyes \\
2026 & \citet{yan2026efficientprovablysecurelinguistic} & \cellyes & \cellyes & \cellno & \cellno & \cellno & \cellno & \cellno & \cellyes & \cellno & \cellno \\
2026 & \citet{11329500} & \cellyes & \cellno & \cellno & \cellno & \cellno & \cellno & \cellno & \cellno & \cellno & \cellyes \\
2026 & \citet{11363394} & \cellyes & \cellno & \cellno & \cellno & \cellno & \cellno & \cellno & \cellno & \cellno & \cellno \\
2026 & \citet{wang2026retoksyncselfsynchronizingtokenizationdisambiguation} & \cellyes & \cellyes & \cellno & \cellno & \cellno & \cellno & \cellno & \cellno & \cellno & \cellyes \\
2026 & \citet{yan2026anchoredslidingwindowrobust} & \cellyes & \cellyes & \cellno & \cellyes & \cellyes & \cellno & \cellno & \cellyes & \cellno & \cellyes \\

\bottomrule[1.0pt]
\end{tabular}}
\caption{\rev{Adopted metrics (excluding steganalysis) to evaluate steganographic security, including the imperceptibility, and the text quality of stegotexts based on generative linguistic steganographic methods.}}
\label{table:GLS_evaluation_security}
\end{table*}
\rowcolors{0}{}{}

\rowcolors{2}{gray!10}{white}
\begin{table*}[!t]
\renewcommand{\arraystretch}{0.6}
\centering
\scalebox{0.63}{
\begin{tabular}{m{1cm}|m{5.1cm}|>{\centering\arraybackslash}m{2.5cm}|>{\centering\arraybackslash}m{3.5cm}|>{\centering\arraybackslash}m{2.5cm}|>{\centering\arraybackslash}m{3cm}|>{\centering\arraybackslash}m{2.5cm}}
\toprule[1.0pt]
\textbf{Year} & \textbf{Reference} &  \textbf{Bits per token} & \textbf{Entropy utilization} & \textbf{Embedding rate} & \textbf{Operation speed} & \textbf{Error rate} \\
\midrule[1.0pt]
2017 & \citet{fang-etal-2017-generating} &  \cellyes & \cellno & \cellno & \cellno & \cellno \\
2017 & \citet{10.1145/3082031.3083240} & \cellno & \cellno & \cellyes & \cellno & \cellno \\

\hline
2018 & \citet{yang2018automaticallygeneratesteganographictext} & \cellyes & \cellno & \cellyes & \cellno & \cellno \\

\hline
2019 & \citet{8470163} & \cellyes & \cellno & \cellyes & \cellyes & \cellno \\
2019 & \citet{ziegler-etal-2019-neural} & \cellyes & \cellno & \cellno & \cellno & \cellno \\
2019 & \citet{dai-cai-2019-towards} & \cellno & \cellno & \cellno & \cellno & \cellno \\
2019 & \citet{9023313} & \cellyes & \cellno & \cellno & \cellno & \cellno \\
2019 & \citet{Text_steganography_on_RNN-Generated-lyrics} & \cellyes & \cellno & \cellno & \cellno & \cellno \\

\hline
2020 & \citet{yang2020graphstegasemanticcontrollablesteganographic}  & \cellyes & \cellno & \cellyes & \cellno & \cellno \\
2020 & \citet{10.1007/978-3-030-43575-2_2} & \cellyes & \cellno & \cellno & \cellno & \cellno \\
2020 & \citet{shen-etal-2020-near}  & \cellyes & \cellno & \cellno & \cellyes & \cellno \\
2020 & \citet{doi:10.2352/ISSN.2470-1173.2020.4.MWSF-291}  & \cellyes & \cellno & \cellno & \cellno & \cellno \\
2020 & \citet{doi:10.1177/1550147720914257}  & \cellno & \cellno & \cellyes & \cellyes & \cellno \\
2020 & \citet{math8091558}  & \cellno & \cellno & \cellyes & \cellyes & \cellno \\

\hline
2021 & \citet{9193914}  & \cellyes & \cellno & \cellno & \cellno & \cellno \\
2021 & \citet{zhang-etal-2021-provably} & \cellyes & \cellno & \cellno & \cellno & \cellno \\
2021 & \citet{9353234} & \cellno & \cellno & \cellno & \cellno & \cellno \\
2021 & \citet{10.1145/3460120.3484550} & \cellyes & \cellno & \cellno & \cellyes & \cellno \\
2021 & \citet{9280343} & \cellno & \cellno & \cellno & \cellno & \cellno \\
2021 & \citet{10.1145/3418598} & \cellyes & \cellno & \cellno & \cellno & \cellno \\
2021 & \citet{yang2021generation} & \cellyes & \cellno & \cellno & \cellyes & \cellno \\
2021 & \citet{WANG20211375} & \cellyes & \cellno & \cellno & \cellyes & \cellno \\

\hline
2022 & \citet{9975301} &  \cellyes & \cellno & \cellyes & \cellno & \cellno \\
2022 & \citet{yu2022mts} & \cellyes & \cellno & \cellno & \cellyes & \cellno \\
2022 & \citet{9430708} & \cellyes & \cellno & \cellyes & \cellno & \cellno \\
2022 & \citet{9714779} & \cellno & \cellno & \cellno & \cellno & \cellno \\
2022 & \citet{varol2022lzw} & \cellyes & \cellno & \cellyes & \cellyes & \cellno \\
2022 & \citet{9864312}& \cellno & \cellno & \cellyes & \cellno & \cellno \\
2022 & \citet{nozaki-murawaki-2022-addressing} & \cellyes & \cellno & \cellno & \cellno & \cellyes \\

\hline
2023 & \citet{SUN2023157} &  \cellno & \cellno & \cellyes & \cellno & \cellno \\
2023 & \citet{10191218} & \cellyes & \cellno & \cellno & \cellno & \cellyes \\
2023 & \citet{10115433} & \cellyes & \cellno & \cellno & \cellno & \cellno \\
2023 & \citet{10179287} & \cellyes & \cellyes & \cellno & \cellyes & \cellno \\
2023 & \citet{10.1007/978-3-031-44696-2_62} & \cellyes & \cellno & \cellno & \cellno & \cellno \\
2023 & \citet{10095722} & \cellyes & \cellno & \cellno & \cellno & \cellno \\
2023 & \citet{witt2023perfectly} & \cellno & \cellyes & \cellno & \cellyes & \cellyes \\
2023 & \citet{10215094} & \cellyes & \cellno & \cellno & \cellyes & \cellyes \\
2023 & \citet{10191984} & \cellno & \cellno & \cellno & \cellno & \cellyes \\
2023 & \citet{10075392} & \cellyes & \cellno & \cellno & \cellno & \cellyes \\

\hline
2024 & \citet{10447545} & \cellno & \cellno & \cellno & \cellyes & \cellno \\
2024 & \citet{10.1145/3664647.3680562} & \cellyes & \cellno & \cellno & \cellno & \cellno \\
2024 & \citet{10374202} & \cellyes & \cellno & \cellno & \cellno & \cellno \\
2024 & \citet{lin-etal-2024-zero} & \cellyes & \cellno & \cellno & \cellno & \cellno \\
2024 & \citet{10669737} & \cellyes & \cellno & \cellno & \cellno & \cellno \\
2024 & \citet{10462497} &  \cellyes & \cellno & \cellno & \cellno & \cellno \\
2024 & \citet{10551807} & \cellyes & \cellno & \cellno & \cellyes & \cellyes \\
2024 & \citet{10446027} & \cellno & \cellno & \cellno & \cellno & \cellno \\
2024 & \citet{10687958} & \cellno & \cellno & \cellno & \cellno & \cellno \\
2024 & \citet{9778236} & \cellyes & \cellno & \cellyes & \cellyes & \cellno \\
2024 & \citet{10.1007/978-981-99-8148-9_44} & \cellyes & \cellno & \cellno & \cellno & \cellno \\
2024 & \citet{qin2024adlmstegauniversal} & \cellno & \cellno & \cellno & \cellno & \cellno \\
2024 & \citet{lin2024novel} & \cellyes & \cellno & \cellno & \cellno & \cellno \\

2024 & \citet{10831652} & \cellno & \cellno & \cellyes & \cellno & \cellno \\
2024 & \citet{10.1145/3664476.3670930} & \cellyes & \cellno & \cellno & \cellno & \cellyes \\
2024 & \citet{10.1007/978-981-99-8073-4_4} & \cellyes & \cellno & \cellno & \cellno & \cellno \\
2024 & \citet{10.1145/3658664.3659657} & \cellyes & \cellno & \cellno & \cellno & \cellno \\
2024 & \citet{huang2024odstegallmbasednearimperceptiblesteganography} & \cellyes & \cellno & \cellno & \cellno & \cellyes \\
2024 & \citet{10831370} & \cellyes & \cellno & \cellno & \cellno & \cellyes \\
2024 & \citet{bai2024semanticsteganographyframeworkrobust} & \cellyes & \cellno & \cellno & \cellno & \cellyes \\

\hline
2025 & \citet{WANG2025113101} &  \cellyes & \cellno & \cellno & \cellno & \cellno \\
2025 & \citet{liao2025framework} & \cellyes & \cellyes & \cellno & \cellyes & \cellno \\
2025 & \citet{WOZNIAK2025113960} & \cellyes & \cellno & \cellno & \cellno & \cellno \\
2025 & \citet{10888762} & \cellyes & \cellno & \cellno & \cellno & \cellno \\
2025 & \citet{10919187} & \cellno & \cellno & \cellno & \cellno & \cellno \\
2025 & \citet{app15179663} & \cellyes & \cellno & \cellno & \cellno & \cellno \\
2025 & \citet{technologies13070264} & \cellyes & \cellno & \cellno & \cellno & \cellno \\
2025 & \citet{10.1162/coli.a.22} & \cellyes & \cellno & \cellno & \cellno & \cellno \\
2025 & \citet{computers14050165} & \cellno & \cellno & \cellyes & \cellyes & \cellyes \\
2025 & \citet{zhou2025autostegaagentdrivenlifelongstrategy} & \cellyes & \cellno & \cellno & \cellno & \cellno \\
2025 & \citet{10919130} & \cellyes & \cellyes & \cellno & \cellyes & \cellno \\
2025 & \citet{pang2025frestegaplugandplaymethodboosting} & \cellyes & \cellno & \cellno & \cellyes & \cellno \\
2025 & \citet{wang2025sparsampefficientprovablysecure} & \cellyes & \cellyes & \cellno & \cellyes & \cellyes \\
2025 & \citet{bai2025shimmer} & \cellyes & \cellyes & \cellno & \cellyes & \cellno \\
2025 & \citet{NEURIPS2025_6c8e215a} & \cellyes & \cellno & \cellyes & \cellyes & \cellno \\
2025 & \citet{pang2025provablesecuresteganographybased} & \cellyes & \cellyes & \cellno & \cellyes & \cellno \\
2025 & \citet{huang2025relativelysecurellmbasedsteganographyconstrained} & \cellyes & \cellno & \cellno & \cellno & \cellno \\
2025 & \citet{10.1007/978-981-96-7005-5_12} & \cellyes & \cellno & \cellno & \cellyes & \cellyes \\
2025 & \citet{11017758} & \cellno & \cellno & \cellno & \cellno & \cellyes \\
2025 & \citet{11008605} & \cellyes & \cellno & \cellno & \cellno & \cellno \\
2025 & \citet{yan-murawaki-2025-addressing} & \cellyes & \cellno & \cellno & \cellyes & \cellyes \\
2025 & \citet{11023508} & \cellyes & \cellno & \cellno & \cellyes & \cellyes \\
2025 & \citet{10804596} & \cellyes & \cellyes & \cellno & \cellyes & \cellyes \\
2025 & \citet{qi2025stead} & \cellyes & \cellno & \cellno & \cellyes & \cellyes \\
2025 & \citet{feng2025highcapacitysecuredisambiguationalgorithm} & \cellyes & \cellno & \cellno & \cellno & \cellyes \\
2025 & \citet{perry2025robuststeganographylargelanguage} & \cellno & \cellno & \cellno & \cellno & \cellyes \\
2025 & \citet{10888944} & \cellyes & \cellno & \cellno & \cellno & \cellyes \\

2025 & \citet{10888607} & \cellno & \cellno & \cellno & \cellno & \cellyes \\

2025 & \citet{sym17091416} & \cellyes & \cellno & \cellno & \cellno & \cellyes \\
\hline

2026 & \citet{Xiang_Ou_He_Yang_Liu_2026} & \cellyes & \cellno & \cellyes & \cellyes & \cellyes \\
2026 & \citet{gao2026textsteganographydynamiccodebook} & \cellyes & \cellno & \cellno & \cellyes & \cellyes \\
2026 & \citet{yan2026efficientprovablysecurelinguistic} & \cellyes & \cellyes & \cellno & \cellyes & \cellno \\
2026 & \citet{11329500} & \cellyes & \cellno & \cellyes & \cellyes & \cellyes \\
2026 & \citet{11363394} & \cellyes & \cellno & \cellno & \cellno & \cellyes \\
2026 & \citet{wang2026retoksyncselfsynchronizingtokenizationdisambiguation} & \cellyes & \cellyes & \cellno & \cellyes & \cellyes \\
2026 & \citet{yan2026anchoredslidingwindowrobust} & \cellyes & \cellno & \cellno & \cellyes & \cellyes \\

\bottomrule[1.0pt]
\end{tabular}}
\caption{\rev{Adopted metrics to evaluate efficiency and robustness of the generative linguistic steganographic methods.}}
\label{table:GLS_evaluation_others}
\end{table*}
\rowcolors{0}{}{}

\end{document}